\pdfoutput=1

\documentclass{aa}
\usepackage{placeins}
\usepackage{multicol}
\usepackage{amsmath}
\usepackage{graphicx}
\usepackage{collcell}
\usepackage{txfonts}
\usepackage[colorlinks=true,linkcolor=blue,citecolor=blue,urlcolor=blue]{hyperref}
\usepackage{orcidlink}
\usepackage{booktabs}
\usepackage{threeparttable}
\usepackage{longtable}
\usepackage{array}
\usepackage{xcolor}
\renewcommand{\topfraction}{0.95}
\renewcommand{\dbltopfraction}{0.95}
\renewcommand{\textfraction}{0.05}
\renewcommand{\floatpagefraction}{0.95}
\renewcommand{\dblfloatpagefraction}{0.95}
\newcommand{\software}[1]{\par\addvspace{6pt}\noindent\textit{Software: }#1\par}
\newcommand{\facilities}[1]{\par\addvspace{6pt}\noindent\textit{Facilities: }#1\par}
\newenvironment{acknowledgments}{\begin{acknowledgements}}{\end{acknowledgements}}
\newcommand{\startlongtable}{}

\newenvironment{ManuscriptParameterTable}{}{}
\makeatletter
\renewcommand*{\aa@headfont}{\fontsize{7}{8}\selectfont}
\def\fps@figure{!tp}
\def\fps@table{!tp}
\newcommand{\ManuscriptTableInput}[1]{%
  \begin{table*}[!t]
  \let\onecolumngrid@push\relax
  \let\onecolumngrid@pop\relax
  \input{#1}%
  \end{table*}%
}
\newcommand{\restartappendixnumbering}{%
  \setcounter{table}{0}\setcounter{figure}{0}\setcounter{equation}{0}%
  \renewcommand{\thetable}{\thesection\arabic{table}}%
  \renewcommand{\thefigure}{\thesection\arabic{figure}}%
  \renewcommand{\theequation}{\thesection\arabic{equation}}%
}
\newcommand{\resetapptablenumbers}{%
  \setcounter{table}{0}\setcounter{figure}{0}\setcounter{equation}{0}%
}
\ExplSyntaxOn
\cs_new:Npn \ManuscriptColumnCount #1 { \tl_count:n {#1} }
\ExplSyntaxOff
\newenvironment{deluxetable*}[1]{%
  \begin{table*}[!tp]\centering
  \setlength{\extrarowheight}{1.5pt}%
  \def\aa@columns{#1}%
  \edef\aa@ncolumns{\ManuscriptColumnCount{#1}}%
  \let\aa@tablehead\@empty
}{\end{table*}}

\newcommand{\tablewidth}[1]{}

\newcommand{\tablehead}[1]{\def\aa@tablehead{#1}}
\newcommand{\colhead}[1]{\multicolumn{1}{c}{#1}}
\newcommand{\startdata}{%
  \expandafter\tabular\expandafter{\aa@columns}%
  \hline\hline
  \ifx\aa@tablehead\@empty\else\aa@tablehead\\\hline\fi
}

\newcommand{\sidehead}[1]{%
  \noalign{\vskip 0.1ex}%
  \multicolumn{\aa@ncolumns}{l}{#1}\\%
}

\makeatother

\renewcommand{\textfraction}{0.2}
\renewcommand{\topfraction}{0.8}
\renewcommand{\dbltopfraction}{0.8}
\renewcommand{\floatpagefraction}{0.7}
\renewcommand{\dblfloatpagefraction}{0.7}

\definecolor{ClassificationSingle}{HTML}{238B45}
\definecolor{ClassificationSBTwo}{HTML}{C62828}
\definecolor{ClassificationInconclusive}{HTML}{2166AC}

\makeatletter
\let\aog@Hy@Warning\Hy@Warning
\def\Hy@Warning#1{%
  \def\aog@temp{#1}%
  \def\aog@emptyanchor{Ignoring empty anchor}%
  \ifx\aog@temp\aog@emptyanchor
  \else
    \aog@Hy@Warning{#1}%
  \fi
}
\makeatother

\begin{document}

\title{Confirmation of 11 brown-dwarf companions to low-mass stars and an estimate of their intrinsic occurrence}

\author{
  Kamil K. Kalinowski\,\orcidlink{0000-0001-9825-7418}\inst{1}\email{kamil@phys.au.dk}\thanks{Corresponding author: Kamil K. Kalinowski\\\texttt{kamil@phys.au.dk}}
  \and Simon H. Albrecht\,\orcidlink{0000-0003-1762-8235}\inst{1}\email{albrecht@phys.au.dk}
  \and Marcus L. Marcussen\,\orcidlink{0000-0003-2173-0689}\inst{1}\email{marcus@phys.au.dk}
  \and Joshua N. Winn\,\orcidlink{0000-0002-4265-047X}\inst{2}\email{jnwinn@princeton.edu} \and Elise Koo\,\orcidlink{0009-0007-0740-0954}\inst{3}\email{e.j.m.koo@uva.nl}
  \and Mia S. Lundkvist\,\orcidlink{0000-0002-8661-2571}\inst{1}\email{lundkvist@phys.au.dk}
  \and Kevin C. Schlaufman\,\orcidlink{0000-0001-5761-6779}\inst{4}\email{kschlaufman@jhu.edu}
  \and Guðmundur Stefánsson\,\orcidlink{0000-0001-7409-5688}\inst{5}\email{gstefansson@schmidtsciences.org}
}
\institute{Department of Physics and Astronomy, Aarhus University, Ny Munkegade 120, DK-8000 Aarhus C, Denmark
  \and Department of Astrophysical Sciences, Princeton University, 4 Ivy Lane, Princeton, NJ 08544, USA \and Anton Pannekoek Institute for Astronomy, University of Amsterdam, Science Park 904, 1098 XH Amsterdam, The Netherlands
  \and William H. Miller III Department of Physics and Astronomy, Johns Hopkins University, 3400 N. Charles Street, Baltimore, MD 21218, USA
  \and Astrophysics \& Space Institute, Schmidt Sciences, 155 W 23rd St, New York, NY 10011, USA
}

\abstract{

Gaia astrometry has led to the identification of candidate substellar companions at AU-scale
separations around late-type stars. By performing NIRPS/HARPS spectroscopy of 53 of these candidates, we have validated 11 as substellar companions and identified 37 as false positives, leaving 5 inconclusive cases. The false positives are unresolved binary stars for which the photocentre exhibits a low-amplitude, substellar-like astrometric signal because of the partial cancellation of the stars' oppositely-directed orbital motion.
Combining these results with the literature, we perform a study of the intrinsic occurrence of substellar objects. After correcting for false positives and survey completeness, we infer an occurrence rate of $0.26^{+0.09}_{-0.07}\%$ for $20$--$80\,M_{\mathrm{Jup}}$ companions with orbital periods of $100$--$1000$ days around $0.4$--$0.6\,M_\odot$ stars. Brown dwarf companions appear to be less common than giant planets around M dwarfs, with an occurrence rate comparable to estimates for Sun-like stars.}


\date{}
\maketitle
\nolinenumbers

\section{Introduction}
\label{sec:introduction}

\begin{figure*}[t]
	\centering
	\setlength{\VSpaceBeforeFigCaption}{3.24pt}
	\includegraphics[width=\textwidth,trim=0bp 186bp 0bp 0bp,clip]{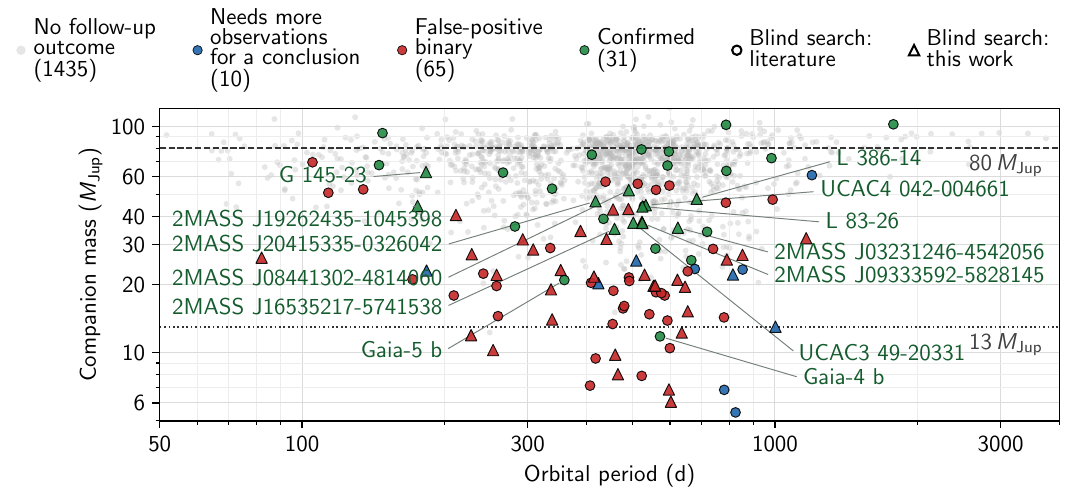}%

	\nointerlineskip
	\includegraphics[width=\textwidth,trim=31pt 0bp 0bp 48bp,clip]{graphics/spectroscopic_inventory_period_mass_wide.pdf}
	\caption{\textbf{Our sample in context.}
		Substellar companion candidates from the blind-search component \citep{Halbwachs+2023} of the Gaia DR3 non-single-star catalogue \citep{GaiaCollaborationArenou+2023} are shown as a function of the orbital period and
		companion mass. Classifications combine this work
		with published follow-up
		\citep{MarcussenAlbrecht2023,Unger2023,Winterhalder2024,
			SahlmannGomez2025,Stefansson2025,Barbato2026,MarcussenEtAl2026,Chevalier2023,Kounkel2021}.
            Green includes companions confirmed with spectrographs, red marks stellar-companion false positives, blue marks
		inconclusive outcomes, and grey marks candidates without an outcome. For the confirmed candidates we show the masses and periods from the follow-up papers (eg. based on RVs). For the remaining ones, the masses of host stars, and the companion masses inferred under the star–dark-companion scenario that follow, were computed with the synthetic mass–absolute-magnitude relation in the Gaia \(G\) band relation of \citet{ChulkovMalkov2022}, while the periods are retained from the Gaia DR3 orbital fits.}
	\label{fig:combined_vetting_sample}
\end{figure*}

The astrometric method for exoplanet detection is especially sensitive to massive planets on relatively wide orbits. Thus, it is complementary to radial-velocity (RV) and transit methods, which are more sensitive to planets with smaller orbits. The Gaia mission \citep{GaiaMission2016} is ultimately expected to deliver thousands of astrometric exoplanet candidates  \citep{Perryman2014,LammersWinn2026}, but the data are being released gradually. The third data release (DR3), which took place on June 13, 2022 \citep{GaiaDR3Summary2023,HollOrbit2023,GaiaCollaborationArenou+2023,MarcussenAlbrecht2023}, was based on the first 33 months (2.8 yr) of observations. It provided $\sim1500$ brown dwarf candidates and a smaller number of exoplanet candidates. %
Figure~\ref{fig:combined_vetting_sample} shows the inventory of DR3 substellar companion candidates and their follow-up classifications, based on both previous work and
the work described in this paper.

Gaia astrometric measurements of an unresolved system refer to the system's center-of-light or ``photocentre.'' For a star-planet system, where the planet's luminosity is negligible, this is simply the star's sky position. However, in systems of multiple luminous sources, the photocentre is approximately the flux-weighted mean position. For binary stars whose components have similar masses and luminosities, the stars' oppositely directed motion around the center of mass causes the photocentre motion to have a smaller amplitude than the motion of either star by itself. This small amplitude
can be mistaken for the motion of a single star induced
by a substellar companion. Indeed, binary stars have proven to be a major source of astrophysical false positives in astrometric searches for substellar companions \citep{VanDeKamp1975,HollOrbit2023}.

High resolution spectroscopy can reveal false positives, while radial velocity (RV) measurements allow us to test the astrometric orbital solutions. The community has therefore undertaken an extensive spectroscopic follow-up of Gaia DR3 substellar-companion candidates using high-resolution spectrographs.
These efforts confirmed and characterized some of Gaia’s first astrometrically
detected planets, while revealing substantial contamination of the candidate
sample with unresolved stellar binaries \citep[e.g.,][]{Winn2022,
	MarcussenAlbrecht2023,Sozzetti2023,Stefansson2025,Barbato2026,MarcussenEtAl2026,Venner2026,Pinamonti2026}.
Some of these works  also uncovered cases in which the RV signals implied by the
published Gaia DR3 astrometric solutions were inconsistent with the observed RVs. Even apparent agreement can be misleading
when a binary's spectral lines remain blended:
the astrometry and RVs of HD~68638 were consistent
with a brown-dwarf companion \citep{Unger2023}, whereas
subtle time variations in its apparent stellar line
profile revealed its stellar binary nature
\citep{MarcussenAlbrecht2023}.

Brown-dwarf companions are much less common than planetary
or stellar companions on close orbits around Sun-like stars,
a deficit known as the brown-dwarf desert \citep{GretherLineweaver2006}; recent estimates place their
occurrence within 10~au below 1\% \citep{Santerne2016, Unger2023}. Measuring their frequency around lower-mass stars tests
whether this scarcity depends on host mass. An initial Gaia DR3 estimate suggested a brown-dwarf
occurrence of approximately 0.3\% around M dwarfs within
100~pc for periods below approximately 1000~d, however this was done with significant assumptions of complete detection and no false positives
\citep{GaiaCollaborationArenou+2023}, motivating a measurement that
accounts for both incompleteness and contamination.

Despite the follow-up efforts undertaken so far, most Gaia DR3
substellar companion candidates remain unclassified.
Here we extend the prior work with a NIRPS and HARPS
spectroscopic survey of 53 candidates around low-mass stars. To help identify binaries with blended
spectral lines, we develop a vetting approach that goes beyond comparing observed RVs with those predicted by astrometric orbits, and visual inspection of spectral cross-correlation functions. We model the CCF as a single star and as a blended binary, allowing the evidence for these hypotheses to be compared. For companions we confirm, as a next step we combine the RV measurements with Gaia
astrometry to refine their orbits. Finally, we combine our classifications with published
follow-up results to estimate the occurrence of
brown dwarfs at intermediate orbital periods around
low-mass stars, accounting for false positives and
survey incompleteness.

\section{DR3 Candidates and data}
\label{sec:data}

\subsection{Sample}
\label{subsec:data_sources}

We selected late-type dwarfs with companion
candidates for which the Gaia DR3 catalog mass is compatible with a brown dwarf or exoplanet (lower bounds on the mass from 5.6 to
47$\,M_{\rm Jup}$) from \texttt{gaiadr3.nss\_two\_body\_orbit}
 using the masses from \texttt{gaiadr3.binary\_masses} \citep{GaiaCollaborationArenou+2023}. Our sample comprises 51 candidates from the Gaia DR3
blind search and two from the targeted search:
Ross~778 and L~499-75. Here, ``blind search'' refers to the general Gaia DR3
astrometric-binary processing of \citet{Halbwachs+2023} with a uniform selection function.
This is
distinct from the separate targeted search applied to
an external list of preselected stars \citep{HollOrbit2023}. Table \ref{tab:nirps_observations} shows the sample which is also illustrated in Figure~\ref{fig:combined_vetting_sample}.

\subsection{Ground-based spectra}
\label{subsec:spectra}

We obtained spectra with the Near-InfraRed Planet Searcher \citep[NIRPS;][]{Bouchy+2017, Wildi+2017} and the High Accuracy Radial velocity Planet Searcher \citep[HARPS;][]{Mayor+2003} on the ESO 3.6-m telescope at La Silla Observatory (ESO programmes 114.2758, 115.27ZE, 116.296E and 117.2AJC). We observed with both spectrographs at the same time, NIRPS in the near-infrared at a resolving power $R \simeq 70{,}000$ (high-efficiency mode) and HARPS in the visible at $R \simeq 115{,}000$ (high-accuracy mode).
We reduced the spectra of both spectrographs with their standard ESO pipelines (\texttt{NIRPS DRS v3.3.0}, \texttt{HARPS DRS v3.8}). We measured the NIRPS RVs with the line-by-line method \citep{Artigau2022}, which measures the Doppler shift of each spectral line against a template spectrum.
For HARPS, we used serval code \citep{Zechmeister2018SERVAL} that matches the spectrum in each order to a template to extract the RVs. We excluded from the orbital fits the 13 HARPS exposures with RV uncertainty \(>200\,\mathrm{m\,s^{-1}}\).
We also excluded HARPS observations obtained after the
May--June 2026 instrument upgrade, since too few post-upgrade
epochs were available to constrain the additional RV
zero-point offset required to combine them with the
pre-upgrade measurements.
We treated RVs from both spectrographs as independent measurements. The full set of HARPS and NIRPS RVs and associated uncertainties are listed in Table~\ref{tab:radial_velocities}.\footnote{The full version of Table~\ref{tab:radial_velocities} will be available in machine-readable form at the CDS upon journal publication.}

\section{Vetting and system characterization}
\label{sec:methods}

Here we determine which of our 53 candidate systems are false positives, before we fit a Keplerian orbit to the RV data and Gaia DR3 astrometric orbital constraints for the 11 confirmed companions. We then determine the masses of the host stars and their companions in the confirmed systems.

\subsection{A forward model to identify unresolved binaries}
\label{sec:forwardmodel}

\subsubsection{Photocentre degeneracy}

For an unresolved binary, Gaia measures the motion of
the photocentre---the mean position of the two components
weighted by their Gaia $G$-band fluxes.
\begin{equation}
	\beta_G=F_{G,2}/(F_{G,1}+F_{G,2})
	\label{lightfrac}
\end{equation}
is the Gaia-band light fraction of the secondary, where $f_M=M_2/(M_1+M_2)$ is its mass fraction and $a$ is the relative orbital semimajor axis. The barycentric semimajor axes are
$a_1=f_Ma$ and $a_2=(1-f_M)a$, while the
angular photocentre semimajor axis is
\begin{equation}
	a_{\text{phot}}=\varpi\frac{a}{\mathrm{au}}
\left|f_M-\beta_G\right|,
	\label{a0mod}
\end{equation}
where $\varpi$ is the parallax \citep{2020MNRAS.495..321P}.

The same small photocentre orbit can arise from a faint, low-mass companion scenario, ($\beta_G\simeq0$, $f_M\ll1$), for which
the photocentre coincides with the primary, or a near-equal-mass, near-equal-flux binary scenario ($f_M\simeq\beta_G\simeq1/2$).
The low-mass companion scenario predicts that the source's spectrum will contain a single set of spectral lines, and a CCF that stays consistent in shape as its RV varies.
The nearly-equal binary scenario predicts that two stellar spectra will be superposed, causing the CCF to have two separate peaks that exhibit opposing RV shifts;
if the shifts are small, the CCF may appear single-peaked
but with a shape that exhibits time-variable distortions.

\subsubsection{Comparison between observed and modelled CCFs}
\label{subsubsec:ccf_profile_comparison}

We construct synthetic CCFs for stellar binary and star--dark-companion scenarios, constrained by the Gaia DR3 astrometric orbital solution and the system photometry. We compare their predicted profile shapes and temporal variations with the observed CCFs to assess whether either scenario is compatible with the data.

Its detailed implementation is described in
Appendix~\ref{app:binary_ccf_model}.
In short, we use each system's absolute $K_s$-band magnitude and
the empirical mass--luminosity relation of \citet{Mann2019}
to constrain the stellar masses. In the binary scenario,
the two stars must together reproduce the observed
magnitude. In the star--dark-companion model, all of
the light is attributed to the host star.
A solar-metallicity, 5-Gyr isochrone
\citep{Baraffe2015} supplies their temperatures and radii.
These and the surface gravities calculated from their
masses and radii select the PHOENIX synthetic spectra
\citep{Husser2013}.
For binary scenario, the radii and spectra determine
the Gaia-band light fraction $\beta_G$. Holding the DR3 period, eccentricity, phase, orbital
orientation and parallax fixed, we select the component
masses that satisfy the $K_s$ constraint and best
match the measured photocentre semimajor axis. The selected masses and DR3 orbit predict the binary component
velocities at each observing epoch.

We construct synthetic CCFs using the correlation masks from the NIRPS pipeline.
The star--dark-companion model contains one stellar PHOENIX spectrum;
the luminous model combines two PHOENIX spectra. Their relative
CCF contributions are determined by the synthetic spectra
and the squared stellar radii.
We fit the projected rotational velocity $v_{\mathrm{rot}}$
and an additional Gaussian velocity width $\sigma$
for each luminous component, together with a common
velocity offset $\gamma_{\rm CCF}$ and an overall CCF amplitude.
These fitted parameters are shared across all observing nights.
Only the CCF baseline, modelled as a linear function of velocity with a separate intercept and slope for each night, varies nightly.

For systems in which the predicted binary components would remain blended, we additionally assess predictive performance of models of both scenarios by refitting each model after omitting one observing night and calculating the residual RMS of the omitted CCF (Appendix~\ref{app:loo}). We define $R_k$ as the ratio of the held-out CCF residual RMS for the  double-lined spectroscopic binary (SB2) model to that for the star--dark-companion model on omitted night $k$; stellar binary is favoured when $R_k<1$.

\begin{figure*}[t]
	\centering
	\setlength{\VSpaceBeforeFigCaption}{-6.79pt}
	\includegraphics[width=\textwidth]{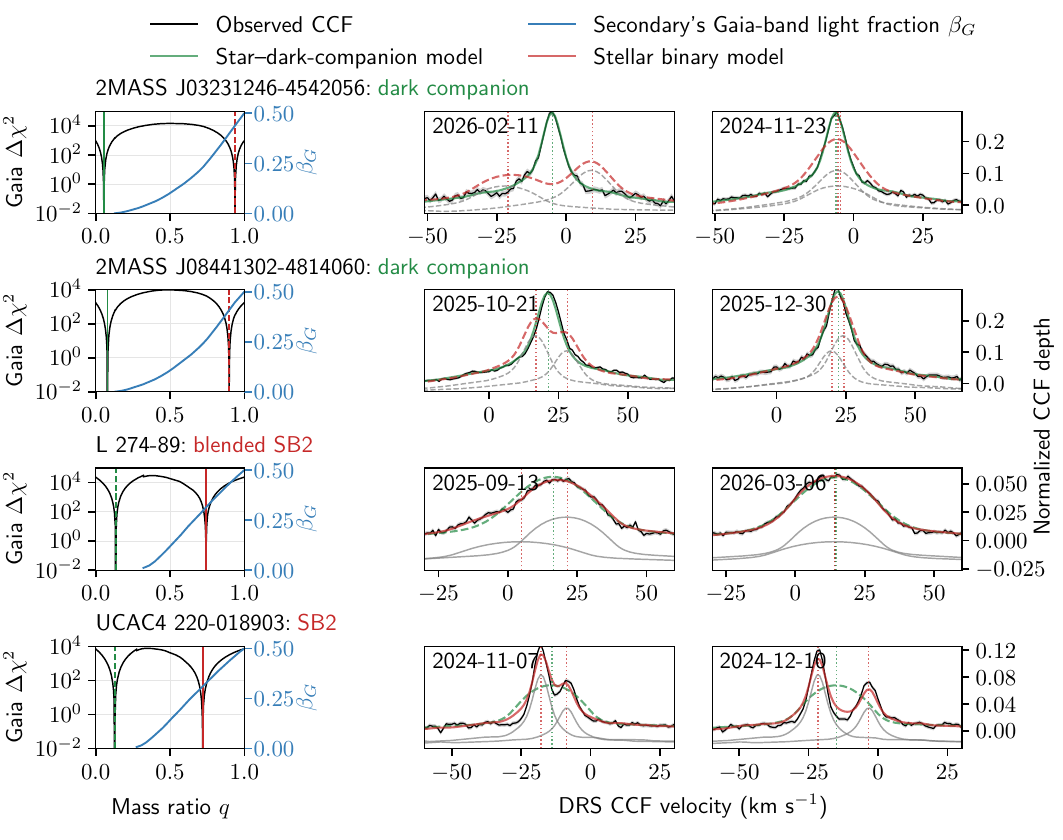}
	\caption{\textbf{Forward modelling of the CCF time series distinguishes dark companions from stellar binaries even when the component lines remain blended.}
	Observed CCFs of 2MASS~J03231246-4542056 and 2MASS~J08441302-4814060 are more consistent with a star--dark-companion model, while observations of L~274-89 and UCAC4~220-018903 are more consistent with stellar binary model.
	\textit{Left column:}
	$\Delta\chi^2_{\rm G}(q)
		\equiv\chi^2_{\rm G}(q)-\min_q\chi^2_{\rm G}(q)$
	measures the mismatch between the
	Gaia DR3 orbit and the photocentre orbit predicted at
	mass ratio $q$ (Appendix~\ref{app:binary_ccf_model}). The 2 minima identify the degenerate solutions: a dark companion orbiting a single star and a luminous binary.
	The blue curves show the Gaia-band light fraction $\beta_G$.
	\textit{CCF panels:} Black curves and grey bands show the observed profiles
	and uncertainties.
	Red and green curves show the luminous-binary synthetic CCF model and
	the star--dark-companion model, respectively.
	Two representative nights are displayed per target; all retained nights
	enter the fits (Appendix~\ref{app:loo}).
	Solid and dashed curves denote the favoured and alternative scenarios,
	respectively, also for the solution markers in the left column.
	Dotted vertical lines mark the predicted component velocities.
	Grey curves show the individual SB2 components, offset by $-0.02$ in CCF
	depth and following the same solid/dashed convention.}
	\label{fig:l27489_binary_ccf}
\end{figure*}

\begin{table*}[p]
\normalsize
\centering
\begingroup
\let\startlongtable\relax
\begin{ManuscriptParameterTable}
\begingroup
\newcommand{\OrbitTableMeasure}[2]{%
  \setbox0=\vbox{\normalsize#2}%
  \expandafter\edef\csname OrbitTableWidth#1\endcsname{\the\wd0}%
}
\newcommand{\OrbitTableEstimate}[4]{%
  \makebox[\csname OrbitTableWidthValue#1\endcsname][r]{$#2$}%
  \makebox[\csname OrbitTableWidthError#1\endcsname][l]{${}_{-#3}^{+#4}$}%
}
\OrbitTableMeasure{Value2}{\hbox{$0.05236$}\hbox{$0.07021$}\hbox{$0.3501$}\hbox{$0.4463$}\hbox{$0.66$}\hbox{$0.7278$}\hbox{$1.0910$}\hbox{$1.1193$}\hbox{$1.2$}\hbox{$1.339$}\hbox{$1.354$}\hbox{$1.85$}\hbox{$10.633$}\hbox{$10.642$}\hbox{$17.487$}\hbox{$190.4$}\hbox{$2.2079$}\hbox{$2.65$}\hbox{$220.74$}\hbox{$241.1$}\hbox{$266.8$}\hbox{$34.5919$}\hbox{$34.6086$}\hbox{$346.8$}\hbox{$37.39$}\hbox{$44.83$}\hbox{$5.973$}\hbox{$524.0$}\hbox{$533.19$}\hbox{$7.4$}\hbox{$75.7$}\hbox{$85$}\hbox{$90.1$}\hbox{$95.1$}}
\OrbitTableMeasure{Error2}{\hbox{${}_{-0.00070}^{+0.00077}$}\hbox{${}_{-0.00095}^{+0.0011}$}\hbox{${}_{-0.0014}^{+0.0015}$}\hbox{${}_{-0.0035}^{+0.0036}$}\hbox{${}_{-0.0042}^{+0.0046}$}\hbox{${}_{-0.0048}^{+0.0050}$}\hbox{${}_{-0.0067}^{+0.0066}$}\hbox{${}_{-0.0088}^{+0.0088}$}\hbox{${}_{-0.0090}^{+0.0091}$}\hbox{${}_{-0.0093}^{+0.0091}$}\hbox{${}_{-0.013}^{+0.013}$}\hbox{${}_{-0.014}^{+0.018}$}\hbox{${}_{-0.022}^{+0.021}$}\hbox{${}_{-0.023}^{+0.023}$}\hbox{${}_{-0.039}^{+0.041}$}\hbox{${}_{-0.29}^{+0.29}$}\hbox{${}_{-0.37}^{+0.35}$}\hbox{${}_{-0.52}^{+1.0}$}\hbox{${}_{-0.62}^{+7.0}$}\hbox{${}_{-0.74}^{+0.78}$}\hbox{${}_{-0.85}^{+1.4}$}\hbox{${}_{-0.90}^{+0.97}$}\hbox{${}_{-1.2}^{+19}$}\hbox{${}_{-1.3}^{+1.1}$}\hbox{${}_{-1.5}^{+1.5}$}\hbox{${}_{-1.6}^{+1.6}$}\hbox{${}_{-17}^{+22}$}\hbox{${}_{-2.6}^{+2.6}$}\hbox{${}_{-3.9}^{+3.3}$}\hbox{${}_{-5.2}^{+5.2}$}\hbox{${}_{-5.9}^{+4.8}$}\hbox{${}_{-6.0}^{+6.0}$}\hbox{${}_{-7.0}^{+5.3}$}}
\OrbitTableMeasure{Value3}{\hbox{$-6.144$}\hbox{$-6.185$}\hbox{$0.0526$}\hbox{$0.06411$}\hbox{$0.19$}\hbox{$0.24$}\hbox{$0.425$}\hbox{$0.454$}\hbox{$0.4618$}\hbox{$1.0178$}\hbox{$1.150$}\hbox{$1.235$}\hbox{$1.31$}\hbox{$1.4088$}\hbox{$1.42$}\hbox{$1.5923$}\hbox{$115.61$}\hbox{$150.5$}\hbox{$157$}\hbox{$21.593$}\hbox{$242.2$}\hbox{$27$}\hbox{$316.3$}\hbox{$35.7$}\hbox{$37.52$}\hbox{$44.8$}\hbox{$501.71$}\hbox{$6.548$}\hbox{$623.1$}\hbox{$64.7$}\hbox{$67.6546$}\hbox{$67.6579$}\hbox{$93.1$}\hbox{$99.7$}}
\OrbitTableMeasure{Error3}{\hbox{${}_{-0.00052}^{+0.00054}$}\hbox{${}_{-0.0015}^{+0.0015}$}\hbox{${}_{-0.0018}^{+0.0018}$}\hbox{${}_{-0.0023}^{+0.0023}$}\hbox{${}_{-0.0032}^{+0.0035}$}\hbox{${}_{-0.0033}^{+0.0033}$}\hbox{${}_{-0.0081}^{+0.0081}$}\hbox{${}_{-0.010}^{+0.010}$}\hbox{${}_{-0.012}^{+0.012}$}\hbox{${}_{-0.015}^{+0.017}$}\hbox{${}_{-0.016}^{+0.019}$}\hbox{${}_{-0.017}^{+0.017}$}\hbox{${}_{-0.019}^{+0.018}$}\hbox{${}_{-0.025}^{+0.027}$}\hbox{${}_{-0.037}^{+0.037}$}\hbox{${}_{-0.17}^{+1.3}$}\hbox{${}_{-0.21}^{+1.9}$}\hbox{${}_{-0.23}^{+0.21}$}\hbox{${}_{-0.23}^{+0.23}$}\hbox{${}_{-0.23}^{+0.56}$}\hbox{${}_{-0.31}^{+0.70}$}\hbox{${}_{-0.60}^{+0.61}$}\hbox{${}_{-1.1}^{+1.1}$}\hbox{${}_{-1.2}^{+1.2}$}\hbox{${}_{-1.3}^{+1.3}$}\hbox{${}_{-2.4}^{+2.3}$}\hbox{${}_{-25}^{+32}$}\hbox{${}_{-27}^{+30}$}\hbox{${}_{-3.0}^{+3.1}$}\hbox{${}_{-4.0}^{+3.9}$}\hbox{${}_{-4.3}^{+4.4}$}}
\OrbitTableMeasure{Value4}{\hbox{$-142.367$}\hbox{$-142.414$}\hbox{$0.0766$}\hbox{$0.0812$}\hbox{$0.60$}\hbox{$0.617$}\hbox{$0.626$}\hbox{$0.6911$}\hbox{$1.0235$}\hbox{$1.0552$}\hbox{$1.08$}\hbox{$1.36$}\hbox{$1.481$}\hbox{$12$}\hbox{$152.1$}\hbox{$158.9$}\hbox{$17.836$}\hbox{$191.2$}\hbox{$2.051$}\hbox{$26.4702$}\hbox{$26.4760$}\hbox{$293.1$}\hbox{$3.2$}\hbox{$311.36$}\hbox{$4.7$}\hbox{$44.2$}\hbox{$490.71$}\hbox{$50$}\hbox{$52.2$}\hbox{$524.3$}\hbox{$59.4$}\hbox{$7.624$}\hbox{$84.0$}\hbox{$90.5$}}
\OrbitTableMeasure{Error4}{\hbox{${}_{-0.0025}^{+0.0026}$}\hbox{${}_{-0.0057}^{+0.0054}$}\hbox{${}_{-0.0074}^{+0.0068}$}\hbox{${}_{-0.0084}^{+0.0084}$}\hbox{${}_{-0.0088}^{+0.0088}$}\hbox{${}_{-0.010}^{+0.0075}$}\hbox{${}_{-0.011}^{+0.0096}$}\hbox{${}_{-0.018}^{+0.017}$}\hbox{${}_{-0.036}^{+0.037}$}\hbox{${}_{-0.037}^{+0.031}$}\hbox{${}_{-0.041}^{+0.042}$}\hbox{${}_{-0.045}^{+0.047}$}\hbox{${}_{-0.047}^{+0.044}$}\hbox{${}_{-0.066}^{+0.054}$}\hbox{${}_{-0.077}^{+0.076}$}\hbox{${}_{-0.24}^{+0.28}$}\hbox{${}_{-0.27}^{+0.68}$}\hbox{${}_{-0.53}^{+0.66}$}\hbox{${}_{-0.57}^{+8.6}$}\hbox{${}_{-0.83}^{+0.86}$}\hbox{${}_{-1.1}^{+1.2}$}\hbox{${}_{-1.3}^{+2.5}$}\hbox{${}_{-1.6}^{+1.7}$}\hbox{${}_{-1.9}^{+2.1}$}\hbox{${}_{-11}^{+17}$}\hbox{${}_{-2.8}^{+3.0}$}\hbox{${}_{-3.1}^{+12}$}\hbox{${}_{-3.7}^{+3.5}$}\hbox{${}_{-33}^{+24}$}\hbox{${}_{-4.2}^{+4.1}$}\hbox{${}_{-4.3}^{+4.6}$}\hbox{${}_{-5.0}^{+5.5}$}\hbox{${}_{-5.2}^{+5.4}$}\hbox{${}_{-5.5}^{+5.4}$}}
\OrbitTableMeasure{Value5}{\hbox{$-10.1598$}\hbox{$-10.1726$}\hbox{$0.07401$}\hbox{$0.1276$}\hbox{$0.3150$}\hbox{$0.4252$}\hbox{$0.55$}\hbox{$0.57$}\hbox{$0.6832$}\hbox{$0.8387$}\hbox{$0.9226$}\hbox{$1.58$}\hbox{$1.5913$}\hbox{$101.7$}\hbox{$116.7$}\hbox{$12.285$}\hbox{$132.3$}\hbox{$15.3$}\hbox{$155.99$}\hbox{$175.47$}\hbox{$2.100$}\hbox{$2.386$}\hbox{$2.49$}\hbox{$235.9$}\hbox{$255.3$}\hbox{$26.6$}\hbox{$26.69$}\hbox{$30.2221$}\hbox{$30.2243$}\hbox{$316.59$}\hbox{$38.698$}\hbox{$417.35$}\hbox{$44.5$}\hbox{$46.63$}}
\OrbitTableMeasure{Error5}{\hbox{${}_{-0.00081}^{+0.00084}$}\hbox{${}_{-0.0034}^{+0.0034}$}\hbox{${}_{-0.0039}^{+0.0039}$}\hbox{${}_{-0.0040}^{+0.0040}$}\hbox{${}_{-0.0050}^{+0.0049}$}\hbox{${}_{-0.0053}^{+0.0054}$}\hbox{${}_{-0.0062}^{+0.0067}$}\hbox{${}_{-0.0069}^{+0.0078}$}\hbox{${}_{-0.0074}^{+0.0072}$}\hbox{${}_{-0.0074}^{+0.0073}$}\hbox{${}_{-0.0090}^{+0.0088}$}\hbox{${}_{-0.0091}^{+0.0092}$}\hbox{${}_{-0.023}^{+0.023}$}\hbox{${}_{-0.024}^{+0.026}$}\hbox{${}_{-0.064}^{+0.063}$}\hbox{${}_{-0.065}^{+0.065}$}\hbox{${}_{-0.12}^{+0.12}$}\hbox{${}_{-0.32}^{+0.30}$}\hbox{${}_{-0.39}^{+0.80}$}\hbox{${}_{-0.40}^{+0.41}$}\hbox{${}_{-0.51}^{+7.9}$}\hbox{${}_{-0.53}^{+7.5}$}\hbox{${}_{-0.67}^{+1.2}$}\hbox{${}_{-0.78}^{+0.72}$}\hbox{${}_{-0.84}^{+0.86}$}\hbox{${}_{-0.86}^{+0.92}$}\hbox{${}_{-1.5}^{+1.5}$}\hbox{${}_{-1.6}^{+1.5}$}\hbox{${}_{-1.6}^{+1.7}$}\hbox{${}_{-2.3}^{+2.2}$}\hbox{${}_{-5.2}^{+6.4}$}\hbox{${}_{-5.3}^{+5.3}$}\hbox{${}_{-5.7}^{+5.5}$}\hbox{${}_{-6.3}^{+8.3}$}}
\OrbitTableMeasure{Value6}{\hbox{$-16.0651$}\hbox{$-16.0691$}\hbox{$0.08168$}\hbox{$0.1280$}\hbox{$0.23$}\hbox{$0.2857$}\hbox{$0.4070$}\hbox{$0.4894$}\hbox{$1.145$}\hbox{$1.2504$}\hbox{$1.46$}\hbox{$1.6$}\hbox{$1.7805$}\hbox{$134.8$}\hbox{$17.5$}\hbox{$18.269$}\hbox{$182.818$}\hbox{$2.7194$}\hbox{$2.758$}\hbox{$21.3$}\hbox{$212.4$}\hbox{$219.66$}\hbox{$243.91$}\hbox{$26.628$}\hbox{$27.10$}\hbox{$3.31$}\hbox{$302.79$}\hbox{$35$}\hbox{$4.6035$}\hbox{$4.613$}\hbox{$47.83$}\hbox{$62.7$}\hbox{$683.01$}\hbox{$90.86$}}
\OrbitTableMeasure{Error6}{\hbox{${}_{-0.00066}^{+0.00067}$}\hbox{${}_{-0.0017}^{+0.0017}$}\hbox{${}_{-0.0024}^{+0.0024}$}\hbox{${}_{-0.0032}^{+0.0032}$}\hbox{${}_{-0.0038}^{+0.0038}$}\hbox{${}_{-0.0039}^{+0.0030}$}\hbox{${}_{-0.0045}^{+0.0045}$}\hbox{${}_{-0.0051}^{+0.0052}$}\hbox{${}_{-0.0060}^{+0.0063}$}\hbox{${}_{-0.0062}^{+0.0062}$}\hbox{${}_{-0.0098}^{+0.0098}$}\hbox{${}_{-0.010}^{+0.010}$}\hbox{${}_{-0.014}^{+0.012}$}\hbox{${}_{-0.023}^{+0.022}$}\hbox{${}_{-0.047}^{+0.047}$}\hbox{${}_{-0.049}^{+0.049}$}\hbox{${}_{-0.096}^{+0.097}$}\hbox{${}_{-0.20}^{+1.8}$}\hbox{${}_{-0.26}^{+0.25}$}\hbox{${}_{-0.26}^{+0.26}$}\hbox{${}_{-0.33}^{+0.72}$}\hbox{${}_{-0.52}^{+0.53}$}\hbox{${}_{-0.70}^{+0.68}$}\hbox{${}_{-0.74}^{+0.75}$}\hbox{${}_{-0.75}^{+0.76}$}\hbox{${}_{-0.82}^{+0.80}$}\hbox{${}_{-0.89}^{+1.5}$}\hbox{${}_{-1.6}^{+19}$}\hbox{${}_{-10}^{+14}$}\hbox{${}_{-2.6}^{+2.3}$}\hbox{${}_{-2.9}^{+3.2}$}\hbox{${}_{-3.0}^{+3.1}$}\hbox{${}_{-3.6}^{+4.5}$}\hbox{${}_{-3.8}^{+3.8}$}}
\OrbitTableMeasure{Value1}{\hbox{$-7.326$}\hbox{$-7.3327$}\hbox{$0.0789$}\hbox{$0.847$}\hbox{$0.87$}\hbox{$0.8751$}\hbox{$1.271$}\hbox{$1.58$}\hbox{$124.3$}\hbox{$18.423$}\hbox{$2.28$}\hbox{$2.5$}\hbox{$221.4$}\hbox{$31.1$}\hbox{$35.3$}\hbox{$457.72$}\hbox{$85.2$}}
\OrbitTableMeasure{Error1}{\hbox{${}_{-0.0038}^{+0.0052}$}\hbox{${}_{-0.0073}^{+0.0073}$}\hbox{${}_{-0.012}^{+0.0089}$}\hbox{${}_{-0.013}^{+0.010}$}\hbox{${}_{-0.018}^{+0.024}$}\hbox{${}_{-0.055}^{+0.053}$}\hbox{${}_{-0.060}^{+0.084}$}\hbox{${}_{-0.16}^{+0.30}$}\hbox{${}_{-0.73}^{+1.5}$}\hbox{${}_{-0.83}^{+9.0}$}\hbox{${}_{-1.2}^{+0.98}$}\hbox{${}_{-1.7}^{+2.4}$}\hbox{${}_{-1.9}^{+2.6}$}\hbox{${}_{-2.5}^{+14}$}\hbox{${}_{-3.2}^{+1.9}$}\hbox{${}_{-5.8}^{+7.5}$}\hbox{${}_{-7.3}^{+5.9}$}}
\setlength{\tabcolsep}{3pt}
\renewcommand{\arraystretch}{1.25}
\caption{Adopted joint orbital fits and their priors.\label{tab:hybrid_selected_fits}}
\begin{tabular}{@{}llcccccc@{}}
\hline\hline
\colhead{Parameter} & \colhead{Prior} & \colhead{\shortstack{2MASS\\J09333592\\-5828145}} & \colhead{\shortstack{2MASS\\J03231246\\-4542056}} & \colhead{\shortstack{2MASS\\J08441302\\-4814060}} & \colhead{\shortstack{2MASS\\J20415335\\-0326042}} & \colhead{G~145-23} & \colhead{} \\
\multicolumn{8}{l}{\textit{Orbital parameters}}\\
$P$ (d) & Flat, $>0$\hspace{4pt} & \OrbitTableEstimate{2}{524.0}{1.3}{1.1} & \OrbitTableEstimate{3}{623.1}{1.1}{1.1} & \OrbitTableEstimate{4}{490.71}{0.83}{0.86} & \OrbitTableEstimate{5}{417.35}{0.32}{0.30} & \OrbitTableEstimate{6}{182.818}{0.096}{0.097} &  \\
$M_0$ (J2016) (deg) & $\mathcal{U}(-3600,3600^\circ)$\hspace{4pt} & \OrbitTableEstimate{2}{241.1}{5.9}{4.8} & \OrbitTableEstimate{3}{44.8}{4.3}{4.4} & \OrbitTableEstimate{4}{293.1}{4.3}{4.6} & \OrbitTableEstimate{5}{255.3}{2.3}{2.2} & \OrbitTableEstimate{6}{21.3}{3.8}{3.8} &  \\
$e$ & $\mathcal{U}(0,1)^a$\hspace{4pt} & \OrbitTableEstimate{2}{0.4463}{0.0090}{0.0091} & \OrbitTableEstimate{3}{0.454}{0.016}{0.019} & \OrbitTableEstimate{4}{0.626}{0.066}{0.054} & \OrbitTableEstimate{5}{0.6832}{0.0053}{0.0054} & \OrbitTableEstimate{6}{0.2857}{0.0039}{0.0030} &  \\
$\omega_1$ (deg) & Uniform on a circle$^a$\hspace{4pt} & \OrbitTableEstimate{2}{266.8}{1.5}{1.5} & \OrbitTableEstimate{3}{316.3}{2.4}{2.3} & \OrbitTableEstimate{4}{191.2}{1.9}{2.1} & \OrbitTableEstimate{5}{155.99}{0.40}{0.41} & \OrbitTableEstimate{6}{243.91}{0.82}{0.80} &  \\
$\alpha$ (mas) & Flat, $>0$\hspace{4pt} & \OrbitTableEstimate{2}{0.3501}{0.0042}{0.0046} & \OrbitTableEstimate{3}{0.425}{0.012}{0.012} & \OrbitTableEstimate{4}{0.617}{0.047}{0.044} & \OrbitTableEstimate{5}{0.8387}{0.0062}{0.0067} & \OrbitTableEstimate{6}{1.145}{0.047}{0.047} &  \\
$i$ (deg) & \shortstack[l]{$p(i)\propto\sin i$\\$0^\circ<i<180^\circ$}\hspace{4pt} & \OrbitTableEstimate{2}{90.1}{5.2}{5.2} & \OrbitTableEstimate{3}{64.7}{3.0}{3.1} & \OrbitTableEstimate{4}{152.1}{3.7}{3.5} & \OrbitTableEstimate{5}{101.7}{1.5}{1.5} & \OrbitTableEstimate{6}{134.8}{2.6}{2.3} &  \\
$\Omega$ (deg) & $\mathcal{U}(-3600,3600^\circ)$\hspace{4pt} & \OrbitTableEstimate{2}{95.1}{6.0}{6.0} & \OrbitTableEstimate{3}{242.2}{4.0}{3.9} & \OrbitTableEstimate{4}{90.5}{5.0}{5.5} & \OrbitTableEstimate{5}{235.9}{1.6}{1.5} & \OrbitTableEstimate{6}{212.4}{2.9}{3.2} &  \\
$K_1$ (km s$^{-1}$) & Derived$^b$\hspace{4pt} & \OrbitTableEstimate{2}{1.354}{0.013}{0.013} & \OrbitTableEstimate{3}{1.150}{0.025}{0.027} & \OrbitTableEstimate{4}{1.08}{0.24}{0.28} & \OrbitTableEstimate{5}{2.386}{0.024}{0.026} & \OrbitTableEstimate{6}{2.758}{0.014}{0.012} &  \\
$M_2$ ($M_{\rm Jup}$) & Derived\hspace{4pt} & \OrbitTableEstimate{2}{37.39}{0.74}{0.78} & \OrbitTableEstimate{3}{35.7}{1.2}{1.2} & \OrbitTableEstimate{4}{52.2}{4.2}{4.1} & \OrbitTableEstimate{5}{46.63}{0.84}{0.86} & \OrbitTableEstimate{6}{62.7}{3.0}{3.1} &  \\
$a$ (AU) & Derived\hspace{4pt} & \OrbitTableEstimate{2}{1.1193}{0.0093}{0.0091} & \OrbitTableEstimate{3}{1.235}{0.010}{0.010} & \OrbitTableEstimate{4}{1.0552}{0.0088}{0.0088} & \OrbitTableEstimate{5}{0.9226}{0.0074}{0.0073} & \OrbitTableEstimate{6}{0.4894}{0.0038}{0.0038} &  \\
$f_M$ & Derived\hspace{4pt} & \OrbitTableEstimate{2}{0.05236}{0.00070}{0.00077} & \OrbitTableEstimate{3}{0.0526}{0.0015}{0.0015} & \OrbitTableEstimate{4}{0.0766}{0.0057}{0.0054} & \OrbitTableEstimate{5}{0.07401}{0.00081}{0.00084} & \OrbitTableEstimate{6}{0.1280}{0.0051}{0.0052} &  \\
\multicolumn{8}{l}{\textit{Nuisance parameters}}\\
$\varpi$ (mas) & Flat, $>0$\hspace{4pt} & \OrbitTableEstimate{2}{5.973}{0.022}{0.021} & \OrbitTableEstimate{3}{6.548}{0.017}{0.017} & \OrbitTableEstimate{4}{7.624}{0.018}{0.017} & \OrbitTableEstimate{5}{12.285}{0.023}{0.023} & \OrbitTableEstimate{6}{18.269}{0.049}{0.049} &  \\
$\gamma_{\rm NIRPS}$ (km s$^{-1}$) & Flat on $\mathbb{R}$\hspace{4pt} & \OrbitTableEstimate{2}{10.642}{0.023}{0.023} & \OrbitTableEstimate{3}{-6.185}{0.037}{0.037} & \OrbitTableEstimate{4}{-142.414}{0.036}{0.037} & \OrbitTableEstimate{5}{-10.1726}{0.0090}{0.0088} & \OrbitTableEstimate{6}{4.613}{0.010}{0.010} &  \\
$\gamma_{\rm HARPS}$ (km s$^{-1}$) & Flat on $\mathbb{R}$\hspace{4pt} & \OrbitTableEstimate{2}{10.633}{0.013}{0.013} & \OrbitTableEstimate{3}{-6.144}{0.015}{0.017} & \OrbitTableEstimate{4}{-142.367}{0.041}{0.042} & \OrbitTableEstimate{5}{-10.1598}{0.0074}{0.0072} & \OrbitTableEstimate{6}{4.6035}{0.0060}{0.0063} &  \\
$\sigma_{\rm NIRPS}$ (m s$^{-1}$) & $\mathrm{LU}(0.01,1000)$\hspace{4pt} & \OrbitTableEstimate{2}{85}{17}{22} & \OrbitTableEstimate{3}{157}{25}{32} & \OrbitTableEstimate{4}{3.2}{3.1}{12} & \OrbitTableEstimate{5}{26.6}{6.3}{8.3} & \OrbitTableEstimate{6}{35}{10}{14} &  \\
$\sigma_{\rm HARPS}$ (m s$^{-1}$) & $\mathrm{LU}(0.01,1000)$\hspace{4pt} & \OrbitTableEstimate{2}{1.2}{1.2}{19} & \OrbitTableEstimate{3}{27}{27}{30} & \OrbitTableEstimate{4}{50}{33}{24} & \OrbitTableEstimate{5}{0.57}{0.53}{7.5} & \OrbitTableEstimate{6}{1.6}{1.6}{19} &  \\
$s_{\rm Gaia}$ & $\mathrm{LU}(1,25)$\hspace{4pt} & \OrbitTableEstimate{2}{1.85}{0.52}{1.0} & \OrbitTableEstimate{3}{1.42}{0.31}{0.70} & \OrbitTableEstimate{4}{1.36}{0.27}{0.68} & \OrbitTableEstimate{5}{1.58}{0.39}{0.80} & \OrbitTableEstimate{6}{3.31}{0.89}{1.5} &  \\
\hline
\noalign{\vskip 3pt}
\colhead{Parameter} & \colhead{\shortstack{2MASS\\J16535217\\-5741538}} & \colhead{\shortstack{UCAC4\\042-004661}} & \colhead{\shortstack{UCAC3\\49-20331}} & \colhead{L~83-26} & \colhead{\shortstack{2MASS\\J19262435\\-1045398}} & \colhead{L~386-14} & \colhead{} \\
\multicolumn{8}{l}{\textit{Orbital parameters}}\\
$P$ (d) & \multicolumn{1}{c}{\OrbitTableEstimate{1}{457.72}{1.2}{0.98}} & \OrbitTableEstimate{2}{533.19}{0.37}{0.35} & \OrbitTableEstimate{3}{501.71}{0.23}{0.21} & \OrbitTableEstimate{4}{524.3}{1.1}{1.2} & \OrbitTableEstimate{5}{175.47}{0.12}{0.12} & \OrbitTableEstimate{6}{683.01}{0.26}{0.25} &  \\
$M_0$ (J2016) (deg) & \multicolumn{1}{c}{\OrbitTableEstimate{1}{124.3}{7.3}{5.9}} & \OrbitTableEstimate{2}{190.4}{1.6}{1.6} & \OrbitTableEstimate{3}{99.7}{1.1}{1.1} & \OrbitTableEstimate{4}{84.0}{5.2}{5.4} & \OrbitTableEstimate{5}{132.3}{5.3}{5.3} & \OrbitTableEstimate{6}{27.10}{0.70}{0.68} &  \\
$e$ & \multicolumn{1}{c}{\OrbitTableEstimate{1}{0.847}{0.018}{0.024}} & \OrbitTableEstimate{2}{0.7278}{0.0014}{0.0015} & \OrbitTableEstimate{3}{0.4618}{0.0018}{0.0018} & \OrbitTableEstimate{4}{0.6911}{0.010}{0.0075} & \OrbitTableEstimate{5}{0.3150}{0.0040}{0.0040} & \OrbitTableEstimate{6}{0.4070}{0.0017}{0.0017} &  \\
$\omega_1$ (deg) & \multicolumn{1}{c}{\OrbitTableEstimate{1}{221.4}{3.2}{1.9}} & \OrbitTableEstimate{2}{220.74}{0.29}{0.29} & \OrbitTableEstimate{3}{115.61}{0.23}{0.23} & \OrbitTableEstimate{4}{311.36}{0.53}{0.66} & \OrbitTableEstimate{5}{316.59}{0.78}{0.72} & \OrbitTableEstimate{6}{302.79}{0.26}{0.26} &  \\
$\alpha$ (mas) & \multicolumn{1}{c}{\OrbitTableEstimate{1}{1.271}{0.060}{0.084}} & \OrbitTableEstimate{2}{1.339}{0.014}{0.018} & \OrbitTableEstimate{3}{1.4088}{0.0032}{0.0035} & \OrbitTableEstimate{4}{1.481}{0.045}{0.047} & \OrbitTableEstimate{5}{2.100}{0.065}{0.065} & \OrbitTableEstimate{6}{2.7194}{0.0062}{0.0062} &  \\
$i$ (deg) & \multicolumn{1}{c}{\OrbitTableEstimate{1}{31.1}{1.7}{2.4}} & \OrbitTableEstimate{2}{75.7}{2.6}{2.6} & \OrbitTableEstimate{3}{93.1}{1.3}{1.3} & \OrbitTableEstimate{4}{59.4}{2.8}{3.0} & \OrbitTableEstimate{5}{26.69}{0.86}{0.92} & \OrbitTableEstimate{6}{90.86}{0.74}{0.75} &  \\
$\Omega$ (deg) & \multicolumn{1}{c}{\OrbitTableEstimate{1}{85.2}{5.8}{7.5}} & \OrbitTableEstimate{2}{346.8}{3.9}{3.3} & \OrbitTableEstimate{3}{150.5}{1.3}{1.3} & \OrbitTableEstimate{4}{158.9}{5.5}{5.4} & \OrbitTableEstimate{5}{116.7}{5.7}{5.5} & \OrbitTableEstimate{6}{219.66}{0.52}{0.53} &  \\
$K_1$ (km s$^{-1}$) & \multicolumn{1}{c}{\OrbitTableEstimate{1}{1.58}{0.16}{0.30}} & \OrbitTableEstimate{2}{2.2079}{0.0067}{0.0066} & \OrbitTableEstimate{3}{1.5923}{0.0033}{0.0033} & \OrbitTableEstimate{4}{2.051}{0.037}{0.031} & \OrbitTableEstimate{5}{1.5913}{0.0069}{0.0078} & \OrbitTableEstimate{6}{1.7805}{0.0032}{0.0032} &  \\
$M_2$ ($M_{\rm Jup}$) & \multicolumn{1}{c}{\OrbitTableEstimate{1}{35.3}{1.9}{2.6}} & \OrbitTableEstimate{2}{44.83}{0.90}{0.97} & \OrbitTableEstimate{3}{37.52}{0.60}{0.61} & \OrbitTableEstimate{4}{44.2}{1.6}{1.7} & \OrbitTableEstimate{5}{44.5}{1.6}{1.7} & \OrbitTableEstimate{6}{47.83}{0.75}{0.76} &  \\
$a$ (AU) & \multicolumn{1}{c}{\OrbitTableEstimate{1}{0.8751}{0.0073}{0.0073}} & \OrbitTableEstimate{2}{1.0910}{0.0088}{0.0088} & \OrbitTableEstimate{3}{1.0178}{0.0081}{0.0081} & \OrbitTableEstimate{4}{1.0235}{0.0084}{0.0084} & \OrbitTableEstimate{5}{0.4252}{0.0034}{0.0034} & \OrbitTableEstimate{6}{1.2504}{0.0098}{0.0098} &  \\
$f_M$ & \multicolumn{1}{c}{\OrbitTableEstimate{1}{0.0789}{0.0038}{0.0052}} & \OrbitTableEstimate{2}{0.07021}{0.00095}{0.0011} & \OrbitTableEstimate{3}{0.06411}{0.00052}{0.00054} & \OrbitTableEstimate{4}{0.0812}{0.0025}{0.0026} & \OrbitTableEstimate{5}{0.1276}{0.0039}{0.0039} & \OrbitTableEstimate{6}{0.08168}{0.00066}{0.00067} &  \\
\multicolumn{8}{l}{\textit{Nuisance parameters}}\\
$\varpi$ (mas) & \multicolumn{1}{c}{\OrbitTableEstimate{1}{18.423}{0.055}{0.053}} & \OrbitTableEstimate{2}{17.487}{0.039}{0.041} & \OrbitTableEstimate{3}{21.593}{0.019}{0.018} & \OrbitTableEstimate{4}{17.836}{0.077}{0.076} & \OrbitTableEstimate{5}{38.698}{0.064}{0.063} & \OrbitTableEstimate{6}{26.628}{0.023}{0.022} &  \\
$\gamma_{\rm NIRPS}$ (km s$^{-1}$) & \multicolumn{1}{c}{\OrbitTableEstimate{1}{-7.3327}{0.012}{0.0089}} & \OrbitTableEstimate{2}{34.5919}{0.0035}{0.0036} & \OrbitTableEstimate{3}{67.6546}{0.0023}{0.0023} & \OrbitTableEstimate{4}{26.4702}{0.0074}{0.0068} & \OrbitTableEstimate{5}{30.2221}{0.0050}{0.0049} & \OrbitTableEstimate{6}{-16.0691}{0.0045}{0.0045} &  \\
$\gamma_{\rm HARPS}$ (km s$^{-1}$) & \multicolumn{1}{c}{\OrbitTableEstimate{1}{-7.326}{0.013}{0.010}} & \OrbitTableEstimate{2}{34.6086}{0.0048}{0.0050} & \OrbitTableEstimate{3}{67.6579}{0.0033}{0.0033} & \OrbitTableEstimate{4}{26.4760}{0.011}{0.0096} & \OrbitTableEstimate{5}{30.2243}{0.0091}{0.0092} & \OrbitTableEstimate{6}{-16.0651}{0.0024}{0.0024} &  \\
$\sigma_{\rm NIRPS}$ (m s$^{-1}$) & \multicolumn{1}{c}{\OrbitTableEstimate{1}{0.87}{0.83}{9.0}} & \OrbitTableEstimate{2}{7.4}{7.0}{5.3} & \OrbitTableEstimate{3}{0.19}{0.17}{1.3} & \OrbitTableEstimate{4}{12}{11}{17} & \OrbitTableEstimate{5}{15.3}{5.2}{6.4} & \OrbitTableEstimate{6}{17.5}{3.6}{4.5} &  \\
$\sigma_{\rm HARPS}$ (m s$^{-1}$) & \multicolumn{1}{c}{\OrbitTableEstimate{1}{2.5}{2.5}{14}} & \OrbitTableEstimate{2}{0.66}{0.62}{7.0} & \OrbitTableEstimate{3}{0.24}{0.21}{1.9} & \OrbitTableEstimate{4}{0.60}{0.57}{8.6} & \OrbitTableEstimate{5}{0.55}{0.51}{7.9} & \OrbitTableEstimate{6}{0.23}{0.20}{1.8} &  \\
$s_{\rm Gaia}$ & \multicolumn{1}{c}{\OrbitTableEstimate{1}{2.28}{0.73}{1.5}} & \OrbitTableEstimate{2}{2.65}{0.85}{1.4} & \OrbitTableEstimate{3}{1.31}{0.23}{0.56} & \OrbitTableEstimate{4}{4.7}{1.3}{2.5} & \OrbitTableEstimate{5}{2.49}{0.67}{1.2} & \OrbitTableEstimate{6}{1.46}{0.33}{0.72} &  \\
\hline
\end{tabular}
\par\noindent\textit{Notes.} The table will also be available in machine-readable form at the CDS upon journal publication. Reported values are posterior medians with errors based on the 16th and 84th percentiles.  $\mathcal{U}$ and $\mathrm{LU}$ denote uniform and log-uniform priors.   $^a$The priors on $e$ and $\omega_1$ are induced by sampling $h=\sqrt{e}\cos\omega_1$ and $k=\sqrt{e}\sin\omega_1$ jointly uniformly over $0<h^2+k^2<1$. $^b$We sample the astrometric amplitude $\alpha$ and derive $K_1$ from $\alpha$, $\varpi$, $P$, $e$, and $i$.
\endgroup
\end{ManuscriptParameterTable}
\endgroup
\end{table*}

\subsection{Orbit inference}
For the systems we identify as dark companion hosts, as a next step we fit the keplerian reflex orbit of the host star around the system barycenter, assuming a single companion. We use the same coordinate convention as the one specified in \cite{MarcussenEtAl2026}, with $\omega_1$ denoting
the argument of periastron of the primary.

\subsubsection{Likelihood}
\label{subsec:likelihood}

We compare three fits for each confirmed system: one using the Gaia DR3
orbital solution alone, one using the NIRPS and HARPS RVs alone, and a joint fit. Let $D$
denote the data included in a given fit and $\boldsymbol{\theta}$ its complete
vector of model parameters. The posterior is
\begin{equation}
	p(\boldsymbol{\theta}\mid D)\propto
	p(\boldsymbol{\theta})\,
	\mathcal{L}_{\rm Gaia}(\boldsymbol{\theta})\,
	\mathcal{L}_{\rm RV}(\boldsymbol{\theta}),
\end{equation}
with the unused likelihood factor omitted in each single-dataset fit.

For the $n$th RV measurement $v_n$, obtained at epoch $t_n$ with instrument
$m\in\{\mathrm{NIRPS},\mathrm{HARPS}\}$ and quoted uncertainty $\sigma_n$, we assume independent Gaussian errors
with variance $V_n=\sigma_n^2+\sigma_{{\rm jit},m}^2$. The jitter
$\sigma_{{\rm jit},m}$ represents additional white noise. The model RV is
$v_{\rm obs}(t_n)=v(t_n)+\gamma_{m}$, where $v(t_n)$ is the primary's Keplerian orbital RV and $\gamma_{m}$ is the RV offset
for that instrument. The RV log likelihood is
\begin{equation}
	\ln\mathcal{L}_{\rm RV}=-\frac{1}{2}\sum_n
	\left[
	\frac{[v_n-v_{\rm obs}(t_n)]^2}{V_n}
	+\ln(2\pi V_n)
	\right].
	\label{eq:rv_likelihood}
\end{equation}

Gaia DR3 provides the fitted two-body orbital parameters and parallax, their
uncertainties and correlations, but not the epoch astrometry. Following
Section~3.2 of \citet{Winn2022}, we adopt a multivariate Gaussian likelihood in
$\boldsymbol{y}=(A,B,F,G,e,P,T_{\rm peri},\varpi)^{\mathsf T}$, centred on the
DR3 solution. Here, $P$ is the orbital period, $e$ is the eccentricity, and $A,B,F,G$ are the angular Thiele--Innes coefficients that
reparametrize the photocentre semimajor axis and orbital orientation angles and
are used by Gaia to report the orbit because they enter the astrometric model
linearly. We calculate them from the model's Campbell orbital elements using the
relations in Appendix~\ref{app:methods}. DR3 reports the time of periastron $T_{\rm peri}$ in days relative to its
reference epoch, J2016.0. Using the mean anomaly $M_0$ at the same reference epoch $t_{\rm ref}$, we calculate the model
value as $T_{\rm peri}=-M_0P/(2\pi)$. We reconstruct the
covariance matrix $\boldsymbol{C}_{\rm Gaia}$ for these eight quantities from
the reported uncertainties and correlation coefficients.

In the joint fits, we additionally test whether the published Gaia covariance
must be inflated to accommodate the RVs. We follow \citet{Stefansson2025} and
introduce a covariance-scale factor $s_{\rm Gaia}$, defined as the ratio between
the effective Gaia standard deviations and their published values. We fix
$s_{\rm Gaia}=1$ for the Gaia-only fits.

We denote the vector of published DR3 parameter
values as $\boldsymbol{y}_{\rm DR3}$ and the corresponding model
predictions as $\boldsymbol{y}(\boldsymbol{\theta})$.
Writing $\boldsymbol{r}_{\rm G}=\boldsymbol{y}(\boldsymbol{\theta})
	-\boldsymbol{y}_{\rm DR3}$, the Gaia log likelihood is
\begin{equation}
	\begin{aligned}
		\ln\mathcal{L}_{\rm Gaia}={} & -\frac{1}{2}\bigl[
			s_{\rm Gaia}^{-2}\boldsymbol{r}_{\rm G}^{\mathsf T}
		\boldsymbol{C}_{\rm Gaia}^{-1}\boldsymbol{r}_{\rm G}                                                     \\
		                             & +\ln\det\bigl(2\pi s_{\rm Gaia}^{2}\boldsymbol{C}_{\rm Gaia}\bigr)\bigr].
	\end{aligned}
	\label{eq:gaia_likelihood}
\end{equation}
Like $\ln(2\pi V_n)$ in Equation~\ref{eq:rv_likelihood}, the determinant term is
the normalization of the Gaussian; because $\det(s_{\rm
		Gaia}^{2}\boldsymbol{C}_{\rm Gaia})=s_{\rm Gaia}^{16}\det\boldsymbol{C}_{\rm
		Gaia}$ for the eight quantities in $\boldsymbol{y}$, it penalizes large $s_{\rm
			Gaia}$.

\subsubsection{Sampling, priors and masses}
\label{Sampling}

We sampled the posteriors of the Gaia-only, RV-only, and joint fits with \texttt{emcee} \citep{ForemanMackey+2013}.
To speed up the sampling of the parameter space near $e=0$ we step in
$\sqrt{e}\cos\omega_1$ and $\sqrt{e}\sin\omega_1$. We also step in $\cos i$, where $i$ is the orbital inclination,
to ensure an isotropic prior on the orbital orientation. Our joint astrometry and RV
parameter vector therefore is
\begin{equation}
	\begin{aligned}
		\boldsymbol{\theta}_{\rm joint}={} & (\varpi,P,M_0,\sqrt{e}\cos\omega_1,
		\sqrt{e}\sin\omega_1,\Omega,\cos i,                                                                                \\
		                                   & \alpha,\gamma_{\rm NIRPS},\gamma_{\rm HARPS},                                                                                \\
		                                   & \ln\sigma_{{\rm jit},{\rm NIRPS}},\ln\sigma_{{\rm jit},{\rm HARPS}},\ln s_{\rm Gaia})\,\, ,
	\end{aligned}
\end{equation}
Here, $\Omega$ is the longitude of the ascending node, measured from north through east, and $\alpha=\varpi a_1/\mathrm{au}$ is the angular semimajor axis of the primary's barycentric orbit. Under the dark-companion approximation, $\alpha=a_{\text{phot}}$. The RV-only fits drop $\varpi$, $\Omega$, $\cos i$, and $\ln s_{\rm
	Gaia}$, and sample the primary's RV semi-amplitude $K_1$ instead of $\alpha$. Gaia-only fits drop $\gamma_i$ and
$\ln\sigma_{{\rm jit},i}$ for both instruments, and $\ln s_{\rm Gaia}$. In the
joint and Gaia-only fits, the reported $K_1$ is derived from the sampled
astrometric orbit using Equation~\ref{eq:component_semiamplitudes}.

The priors for the joint fits are summarized in Table~\ref{tab:hybrid_selected_fits}. These include the RV jitter $\sigma_{\rm jit}$ and the Gaia covariance factor
$s_{\rm Gaia}$, which have log-uniform priors, $p(x)\propto1/x$, on
$0.01$--$1000\,\mathrm{m\,s^{-1}}$ and $1$--$25$, respectively. The higher prior
density at smaller values favours little additional dispersion while still
permitting larger RV jitter or Gaia uncertainty inflation when required by the
data.

We fitted all targets allowing for a linear RV trend at first. The fitted trend coefficients
were consistent with zero in all cases, so we fixed them to zero in the adopted fits.

We define \(M_0\) at \(t_{\rm ref}=\) J2016.0 for the Gaia-only and joint fits, matching the Gaia DR3 reference epoch. For the RV-only fits, we use J2025.0, closer to the RV observations, to reduce the correlation between period and phase.

We assessed convergence across four independently initialized chains, requiring
rank-normalized split-$\widehat{R}$ and folded split-$\widehat{R}$ below $1.02$,
and rank-normalized bulk effective sample sizes of at least 400 for every
coordinate \citep{Vehtari2021}.

\begin{figure*}[!tp]
	\centering
	\setlength{\VSpaceBeforeFigCaption}{-3.78pt}
	\includegraphics[width=\textwidth]{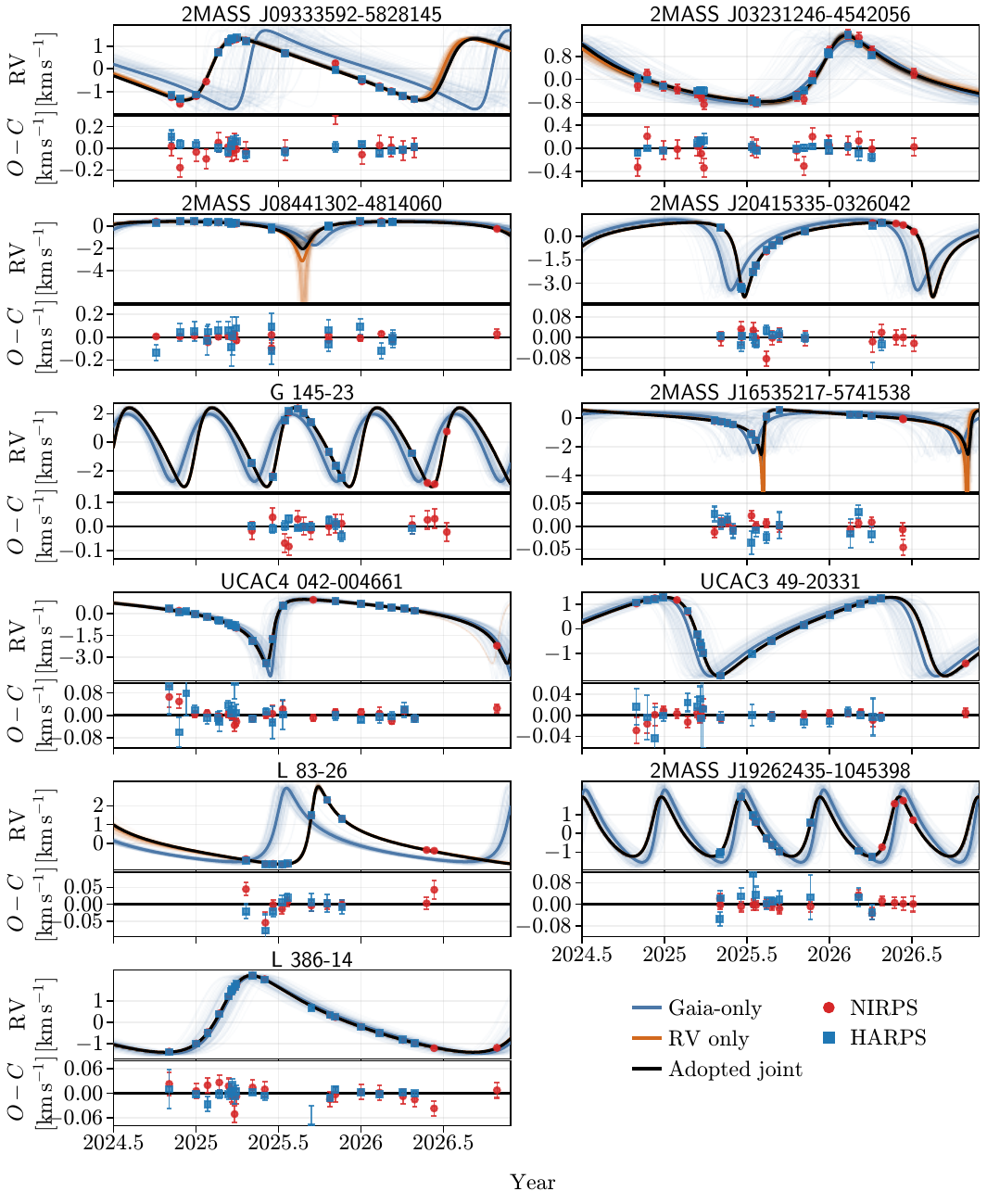}
	\caption{NIRPS and HARPS RV time series for the eleven confirmed brown dwarfs. \emph{Upper subpanels}: maximum-a-posteriori (MAP) joint fit model (bold curves) and 100 random draws in thin curves of Gaia-only, RV only and joint fits.  Red NIRPS and blue HARPS error bars
		combine formal $1\sigma$ uncertainties and adopted-fit posterior-median
		jitter in quadrature. \emph{Lower subpanels}: residuals from the adopted
		joint MAP model.}
	\label{fig:rv_all_systems}
	\label{fig:rv_all_systems_2024p5_2027}
	\label{fig:rv_1787984409900989824}
	\label{fig:rv_4188152322648398592}
	\label{fig:rv_4224582166524567040}
	\label{fig:rv_4846381143359598592}
	\label{fig:rv_5196317344716068224}
	\label{fig:rv_5305690099977167232}
	\label{fig:rv_5329058707789307648}
	\label{fig:rv_5927476672953974784}
	\label{fig:rv_6383984898525990272}
	\label{fig:rv_5533714311754660864}
	\label{fig:rv_5284372837240905344}
\vspace{24pt}
\end{figure*}

\begin{figure*}[t]
	\centering
	\setlength{\VSpaceBeforeFigCaption}{-5.29pt}
	\includegraphics[width=\textwidth,trim=0 14bp 0 0,clip]{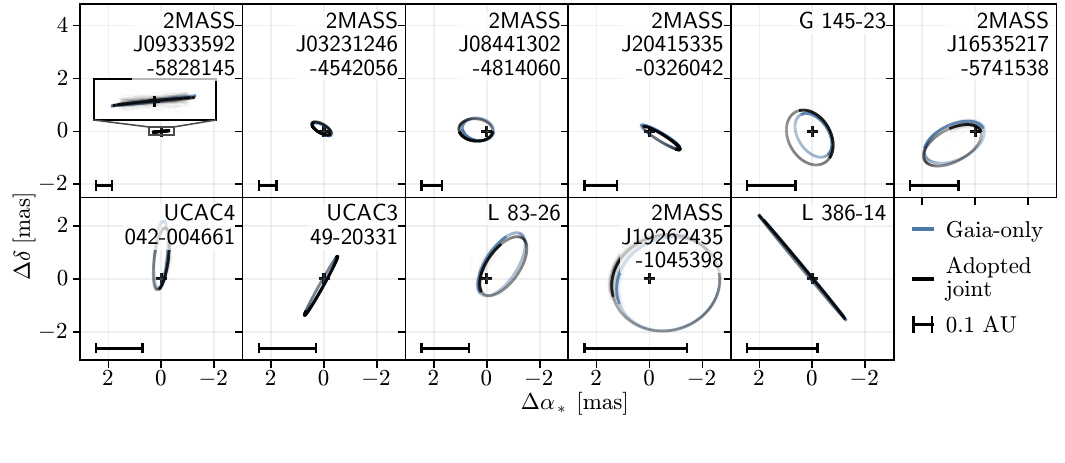}
	\caption{Sky-projected reflex motion of the eleven confirmed brown dwarf host stars.
		Thin curves are 100 posterior draws and thick curves are
		maximum-a-posteriori (MAP) orbits for the Gaia-only and adopted joint
		fits; crosses mark the center of mass. The axes show right-ascension offsets $\Delta\alpha^\ast$, including the cosine of declination, and declination offsets $\Delta\delta$. The line-of-sight coordinate $LOS$ is positive away from the observer. Solid segments are on the near
		side of the sky plane ($LOS\leq0$); lighter segments are on the far side
		($LOS>0$). Black bars show
		$0.1\,\mathrm{AU}$ at each system's distance.}
	\label{fig:reflex_all_systems}
\end{figure*}

We describe posterior uncertainty with 16th and 84th percentiles and use the
posterior median as the central value in Table~\ref{tab:hybrid_selected_fits}. We use the maximum-a-posteriori (MAP) state as the representative orbit in Figures~\ref{fig:rv_all_systems} and \ref{fig:reflex_all_systems}. Appendix~\ref{app:three_fit_corners} compares the posteriors of the Gaia-only, RV-only, and joint fits.

From these posteriors we can derive
\begin{equation}
	\frac{M_2^3}{(M_1+M_2)^2} = \frac{P K_1^3(1-e^2)^{3/2}}{2\pi G \sin^3 i},
\end{equation}
with the stellar and companion masses $M_1$ and $M_2$.
We can only determine this combination, as we only have astrometric and RV data on
the motion of the host, not of the companion. $M_1$ therefore has to come from an
external source. We derived it from the empirical $M_{K_s}$--mass relation of
\citet{Mann2019}, using extinction-corrected near-infrared 2MASS $K_s$ magnitudes
\citep{Skrutskie2006} and the parallaxes of the DR3 orbital solutions; all
eleven values are consistent with the primary masses in
\texttt{gaiadr3.binary\_masses}, but with much smaller uncertainties.

\section{Observational results}
\label{sec:results}

\subsection{Spectroscopic vetting results}
\label{subsec:binary_vetting}

Of the 53 targets, we classify 11 as single-lined systems consistent with dark companions, 37 as stellar binaries, and five as inconclusive, pending further observations. Figure~\ref{fig:l27489_binary_ccf} illustrates representative vetting outcomes. The observed CCF profiles of all 53 targets are shown in Figure~\ref{fig:ccf_classification_grid}.

For ten of our eleven confirmed companions, the stellar binary scenario predicts that the RV separation of the binary components at the NIRPS epoch of greatest predicted separation is 2.3--6.1 times the observed CCF full width at half maximum (FWHM). Yet the observed profiles remain single-peaked. For the remaining system, 2MASS~J08441302-4814060, the greatest predicted separation is only $11\,\mathrm{km\,s^{-1}}$, below the observed CCF FWHM of $15\,\mathrm{km\,s^{-1}}$ at that epoch, so the absence of resolved peaks alone cannot distinguish the two scenarios. Nevertheless the observed line profiles favour a single star with a dark companion over a binary whose spectral lines overlap, with a median $R_k=1.9$ across the omitted nights.

Conversely, the stellar-binary model reproduces the phase-dependent CCF distortions of L~274-89 and yields smaller held-out residuals on all nine nights, with a median $R_k=0.8$ across the omitted nights. We therefore classify this system as a blended stellar binary. The all-epoch comparisons for both L~274-89 and 2MASS~J08441302-4814060 are presented in Figure~\ref{fig:ccf_all_epochs} (Appendix~\ref{app:loo}). The remaining 36 targets classified as binary simply show a double-peaked CCF in at least one epoch by visual inspection.

We confirm the previously published stellar-binary classifications of five systems: G-ASOI-21, G-ASOI-25, and G-ASOI-40 from \citet{Stefansson2025}, and Ross~778 and 2M1515-08 from \citet{MarcussenEtAl2026}.

\subsection{Orbital solutions of the confirmed companions}
\label{subsec:individual_rv_comparisons}

The main orbital parameters from our joint Gaia+RV fits for the eleven validated
companions are listed in Table~\ref{tab:hybrid_selected_fits}, and Appendix~\ref{app:three_fit_corners} shows the posteriors.
Figures~\ref{fig:rv_all_systems} and~\ref{fig:reflex_all_systems} show the RV
time series and sky-projected reflex orbits, respectively.
To check for consistency, we also performed RV-only and Gaia-only fits, whose
results are also shown in these figures. However, all results are from the joint
fits, if not mentioned otherwise.

\begin{itemize}
	\item 2MASS~J09333592-5828145, 2MASS~J03231246-4542056, 2MASS~J20415335-0326042, UCAC4~042-004661 and  L~386-14 show good agreement
	      between the NIRPS RVs and the DR3 astrometric predictions.

	\item %
         For 2MASS~J08441302-4814060, the spectroscopic observations do not
	      sample the periastron passage, providing only limited independent
	      constraints on the orbital parameters. Nevertheless, the measured RVs trace
	      the curvature predicted by the Gaia DR3 orbit, while forward modelling of
	      the CCFs strongly disfavors the corresponding stellar binary scenario
	      (Figure~\ref{fig:l27489_binary_ccf}, second row). The all-epoch profile
	      comparison is shown in Figure~\ref{fig:ccf_all_epochs}.

	\item For G~145-23, %
          the Gaia-only and RV-only
	      solutions differ in mean anomaly at J2025 (\(2.7\sigma\)), eccentricity
	      (\(2.4\sigma\)), and the joint fit favours an inflated Gaia covariance
	      factor. The Robo-AO multiplicity survey reported no resolved companion
	      \citep{Lamman2020}.

	\item For 2MASS~J16535217-5741538, the RVs favour a higher
	      eccentricity than the DR3 astrometric solution (\(3.4\sigma\)), with the
	      joint fit yielding \(e=0.847_{-0.018}^{+0.024}\).

	\item  %
         For UCAC3~49-20331  the limited phase coverage broaden the RV-only posterior. Although the DR3 solution lies at
	      its periphery, the joint fit does not require inflation of the Gaia
	      covariance. The star has a resolved neighbour HD~43920 at a projected separation of
	       \(37''\) (\(\sim1700\,\mathrm{AU}\)). It has a proper-motion difference of
	        \(2.2\,\mathrm{mas\,yr^{-1}}\) and a matching RV, but a \(4.4\sigma\)
	        parallax discrepancy.

	\item L~83-26
        shows the strongest RV--Gaia DR3 tension in our sample: the RV-only and
	      Gaia-only solutions differ by \(4.4\sigma\) in mean anomaly at J2025 and
	      \(3.3\sigma\) in \(\omega_1\), while their periods are consistent.
	      An additional companion omitted from
	      our single-companion model could explain the discrepancy.

	\item 
          For 2MASS~J19262435-1045398, the RV-only and Gaia-only
	      solutions differ in argument of periastron (\(1.9\sigma\)), while
	      their RV semi-amplitudes are consistent. The joint fit yields
	      \(s_{\rm Gaia}=2.49_{-0.67}^{+1.2}\), favouring covariance
	      inflation.

\end{itemize}

Remarkably, none of the 31 confirmed dark companions of the DR3 blind search, from
this work and the literature, has an eccentricity below $0.25$.

\section{Population analysis}
\label{sec:population_analysis}

\begin{figure}[t]
	\centering
	\setlength{\VSpaceBeforeFigCaption}{4pt}
	\includegraphics[width=\columnwidth]{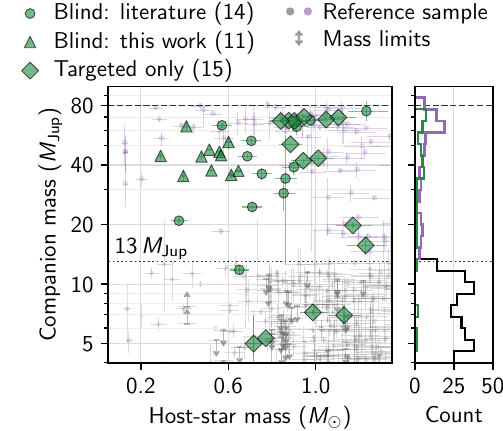}
	\caption{The 11 brown dwarfs found here in the context of known brown dwarfs and exoplanets. 
		Targeted-search systems come from
		\citet{Unger2023,MarcussenEtAl2026}, \citet{Sozzetti2023} (Gaia-3~b),
		and \citet{Pinamonti2026} (Gaia-6~B).
		The reference sample combines NASA Exoplanet Archive planets (gray),
		with reported true masses, retrieved on
		2026 September 21 \citep{Christiansen2025Archive}, and brown dwarfs
		(purple) from \citet{Stevenson+2023}, supplemented by the five substellar
		systems of \citet{Winterhalder2024}. }
	\label{fig:companion_host_mass}
\end{figure}

Gaia has high sensitivity to massive companions around nearby low-mass stars. Figure~\ref{fig:companion_host_mass} places the 11 new brown dwarfs among the detected companions.  This study adds eleven objects to the sparsely represented population of  known companions with masses $35$--$62\,M_{\mathrm{Jup}}$ around ${\sim}0.3$--$0.65\,M_{\odot}$ primaries, expanding the sample available for a demographic analysis.

Host-star metallicity and orbital eccentricity distributions
provide clues to the formation of giant planets and brown
dwarfs. An RV study of predominantly FGK hosts suggests that, at separations of $1$--$50\,\mathrm{au}$, gravitational
fragmentation becomes more important relative to core
accretion above a companion mass of approximately
$27\,M_{\rm Jup}$: lower-mass companions preferentially
orbit metal-rich stars, whereas higher-mass companions
have host metallicities closer to the field-star
distribution \citep{Giacalone2026}. Furthermore, for companions around predominantly FGK
hosts at separations of $1$--$10\,\mathrm{au}$,
\citet{Gilbert2026} find a gradual increase in mean
orbital eccentricity with companion mass, from
approximately 0.2 for Jovian planets to approximately
0.5 for brown dwarfs. 

For a Gaia-selected sample predominantly around M dwarfs,
\citet{Pennell+2026} find stellar-binary-like
eccentricities and host metallicities down to approximately
$20\,M_{\rm Jup}$. 

For giant planets 
\citet{Johnson2010} found that occurrence increases
approximately linearly with host mass over
$0.2$--$1.9\,M_{\odot}$. Their fitted relation implies approximately half the occurrence around $0.5\,M_{\odot}$ stars compared with
solar-mass stars at the same metallicity.
For periods of $100$--$1000\,\mathrm{d}$, \citet{Ribas2023} measured
$3\pm1$ giant planets per 100 M dwarfs, with minimum
masses of approximately $0.3$--$3\,M_{\rm Jup}$.
For the same period interval, \citet{Mignon2025} measured
$0.6^{+1.2}_{-0.2}$ planets per 100 M dwarfs with
minimum masses of $1$--$10\,M_{\rm Jup}$.

A complementary demographic question is how frequently
brown dwarfs orbit low-mass stars, compared with giant
planets. An initial Gaia DR3 estimate suggested a brown-dwarf
occurrence of approximately 0.3\% around M dwarfs within
100~pc for periods below approximately 1000~d, however this was done with significant assumptions of complete detection and no false positives
\citep{GaiaCollaborationArenou+2023}, motivating a measurement that
accounts for both incompleteness and contamination.

We determined the occurrence rate, $f_{\mathrm{BD}}$, of brown dwarfs with
masses of $20$--$80\,M_{\mathrm{Jup}}$ and orbital periods of $100$--$1000$
days. We considered host stars of $0.4$--$0.6\,M_{\odot}$ within
$100\,\mathrm{pc}$ that have no stellar companion unresolved by Gaia

In its
simplest form, $f_{\mathrm{BD}}$ is
\begin{equation}
	f_{\mathrm{BD}} = \frac{N_{\mathrm{BD}}}{N_{\mathrm{stars,\,searched}}}.
	\label{eq:fbd}
\end{equation}
Here $N_{\mathrm{BD}}$ is the number of stars with at least one brown dwarf,
and $N_{\mathrm{stars,\,searched}}$ is the effective number of single stars searched, accounting for detection efficiency. Our methods for accounting
for detection efficiency and unresolved
binaries are described below.

The masses of all stars we used for this population analysis, and the companion masses that follow, were computed with the synthetic relation between mass and absolute Gaia \(G\) band magnitude of \citet{ChulkovMalkov2022} applicable to main sequence stars with a broad range of masses.

\begin{figure*}[t]
	\centering
	\setlength{\VSpaceBeforeFigCaption}{-10.06pt}
	\includegraphics[width=\textwidth]{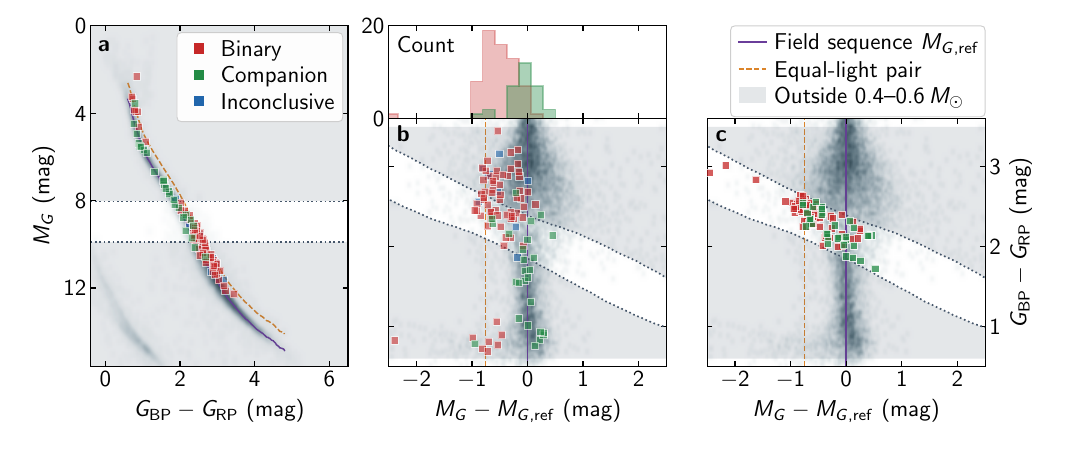}
	\caption{Color--magnitude position for the blind-search
	candidates. The heatmap background shows the field stars of
	Sect.~\ref{sec:false_positive}, and the solid line the reference dwarf sequence.
	The dashed line is brighter by $2.5\log_{10}2=0.753$~mag, as expected for
	an unresolved pair of identical stars. The shaded area corresponds to range in the Gaia G band that corresponds to main sequence masses outside $0.4-0.6\,\text{M}_{\text{Sun}}$ range used in the occurrence rate estimate.
	\textit{(a)} The
	blind-search candidates with reported spectroscopic or interferometric outcomes  of Figure~\ref{fig:combined_vetting_sample} in the $M_G$ versus
	$G_{\rm BP}-G_{\rm RP}$ diagram.
	 \textit{(b)} The same candidates and
	field-star density shown as color versus $\Delta M_G=M_G-M_{G,\mathrm{ref}}$, the
	displacement from the reference median at the same color. The top
 panel shows histograms for definitively
	classified binaries and confirmed companions.  \textit{(c)} Same diagram as (b), different sample: the unclassified candidates in our occurrence-rate parameter space, coloured by their \citet{Way+2026} classifications. Red squares mark systems flagged as stellar binaries and green squares mark systems with
    $E_{\mathrm{Way}} \leq 0.376\,\mathrm{mag}$, the binary threshold.}
	\label{fig:combined_color_magnitude}
\end{figure*}

\subsection{Number of stars searched}
\label{sec:stars_searched}

We limited the sample to 100\,pc because Gaia provides a nearly complete census
of M dwarfs more massive than $\sim0.1\,M_{\odot}$ out to that distance
\citep{GaiaCollaborationSmart+2021}. The sample comes from the Gaia Catalogue of Nearby Stars, excludes white dwarfs and contains
$N_{\mathrm{stars}}=51{,}639$ sources. For each star we determined its contribution to $N_{\mathrm{stars, searched}}$. Two factors enter this calculation: one reduces the count so that stars that have \emph{stellar} companions unresolved by Gaia are excluded from $N_{\mathrm{stars, searched}}$ (section \ref{unresolvedmultiplicity}) and the second one allows us to account for detection completeness (section \ref{detectioncompleteness}).

\subsubsection{Correction for unresolved stellar multiplicity}
\label{unresolvedmultiplicity}

Our occurrence rate of brown dwarfs refers to those that orbit single stars and stars whose stellar companions Gaia resolve. We exclude from this occurrence-rate measurement hosts
with stellar companions unresolved by Gaia, whether
in binaries or higher-order multiple systems.The orbital solutions of \citet{Halbwachs+2023} describe a single Keplerian photocentre orbit. An additional
unresolved stellar companion can dilute the astrometric
signal of a brown dwarf orbiting one component and
introduce further orbital motion, biasing the orbital fits and changing the substellar companion detection probability.
In addition, partially resolved pairs, which Gaia resolves only in some scan directions, were removed as far as possible before the processing \citep{Halbwachs+2023}. Instead of modelling the detection of substellar companions
in these systems, we exclude such hosts from the population
whose occurrence we measure. To make the denominator in \eqref{eq:fbd} refer
to the same host population as the numerator, we weight
each source by the estimated probability that
it has no stellar companion unresolved by Gaia.

We therefore assign each Gaia source an eligibility weight $w_i$, representing
the probability that it does not have a stellar companion unresolved by Gaia. We assign $w_i=1$ to hosts of confirmed substellar companions. Sources classified as stellar binary systems from Gaia NSS solutions or spectroscopic follow-up receive $w_i=0$, and unclassified
companion candidate hosts receive a weight equal to a probability that they are a genuine host rather than a stellar binary, which we will estimate in section \ref{sec:false_positive}.
For the remaining sources, we set $w_i=1-f_{\mathrm{u},i}$, where $f_{\mathrm{u},i}$ is the estimated probability that the source $i$ has a stellar companion unresolved by Gaia.

This probability, $f_{\mathrm{u},i}$, depends on the distance
of the star. We estimated it from the nearby M-dwarf multiplicity classification
compiled by \citet{Winters+2019}, using their 36 systems with primary masses of
$0.4$--$0.6\,M_{\odot}$ within $10\,\mathrm{pc}$ that have been searched
extensively for companions. Of these, 10 have a stellar companion. We added two systems with primary masses of \(0.353\)--\(0.375\,M_{\odot}\) to account for unresolved binaries whose combined light can place them above our \(0.4\,M_{\odot}\) selection limit.  Assuming that this
subsample is representative of our sample, we placed the Winters stars at the
distance of each star.  We retained the component masses published in \citet{Winters+2019}, inferred each component’s individual absolute \(G\) magnitude from inverted \citet{ChulkovMalkov2022} mass--magnitude relation, and assigned masses to unresolved binaries by summing their component fluxes and applying the same relation to the combined magnitude. For multiples among this sample, we converted their projected separations into angles, and
used the probability that Gaia resolves a pair at that angle
(Appendix~\ref{app:funresolved}). We calculated the fractions in distance bins. Averaged over our sample,
$18.8^{+6.1}_{-5.7}\,\%$ of the stars have an unresolved stellar companion. The
uncertainty comes from resampling the 38 reference systems with replacement and recalculating the unresolved fraction for each draw (bootstrapping).
When calculating $f_{\mathrm{u},i}$ for the remaining sources assigned $w_i=1-f_{\mathrm{u},i}$, we subtract the contribution of stellar
binaries already identified through Gaia NSS solutions
or follow-up spectroscopy from the expected
unresolved-binary count.

\subsubsection{Detection completeness}
\label{detectioncompleteness}

We define the detection completeness,
$C_i$, as the probability that
 a brown dwarf in our adopted mass/period
 range, orbiting star $i$, would
 have been detected in the DR3 blind
 search.
In connection with the previous section, our detection-completeness simulations assume a single luminous host star and therefore do not describe the detectability of brown dwarfs in unresolved stellar multiple systems.
We injected 100 brown dwarfs around each star, drawing $M_2$ and
$P$ from log-uniform distributions,
with isotropic orbital orientations and random orbital phases.

For the underlying eccentricity distribution we considered two distributions: a
Rayleigh distribution with $\sigma_e=0.33$, characteristic of close stellar
binaries \citep{Wu2025Eccentricities}, and a $\mathrm{Beta}(0.867,3.03)$
distribution that \citet{Kipping2013} derived for RV-detected exoplanets.
None of the 31 confirmed dark companions of the DR3 blind search has an eccentricity below $0.25$ (Fig.~\ref{fig:rayleigh_completeness}a). This disfavors the $\mathrm{Beta}(0.867,3.03)$ distribution, which predicts many nearly circular orbits. Therefore, we assumed a Rayleigh
distribution.

For each injected system, we simulated and fitted Gaia DR3 astrometric data using
\texttt{gaiamock}, following the procedure described by
\citet{ElBadry2024GaiaMock}. We counted an injection
as recovered if it yielded a Keplerian orbital solution
and passed acceptance and publication criteria intended
to reproduce those of the DR3 blind search
\citep{Halbwachs+2023}. The completeness $C_i$ is
the fraction of the 100 injections around star $i$
that met these criteria.

Each Gaia source therefore contributes \(w_i C_i\) to the effective number of stars that were searched. With this contribution we then determined the number of stars searched,
\begin{equation}
N_{\mathrm{stars,\,searched}} = \sum_{i=1}^{N_{\mathrm{stars}}} w_iC_i .
\label{eq:nsearched}
\end{equation}

\begin{figure*}[t]
\centering
\setlength{\VSpaceBeforeFigCaption}{-8pt}
\includegraphics[width=\textwidth]{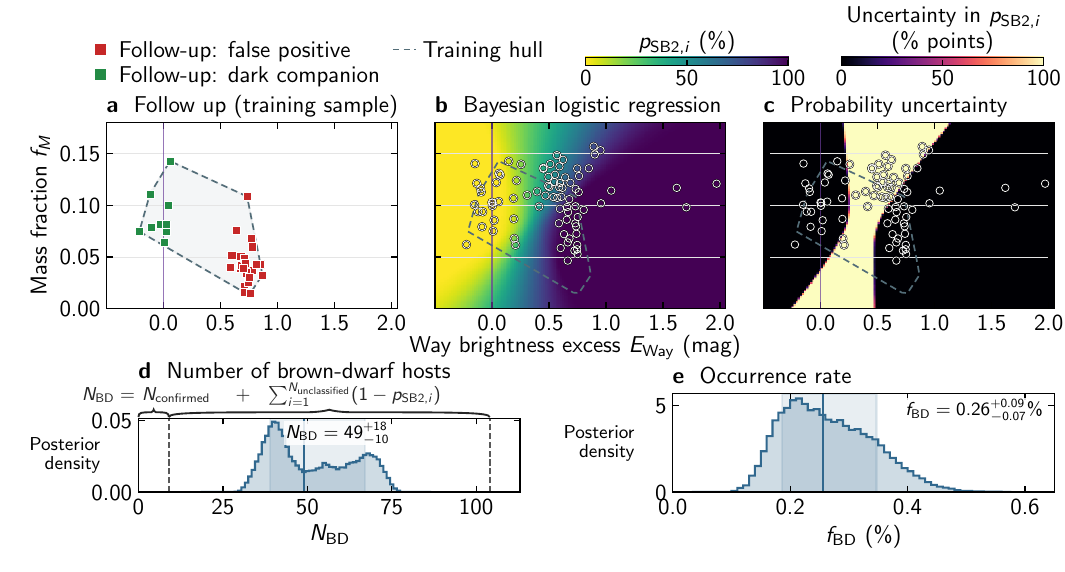}
\caption{Results of our logistic regression to classify companions. Top row: Mass fraction \(f_M\), inferred under the dark-companion assumption, versus brightness excess \(E_{\mathrm{Way}}\) of \cite{Way+2026} for blind search systems with host masses of \(0.4\)--\(0.6\,M_\odot\). \textbf{(a)} The 32 systems with follow-up used to train the logistic regression: 23 false positives and nine dark companions. \textbf{(b)} Posterior mean false-positive probability \(p_{\mathrm{SB2},i}\) from the MCMC logistic fit. \textbf{(c)} Uncertainty in \(p_{\mathrm{SB2},i}\), shown as the width of its central 68\% credible interval in percentage points. Open circles in panels (b) and (c) mark the 82 unclassified candidates with a Way predictor; the 13 candidates without one are not shown. \textbf{(d)} Posterior distribution of the number of brown dwarf hosts, \(N_{\mathrm{BD}}\), including the nine confirmed hosts and the uncertain classifications of all 95 unclassified candidates. Vertical dashed lines mark the limits of 9 and 104 if all unclassified candidates are false positives or dark companions, respectively. \textbf{(e)} Posterior distribution of the occurrence rate \(f_{\mathrm{BD}}\) for the adopted Rayleigh eccentricity distribution.}
\label{fig:mass_fraction_way_li}
\end{figure*}

\subsection{Number of brown-dwarf hosts}
\label{sec:false_positive}

To determine the numerator of Equation~\ref{eq:fbd}, $N_{\text{BD}}$, we started with the
DR3 blind search candidate sample of \citet{Halbwachs+2023}.\footnote{This leaves
	out some of the substellar companion candidates detected by Gaia, such as HD~68638 discussed in the introduction, as these come from the targeted-search
	candidate list \citep{HollOrbit2023}, with a different selection function.}
Of the 116 candidates in our parameter space range, $N_{\mathrm{confirmed}}=9$ have been classified as genuine
brown-dwarf hosts, 12 as false positives due to double-lined spectroscopic
binaries (SB2), and $N_{\mathrm{unclassified}}=95$ have not been followed up or
have insufficient data and remain unclassified.  %
Given the false-positive probabilities
of the unclassified candidates,
the expected number of genuine
brown-dwarf detections is
\begin{equation}
	N_{\mathrm{BD}} = N_{\mathrm{confirmed}} + \sum_{i=1}^{N_{\mathrm{unclassified}}} \left(1-p_{\mathrm{SB2},i}\right),
	\label{eq:nbd}
\end{equation}
where $p_{\mathrm{SB2},i}$ is the probability that the unclassified
candidate $i$ is an unresolved stellar
binary masquerading as a substellar
companion. If all unclassified candidates were genuine
brown-dwarf hosts, $N_{BD}$ would
be 104; if all were false positives,  $N_{BD}$ would be 9.
Our actual estimate of $N_{BD}$ 
depends on our model used to
assign the false-positive probabilities.
	
We expect the false-positive probability $p_{\mathrm{SB2},i}$ of each candidate
to depend chiefly on two observable
quantities: the mass fraction 
$f_M = M_2/(M_1 + M_2)$ that is calculated
under the assumption of a dark
companion, and the brightness
excess of the system relative
to single field stars of the same
colour. As discussed in sect.~\ref{sec:forwardmodel}, if the system is a
false positive, the ratio of the masses of the two stars and the ratio of their
fluxes determine the photocentre amplitude. The
smaller the candidate mass is relative to the host mass, the closer to equal
masses (and equal brightness) the two stars must be. How often a candidate with
a given $f_M$ (assuming a dark companion) is a false positive also depends on the occurrence rates of such binaries and of brown dwarfs. %

In many false positive cases we therefore expect the two stars have similar fluxes, and the pair is
brighter, by up to 0.75\,mag, than a single star of the same color. However for the
least massive hosts, a brown-dwarf mass is a larger fraction of the host mass,
and the two stars can differ considerably in mass. The secondary then adds
little light, but it is also redder than the primary, and the pair still lies
above the main sequence at its color, although by less than
0.75\,mag.

To quantify the elevation of candidates above the main sequence, we constructed
a reference sequence of field stars (Fig.~\ref{fig:combined_color_magnitude}) from Gaia Catalogue of Nearby Stars.
We used stars within 50~pc with a renormalized unit weight error
(RUWE) below 1.2, that is, with astrometry consistent with a single star,
excluding white dwarfs and stars with $G_{\rm BP}>20.3$. We excluded
the latter to avoid the known faint-source bias in BP photometry
\citep{Riello2021}. The remaining 19,313 field stars define the reference
sequence, which is the running median of $M_G$ at each $G_{\rm BP}-G_{\rm RP}$
color. The offset of a candidate from this sequence is $\Delta M_G = M_G -
	M_{G,\mathrm{ref}}$, where $M_{G,\mathrm{ref}}$ is the median $M_G$ of the
reference sequence at the color of the candidate. As can be seen in
Fig.~\ref{fig:combined_color_magnitude}b, most confirmed companions lie close to
this sequence, whereas many of the false positives lie above it. It also suggests a dependence of false positive fraction on color: for the bluest candidates the elevation above the sequence predicts a false
positive better than for the reddest ones, and at colors of about 1.5\,mag all
candidates are confirmed companions.
The effectiveness of brightness
excess as a false-positive diagnostic
therefore appears to depend on
stellar colour. To account for that, we decided to use the information
contained in the Gaia low-resolution
BP/RP spectra (XP). Specifically,
we adopted the metric
of \cite{Way+2026}, which we
denote $E_{\mathrm{Way}}$, which quantifies how much brighter a system is than 
expected for a single star, given
the shape of its Gaia RP spectrum.

\begin{figure*}[t]
	\centering
	\setlength{\VSpaceBeforeFigCaption}{0.49pt}
	\includegraphics[width=\textwidth]{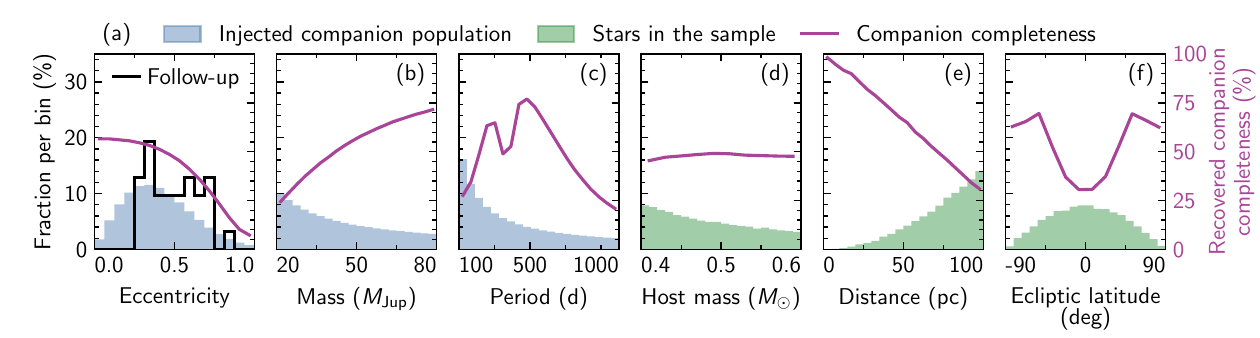}
	\caption{Gaia DR3 companion completeness.
Panels (a)--(c) show the injected fraction per bin (blue, left axis) and the fraction recovered (purple, right axis) versus companion eccentricity, mass, and period. Panel (a) also shows the eccentricities of all 31 confirmed dark companions from the DR3 blind search (solid black). These comprise 26 RV-based and five Gaia+GRAVITY-based eccentricities \citep[this work;][]{Latham2002,Unger2023,GaiaCollaborationArenou+2023,Stefansson2025,Barbato2026,MarcussenEtAl2026,Winterhalder2024}, rather than values from astrometry-only DR3 solutions.
Panels (d)--(f) show the distribution of the 51,639 census stars (green, left axis) and the mean per-star recovery fraction (purple, right axis) in bins of host mass, distance, and ecliptic latitude, respectively.}
	\label{fig:rayleigh_completeness}
\end{figure*}

We fitted a logistic model separating true companions from false positives with MCMC in the $f_M$ and $E_{\mathrm{Way}}$ predictor space,
\begin{equation}
    p_{\mathrm{SB2},i} =
    \frac{1}{1+\exp[-(\alpha+\beta_M z_{M,i}+\beta_E z_{E,i})]},
    \label{eq:logistic_companion_probability}
\end{equation}
where $z_{M,i}$ and $z_{E,i}$ are the standardized values of $f_M$ and $E_{\mathrm{Way}}$, respectively, $\alpha$ is the logistic intercept, and $\beta_M$ and $\beta_E$ are the corresponding slopes. The likelihood, priors, and sampling procedure are described in Appendix~\ref{app:logistic_classifier}.
The training sample comprised 32 classified blind-search systems within 100\,pc with host masses
of $0.4$--$0.6\,M_\odot$ (23 false positives and
nine confirmed companions). We included classified systems over the full range of companion masses and periods in the training, and applied the resulting fit to unclassified candidates in the \(20\)--\(80\,M_{\rm Jup}\) and \(100\)--\(1000\)-day interval.
The training sample and the inferred false-positive probabilities and their uncertainties are shown in Fig.~\ref{fig:mass_fraction_way_li}a--c.
We obtained \(\sum_{i=1}^{N_{\mathrm{unclassified}}}(1-p_{\mathrm{SB2},i})=40^{+18}_{-10}\) brown-dwarf hosts among the unclassified candidates.

We assume that, at fixed $f_M$ and $E_{\mathrm{Way}}$, the false-positive probability is the same for classified and unclassified candidates. Not so many of the reddest targets have been followed-up, so the occurrence rate host mass range of $0.4-0.6$ has been chosen so that the samples are indeed comparable in colour, signal to noise in Gaia XP and apparent brightness in $G$. For some targets however, the inference still depends on extrapolation, so we have to assume that the logistic relationship continues into regions with no classified examples.

Restricting the host masses to $0.4$--$0.6\,M_\odot$
also limits the dark-companion mass fraction among  the unclassified candidates to
$f_M\lesssim0.16$, compared with $f_M\lesssim0.28$
if hosts down to $0.2\,M_\odot$ were included.
Larger dark-companion mass fractions can correspond
to more unequal stellar binaries,
whose fainter secondaries are less readily identified
by the overluminosity method of \citet{Way+2026}. 

A potential bias in the training sample is that stellar binaries are easier
to classify conclusively than dark-companion systems: just one spectroscopic observation often already reveals a double-lined spectrum, while confirming that the companion is dark requires many observations over the target's period.
However, none of the 10 candidates with inconclusive
RV follow-up (blue in Fig.~\ref{fig:combined_vetting_sample})
survives the host-mass and distance cuts of the occurrence rate bin, so this source of bias does not appear to affect our training sample.

\subsection{Occurrence rate}\label{sec:rate}

The DR3 blind search would have recovered 47.7\,\% of the brown dwarfs injected
around the stars of our sample for the Rayleigh eccentricity distribution. Equation~\ref{eq:nsearched} gives
$N_{\mathrm{stars,\,searched}}=19,937^{+1,389}_{-1,527}$.
Figure~\ref{fig:rayleigh_completeness} shows the injection distributions and the recovery fraction versus
companion eccentricity, companion mass, orbital period, host mass, host distance, and ecliptic latitude
(Appendix~\ref{app:scanlaw}).

The unclassified candidates contribute an expected number of \(40^{+18}_{-10}\) brown-dwarf hosts. With the 9 confirmed hosts we obtain \(N_{\mathrm{BD}}=49^{+{18}}_{-10}\) (Fig.~\ref{fig:mass_fraction_way_li}d).

With Equation~\eqref{eq:fbd} we obtain $f_{\mathrm{BD}}=0.26^{+0.09}_{-0.07}\,\%$ (Fig.~\ref{fig:mass_fraction_way_li}e). Its uncertainty was calculated through 40,000 Monte Carlo realizations.

In each realization $r$, we drew the regression coefficients
jointly from their posterior and sampled the candidate false-positive probabilities. Adding the nine confirmed hosts gave
a numerator for eq. \eqref{eq:fbd}, $N_{BD,r}$. We paired these probabilities with one realization
of the unresolved-binary fraction as a function of
distance, obtained as one draw from the bootstrap explained in section \ref{unresolvedmultiplicity}.
Using this realization and the same candidate
classifications as used for the numerator, we recalculated the effective
denominator. To include Poisson counting uncertainty, we drew
the underlying expected detection count from its
Gamma distribution conditional on $N_{\mathrm{BD},r}$
and divided this draw by the effective denominator.

As a sensitivity test, we checked that changing the injected eccentricity distribution to the Kipping distribution would change the occurrence to $f_{\mathrm{BD}}=0.23^{+0.08}_{-0.06}\,\%$. Alternatively, restricting the regression training sample to companions with masses of $20$--$80\,M_{\rm Jup}$ and periods of $100$--$1000$ days gave $f_{\rm BD}=0.23^{+0.10}_{-0.06}\%$, while retaining the full training sample but using in the regression only the brightness excess $E_{\rm Way}$, without the mass fraction $f_M$, gave $f_{\rm BD}=0.21^{+0.05}_{-0.04}\%$.

\section{Discussion}
\label{sec:discussion}

\subsection{DR3-RV agreement}

The outcomes of previous RV follow-up of Gaia DR3 substellar candidates have included RV non-detections that rule out the companion implied by the astrometric data
\citep{Venner2026},
orbits that differ substantially in RV amplitude or eccentricity
\citep{MarcussenAlbrecht2023,Sozzetti2023}, and good agreement between the RV and astrometric solutions 
\citep{Stefansson2025,Barbato2026,MarcussenEtAl2026}.
Our NIRPS RVs are broadly consistent with the DR3 orbital constraints of the eleven confirmed companions: six systems show good agreement, while some exhibit tension.
These discrepancies are milder than the factor-of-three and factor-of-fifteen RV-amplitude mismatches reported for HD~40503 \citep{MarcussenAlbrecht2023} and Gaia-3~b \citep{Sozzetti2023}.

\subsection{DR3 exoplanet vetting completeness}

Of the 17 blind-search candidates selected because their DR3 companion-mass
lower bounds fall below $13\,M_{\rm Jup}$, 13 are stellar-binary false positives,
one is a confirmed companion (Gaia-4~b), and three remain inconclusive.
These are ASOI-16 and ASOI-49, observed by \citet{Stefansson2025},
and 2MASS J1131, observed in this work.
ASOI-16 shows broad, variable line profiles suggestive of an unresolved
binary; ASOI-49 and 2MASS J1131 remain single-lined but lack sufficient
orbital-phase coverage for confirmation. Thus, all 17 candidates have
spectroscopic follow-up, but three require further observations to
establish their nature.

\subsection{Occurrence rate in context}

Our occurrence rate of $0.26^{+0.09}_{-0.07}\%$ for $0.4$--$0.6\,M_\odot$ hosts is close to the initial Gaia DR3 estimate of approximately $0.3\%$. Their estimate considers a broader mass range of approximately $0.1$--$0.6\,M_\odot$. In our analysis,
the reduction due to false positives is partly offset by
corrections for detection incompleteness and unresolved stellar companions.

The brown dwarfs appear to orbit M dwarfs less often than giant planets: \citet{Ribas2023} found that $3\% \pm 1\%$ of $0.08$--$0.70\,M_\odot$ stars host a $0.3$--$3\,M_{\mathrm{Jup}}$ planet in the same period interval, while \citet{Mignon2025} found that $0.6^{+1.2}_{-0.2}\%$ $1$--$10\,M_{\mathrm{Jup}}$ planets are found per $0.10$--$0.75\,M_\odot$ star in the same period interval.

Moreover, our measurement is of a similar magnitude to
published brown dwarf occurrence estimates around
predominantly FGK stars. \citet{Santerne2016} measured
$0.29\pm0.17\%$ around $0.7$--$1.4\,M_\odot$ dwarfs
for orbital periods below $400\,\mathrm{d}$.
\citet{Unger2023} reported a tentative occurrence of
$0.8^{+0.3}_{-0.2}\%$ for $13$--$80\,M_{\mathrm{Jup}}$
companions within $10\,\mathrm{au}$ of FGK stars.
More recently, \citet{VanZandt2026} found
$0.2^{+0.3}_{-0.1}$ and $0.5^{+0.4}_{-0.2}$ brown dwarfs
per 100 stars at separations of $0.3$--$1\,\mathrm{au}$
and $1$--$3\,\mathrm{au}$, respectively, for
$13$--$80\,M_{\mathrm{Jup}}$ companions from RVs.

\section{Conclusions and lessons learned for DR4}
\label{sec:conclusions}

We followed up 53 Gaia DR3 candidate companions to late-type
stars with NIRPS. Our main findings are:
\begin{itemize}
	\item We validated 11 brown dwarfs with dynamical masses of
	      $35$--$62\,M_{\rm Jup}$ and orbital semimajor axes of
	      $0.4$--$1.3\,\mathrm{au}$.

	\item Combining this work with published spectroscopic and interferometric
	      follow-up, 96 targets from the DR3 blind search have
	      definitive classifications: 31 dark companions and 65 luminous-binary
	      false positives. The false-positive fraction is therefore $68\%$.

	\item We newly classify seven DR3 exoplanet candidates as stellar
	      binaries, while one remains inconclusive. All 17 blind-search
	      candidates whose DR3 mass estimates allow a planetary
	      companion below $13\,M_{\rm Jup}$ now have spectroscopic
	      follow-up, although three remain inconclusive.

	\item Across the combined sample, candidates whose DR3 solutions permit lower companion masses are more frequently revealed as stellar binaries.

	\item Correcting for unresolved stellar companions and Gaia companion completeness, we obtain an occurrence rate of $0.26^{+0.09}_{-0.07}\%$ for $20$--$80\,M_{\mathrm{Jup}}$ brown dwarfs with $100$--$1000$ day orbital periods around $0.4$--$0.6\,M_\odot$ stars without unresolved stellar companions.

	\item Modelling the full, time-resolved CCF profiles
	      allowed us to classify two systems, 2MASS~J08441302-4814060 and L~274-89, for which the luminous-binary model predicts velocity separations smaller than the observed CCF FWHM, such that the blended binary profiles would resemble those of single stars. The method can reveal a binary impostor through subtle, phase-dependent changes in CCF morphology, or disfavor the binary scenario and thereby strengthen the identification of a dark companion.
\end{itemize}

The larger Gaia DR4 candidate sample will plausibly include more ambiguous systems similar to L 274-89. The CCF forward
modelling demonstrated here provides an additional diagnostic
for assessing their nature, potentially already using the Gaia RVS spectra, despite their lower resolution.

\software{NIRPS DRS (v3.3.0; \citealt{Bouchy+2025}), HARPS DRS (v3.8),
	LBL (v0.67.008; \citealt{Artigau2022}),
	SERVAL (\citealt{Zechmeister2018SERVAL}),
	MADYS (v2.1.1; \citealt{SquicciariniBonavita2022}),
	GaiaMock (\citealt{ElBadry2024GaiaMock}),
	emcee (\citealt{ForemanMackey+2013}),
	ArviZ (\citealt{Kumar2019}),
	Astropy (\citealt{Astropy2022}),
	corner (\citealt{ForemanMackey2016corner})}

\begin{acknowledgments}

	Based on observations collected at the European Southern Observatory under ESO programmes 114.2758, 115.27ZE, 116.296E, and 117.2AJC. We thank François Bouchy and Lucile Mignon for coordinating the NIRPS/HARPS observations.

	This work has made use of data from the European Space Agency (ESA) mission Gaia (\url{https://www.cosmos.esa.int/gaia}), processed by the Gaia Data Processing and Analysis Consortium (DPAC; \url{https://www.cosmos.esa.int/web/gaia/dpac/consortium}). Funding for the DPAC has been provided by national institutions, in particular the institutions participating in the Gaia Multilateral Agreement.

	This research has made use of the SIMBAD database, operated at CDS, Strasbourg, France \citep{Wenger2000}.

	This publication makes use of data products from the Two Micron All Sky Survey, which is a joint project of the University of Massachusetts and the Infrared Processing and Analysis Center/California Institute of Technology, funded by the National Aeronautics and Space Administration and the National Science Foundation.

	This research has made use of the NASA Exoplanet Archive, which is operated by the California Institute of Technology, under contract with the National Aeronautics and Space Administration under the Exoplanet Exploration Program.

    The results presented in this work were obtained at the Centre for Scientific Computing, Aarhus (CSCAA).

	K.K.K. and S.H.A. acknowledge support from the Independent Research Fund Denmark (Danmarks Frie Forsknings fond) through grant 5241-00072B.

\end{acknowledgments}

\facilities{ESO:3.6m (NIRPS, HARPS), Gaia, Exoplanet Archive, CDS, IRSA}

\setcounter{topnumber}{4}
\setcounter{dbltopnumber}{3}
\setcounter{totalnumber}{6}
\renewcommand{\topfraction}{0.95}
\renewcommand{\dbltopfraction}{0.95}
\renewcommand{\textfraction}{0.05}
\renewcommand{\floatpagefraction}{0.95}
\renewcommand{\dblfloatpagefraction}{0.95}
\appendix
\nolinenumbers
\restartappendixnumbering

\onecolumn
\raggedbottom
\section{Observed sample}
\label{app:sample}
\enlargethispage{25pt}

\begingroup
\fontsize{9}{10}\selectfont
\setlength{\tabcolsep}{4pt}
\setlength{\LTcapwidth}{\linewidth}
\setlength{\LTpre}{0pt}
\setlength{\LTpost}{0pt}
\renewcommand{\arraystretch}{1.12}
\newcommand{\TableCompactCell}[1]{#1}
\begin{longtable}{@{}>{\collectcell\TableCompactCell}l<{\endcollectcell}>{\collectcell\TableCompactCell}l<{\endcollectcell}cccccccc@{}}
\caption{Stellar properties and spectroscopic classifications obtained in this work. The $M_1$ values of the eleven single-lined systems use the  $K_s$-band relation of \citet{Mann2019}; all other $M_1$ values use the $G$-band relation of \citet{ChulkovMalkov2022} and assume a dark companion. Lower bound on the companion mass $M_{2,\mathrm{low}}^{\mathrm{DR3}}$ comes from \texttt{gaiadr3.binary\_masses}. $N_{\rm NIRPS}$ is the number of NIRPS spectra used in the analysis, $N_{\rm HARPS}$ is the number of HARPS RV epochs used in the orbital fits, and $\widetilde{\mathrm{S/N}}_{\rm NIRPS}$ is the median NIRPS order-55 S/N.\label{tab:nirps_observations}}\\
\hline\hline

\colhead{Name} & \colhead{Gaia DR3 source ID} & \colhead{$G$} & \colhead{$M_1$} & \colhead{$M_{2,\mathrm{low}}^{\mathrm{DR3}}$} & \colhead{RUWE} & \colhead{$N_{\rm NIRPS}$} & \colhead{$N_{\rm HARPS}$} & \colhead{\shortstack{$\widetilde{\mathrm{S/N}}$\\$_{\rm NIRPS}$}} & \colhead{\shortstack{Classifi-\\cation}} \\
\colhead{} & \colhead{} & \colhead{(mag)} & \colhead{($M_\odot$)} & \colhead{($M_{\rm Jup}$)} & \colhead{} & \colhead{} & \colhead{} & \colhead{} & \colhead{}
\\
\hline
\endfirsthead
\hline\hline

\colhead{Name} & \colhead{Gaia DR3 source ID} & \colhead{$G$} & \colhead{$M_1$} & \colhead{$M_{2,\mathrm{low}}^{\mathrm{DR3}}$} & \colhead{RUWE} & \colhead{$N_{\rm NIRPS}$} & \colhead{$N_{\rm HARPS}$} & \colhead{\shortstack{$\widetilde{\mathrm{S/N}}$\\$_{\rm NIRPS}$}} & \colhead{\shortstack{Classifi-\\cation}} \\
\colhead{} & \colhead{} & \colhead{(mag)} & \colhead{($M_\odot$)} & \colhead{($M_{\rm Jup}$)} & \colhead{} & \colhead{} & \colhead{} & \colhead{} & \colhead{}
\\
\hline
\endhead
\hline
\endfoot
2MASS J09333592-5828145 & 5305690099977167232 & 14 & $0.646^{+0.016}_{-0.016}$ & 39 & 1.6 & 18 & 16 & 23 & \textcolor{ClassificationSingle}{Single} \\
2MASS J03231246-4542056 & 4846381143359598592 & 14 & $0.613^{+0.015}_{-0.015}$ & 33 & 1.9 & 19 & 17 & 20 & \textcolor{ClassificationSingle}{Single} \\
2MASS J08441302-4814060 & 5329058707789307648 & 14 & $0.601^{+0.015}_{-0.015}$ & 45 & 2.5 & 20 & 18 & 24 & \textcolor{ClassificationSingle}{Single} \\
2MASS J20415335-0326042 & 4224582166524567040 & 13 & $0.559^{+0.014}_{-0.014}$ & 47 & 1.9 & 14 & 11 & 33 & \textcolor{ClassificationSingle}{Single} \\
G 145-23 & 1787984409900989824 & 13 & $0.4084^{+0.0095}_{-0.0094}$ & 47 & 4.7 & 14 & 11 & 36 & \textcolor{ClassificationSingle}{Single} \\
2MASS J16535217-5741538 & 5927476672953974784 & 13 & $0.3942^{+0.0097}_{-0.0096}$ & 33 & 3.1 & 13 & 11 & 34 & \textcolor{ClassificationSingle}{Single} \\
UCAC4 042-004661 & 5196317344716068224 & 12 & $0.567^{+0.014}_{-0.014}$ & 44 & 4.1 & 23 & 22 & 48 & \textcolor{ClassificationSingle}{Single} \\
UCAC3 49-20331 & 5284372837240905344 & 12 & $0.523^{+0.013}_{-0.013}$ & 36 & 4.5 & 19 & 18 & 53 & \textcolor{ClassificationSingle}{Single} \\
L 83-26 & 6383984898525990272 & 13 & $0.477^{+0.012}_{-0.011}$ & 44 & 4.2 & 10 & 8 & 35 & \textcolor{ClassificationSingle}{Single} \\
2MASS J19262435-1045398 & 4188152322648398592 & 13 & $0.2905^{+0.0072}_{-0.0071}$ & 45 & 6.9 & 15 & 11 & 37 & \textcolor{ClassificationSingle}{Single} \\
L 386-14 & 5533714311754660864 & 12 & $0.513^{+0.012}_{-0.012}$ & 46 & 7.7 & 19 & 18 & 77 & \textcolor{ClassificationSingle}{Single} \\
2MASS J03203928-4437347 & 4847265734823582848 & 15 & $0.436^{+0.042}_{-0.041}$ & 25 & 1.4 & 4 & 0 & 7.6 & \textcolor{ClassificationInconclusive}{Inconclusive} \\
2MASS J11312909-4437287 & 5375875638010549376 & 14 & $0.362^{+0.040}_{-0.038}$ & 11 & 1.7 & 2 & 0 & 24 & \textcolor{ClassificationInconclusive}{Inconclusive} \\
2MASS J06585124-3837033 & 5565711440145909504 & 15 & $0.245^{+0.032}_{-0.029}$ & 21 & 2.2 & 2 & 0 & 9.5 & \textcolor{ClassificationInconclusive}{Inconclusive} \\
2MASS J12592816-5158352 & 6081071334868194176 & 13 & $0.608^{+0.044}_{-0.044}$ & 18 & 1.4 & 1 & 0 & 21 & \textcolor{ClassificationInconclusive}{Inconclusive} \\
2MASS J20513852-4117300 & 6677563745912843776 & 16 & $0.335^{+0.038}_{-0.036}$ & 21 & 1.4 & 7 & 0 & 7.5 & \textcolor{ClassificationInconclusive}{Inconclusive} \\
2MASS J15451623+1330376 & 1191074832968581760 & 13 & $0.511^{+0.044}_{-0.043}$ & 21 & 2.4 & 1 & 0 & 44 & \textcolor{ClassificationSBTwo}{SB2} \\
Ross 778 & 1739379433242806144 & 11 & $0.531^{+0.044}_{-0.043}$ & 27 & 2.9 & 1 & 0 & 27 & \textcolor{ClassificationSBTwo}{SB2} \\
2MASS J05081150-2731434 & 2955804675022081792 & 13 & $0.478^{+0.043}_{-0.042}$ & 27 & 6.2 & 1 & 0 & 30 & \textcolor{ClassificationSBTwo}{SB2} \\
LP 545-4 & 3097269490318754176 & 14 & $0.318^{+0.037}_{-0.035}$ & 26 & 5.6 & 1 & 0 & 27 & \textcolor{ClassificationSBTwo}{SB2} \\
2MASS J04360548-0623013 & 3198650544073202688 & 14 & $0.384^{+0.041}_{-0.039}$ & 40 & 4.2 & 3 & 0 & 19 & \textcolor{ClassificationSBTwo}{SB2} \\
2MASS J12341480-2737498 & 3472253945241732736 & 14 & $0.329^{+0.038}_{-0.036}$ & 39 & 9.6 & 1 & 0 & 49 & \textcolor{ClassificationSBTwo}{SB2} \\
LP 754-5 & 4188996885011268608 & 14 & $0.201^{+0.027}_{-0.023}$ & 6.5 & 2.0 & 1 & 0 & 40 & \textcolor{ClassificationSBTwo}{SB2} \\
UCAC4 104-001812 & 4694084076943966080 & 13 & $0.374^{+0.040}_{-0.038}$ & 26 & 2.9 & 3 & 0 & 35 & \textcolor{ClassificationSBTwo}{SB2} \\
2MASS J00403291-6829195 & 4702845638429469056 & 14 & $0.416^{+0.042}_{-0.040}$ & 6.2 & 1.6 & 4 & 0 & 19 & \textcolor{ClassificationSBTwo}{SB2} \\
SCR J0505-5633 & 4764340705296117120 & 15 & $0.313^{+0.037}_{-0.034}$ & 12 & 1.9 & 3 & 0 & 17 & \textcolor{ClassificationSBTwo}{SB2} \\
L 181-8 & 4767186928583729792 & 13 & $0.555^{+0.044}_{-0.044}$ & 31 & 2.1 & 1 & 0 & 24 & \textcolor{ClassificationSBTwo}{SB2} \\
UCAC4 240-004042 & 4842246017566495232 & 13 & $0.348^{+0.039}_{-0.037}$ & 7.2 & 1.9 & 1 & 0 & 19 & \textcolor{ClassificationSBTwo}{SB2} \\
2MASS J03392766-3124220 & 4863699860246818304 & 14 & $0.475^{+0.043}_{-0.042}$ & 32 & 2.1 & 1 & 0 & 24 & \textcolor{ClassificationSBTwo}{SB2} \\
2MASS J00401448-4708178 & 4976018031041457664 & 14 & $0.528^{+0.044}_{-0.043}$ & 21 & 1.6 & 1 & 0 & 19 & \textcolor{ClassificationSBTwo}{SB2} \\
LP 990-98 & 4983571882081864960 & 13 & $0.438^{+0.042}_{-0.041}$ & 11 & 1.6 & 1 & 0 & 29 & \textcolor{ClassificationSBTwo}{SB2} \\
LP 886-63 & 5052449001298518528 & 14 & $0.416^{+0.042}_{-0.040}$ & 19 & 1.8 & 1 & 0 & 24 & \textcolor{ClassificationSBTwo}{SB2} \\
2MASS J03310650-2441518 & 5085864568417061120 & 14 & $0.398^{+0.041}_{-0.040}$ & 15 & 1.8 & 1 & 0 & 30 & \textcolor{ClassificationSBTwo}{SB2} \\
L 98-67 & 5271515801094390912 & 12 & $0.555^{+0.044}_{-0.044}$ & 20 & 3.8 & 1 & 0 & 26 & \textcolor{ClassificationSBTwo}{SB2} \\
SCR J1042-4240 & 5391312403307270784 & 14 & $0.343^{+0.039}_{-0.037}$ & 32 & 9.6 & 1 & 0 & 31 & \textcolor{ClassificationSBTwo}{SB2} \\
APMPM J0710-5704 & 5486916932205092352 & 12 & $0.295^{+0.036}_{-0.033}$ & 9.9 & 3.8 & 2 & 0 & 35 & \textcolor{ClassificationSBTwo}{SB2} \\
UCAC4 220-018903 & 5519057787396445824 & 13 & $0.306^{+0.037}_{-0.034}$ & 41 & 6.1 & 2 & 0 & 36 & \textcolor{ClassificationSBTwo}{SB2} \\
2MASS J07002772-2805150 & 5610436308992623104 & 14 & $0.431^{+0.042}_{-0.041}$ & 22 & 1.9 & 1 & 0 & 24 & \textcolor{ClassificationSBTwo}{SB2} \\
UCAC4 313-025977 & 5612039087715504640 & 14 & $0.267^{+0.034}_{-0.031}$ & 13 & 4.7 & 9 & 0 & 34 & \textcolor{ClassificationSBTwo}{SB2} \\
2MASS J07294599-2225465 & 5618776310850226432 & 15 & $0.410^{+0.042}_{-0.040}$ & 20 & 1.8 & 3 & 0 & 10 & \textcolor{ClassificationSBTwo}{SB2} \\
2MASS J12420716-5055061 & 6079316686107743488 & 15 & $0.336^{+0.038}_{-0.036}$ & 19 & 1.8 & 2 & 0 & 17 & \textcolor{ClassificationSBTwo}{SB2} \\
2MASS J15154785-0858009 & 6319620720489927680 & 13 & $0.471^{+0.043}_{-0.042}$ & 22 & 2.3 & 3 & 0 & 50 & \textcolor{ClassificationSBTwo}{SB2} \\
L 1-114 & 6342331068739840000 & 12 & $0.452^{+0.043}_{-0.042}$ & 34 & 5.1 & 1 & 0 & 42 & \textcolor{ClassificationSBTwo}{SB2} \\
2MASS J23105055-7027331 & 6381440834777420928 & 15 & $0.277^{+0.034}_{-0.032}$ & 19 & 2.1 & 3 & 0 & 8.0 & \textcolor{ClassificationSBTwo}{SB2} \\
2MASS J18211716-7235322 & 6418925831870553472 & 15 & $0.352^{+0.039}_{-0.037}$ & 14 & 1.5 & 4 & 0 & 8.9 & \textcolor{ClassificationSBTwo}{SB2} \\
UCAC4 192-200468 & 6521749994635476992 & 14 & $0.429^{+0.042}_{-0.041}$ & 14 & 1.8 & 1 & 0 & 28 & \textcolor{ClassificationSBTwo}{SB2} \\
L 499-75 & 6574387911224287488 & 11 & $0.477^{+0.043}_{-0.042}$ & 47 & 5.2 & 2 & 0 & 36 & \textcolor{ClassificationSBTwo}{SB2} \\
L 274-89 & 6644841042920381184 & 13 & $0.276^{+0.034}_{-0.031}$ & 37 & 12 & 9 & 0 & 46 & \textcolor{ClassificationSBTwo}{Blended SB2} \\
2MASS J20045745-4257230 & 6685861691447769600 & 14 & $0.459^{+0.043}_{-0.042}$ & 17 & 1.7 & 4 & 0 & 13 & \textcolor{ClassificationSBTwo}{SB2} \\
2MASS J20134731-3758222 & 6694115931396057728 & 13 & $0.342^{+0.039}_{-0.037}$ & 8.4 & 2.6 & 6 & 0 & 43 & \textcolor{ClassificationSBTwo}{SB2} \\
UCAC4 280-221518 & 6781298098147816192 & 14 & $0.297^{+0.036}_{-0.033}$ & 5.6 & 1.7 & 1 & 0 & 29 & \textcolor{ClassificationSBTwo}{SB2} \\
2MASS J20431614-2948589 & 6795231865409810176 & 13 & $0.313^{+0.037}_{-0.035}$ & 24 & 4.6 & 1 & 0 & 45 & \textcolor{ClassificationSBTwo}{SB2} \\
2MASS J20545886-1315163 & 6888630155507448704 & 16 & $0.268^{+0.034}_{-0.031}$ & 24 & 1.7 & 9 & 0 & 5.7 & \textcolor{ClassificationSBTwo}{SB2} \\
\end{longtable}
\normalsize
\section{Radial velocities}
\resetapptablenumbers
\begin{table}[h!]
\centering
\caption{HARPS and NIRPS radial velocities used in the orbital fits.}
\label{tab:radial_velocities}
\begin{tabular}{lrrrl}
\hline\hline
Target & Time & RV & $\sigma_{\rm RV}$ & Instrument \\
 & BJD$_{\rm TDB}-2\,400\,000$ & (km\,s$^{-1}$) & (km\,s$^{-1}$) & \\
\hline
2MASS~J09333592-5828145 & 60621.85447 & 9.46355 & 0.05809 & HARPS \\
2MASS~J09333592-5828145 & 60621.85483 & 9.39125 & 0.02457 & NIRPS \\
2MASS~J09333592-5828145 & 60640.78910 & 9.29738 & 0.04135 & HARPS \\
2MASS~J09333592-5828145 & 60640.78927 & 9.10014 & 0.02222 & NIRPS \\
\multicolumn{5}{l}{\ldots} \\
2MASS~J03231246-4542056 & 60616.70976 & -6.10467 & 0.04486 & HARPS \\
2MASS~J03231246-4542056 & 60616.70984 & -6.38947 & 0.02555 & NIRPS \\
\hline
\end{tabular}
\par\smallskip
\footnotesize\textit{Notes.} The full table will be available in machine-readable form at the CDS upon journal publication.
\end{table}

\endgroup

\begingroup
\setlength{\columnsep}{30pt}
\setlength{\linenumbersep}{3pt}
\begin{multicols}{2}
\section{Thiele--Innes coefficients}
\label{app:methods}
\resetapptablenumbers

The Thiele--Innes coefficients used in the DR3 likelihood are
\begin{equation}
	\begin{aligned}
		A & =\alpha(\cos\omega_1\cos\Omega-\sin\omega_1\sin\Omega\cos i),  \\
		B & =\alpha(\cos\omega_1\sin\Omega+\sin\omega_1\cos\Omega\cos i),  \\
		F & =-\alpha(\sin\omega_1\cos\Omega+\cos\omega_1\sin\Omega\cos i), \\
		G & =-\alpha(\sin\omega_1\sin\Omega-\cos\omega_1\cos\Omega\cos i).
	\end{aligned}
\end{equation}

\resetapptablenumbers
\section{Forward modelling of the CCFs}

\subsection{Implementation}
\label{app:binary_ccf_model}

In a two-body system, the barycentric semimajor axes and RV
semi-amplitudes of the components are:
\begin{align}
 a_1&=\frac{q}{1+q}a,
 &a_2&=\frac{1}{1+q}a,\\
 K_1&=\frac{2\pi a_1\sin i}{P\sqrt{1-e^2}},
 &K_2&=\frac{2\pi a_2\sin i}{P\sqrt{1-e^2}}
       =\frac{K_1}{q},
 \label{eq:component_semiamplitudes}
\end{align}
where $q$ is a mass ratio $M_2/M_1$. At observing time
$t$, we calculate the true anomaly
$\nu$.
The component velocities are
\begin{equation}
 \begin{aligned}
 v_1(t)&=\gamma_{\rm CCF}+\eta K_1
  [\cos(\nu(t)+\omega_1)+e\cos\omega_1],\\
 v_2(t)&=\gamma_{\rm CCF}-\eta K_2
  [\cos(\nu(t)+\omega_1)+e\cos\omega_1],
 \end{aligned}
 \label{eq:component_velocities}
\end{equation}
where $\gamma_{\rm CCF}$ is a common velocity offset, and
$\eta=\pm1$ stands for the astrometric node ambiguity.\footnote{There is a $180^\circ$ ambiguity in the argument of periapsis, $\omega_1$, in astrometry alone, because the projected orbit does
not by itself distinguish the ascending from the descending node.}

To predict these velocities, we first determine the component masses from the Gaia orbit and photometry. For given Gaia DR3 companion candidate, we take its period $P$, eccentricity $e$, periastron epoch
$T_{\rm peri}$, parallax, and Thiele-Innes coefficients from its astrometric fit and derive the
photocentre semimajor axis $a_{\text{phot}}$, inclination $i$, and
argument of periastron $\omega_{1}$. We obtain the apparent
$K_s$ magnitude from 2MASS \citep{Skrutskie2006}, using the
Gaia DR3--2MASS cross-match, and convert it to an absolute
magnitude using the DR3 parallax.

If the secondary is a dark companion, the photocentre coincides with
the luminous primary. We assign the $K_s$
light to the primary and infer its mass using the
empirical mass--absolute-$K_s$-magnitude relation
of \citet{Mann2019}.
With the primary mass fixed and $\beta_G=0$, we vary
$q$. At each trial $q$, the
component masses and DR3 period determine $a$
through Kepler's third law. Equation~\eqref{a0mod}
then gives the predicted photocentre semimajor axis
$a_{\text{phot},\rm mod}(q)$, where ``mod'' denotes the model prediction. Holding the eccentricity, time of periastron, orientation, period and parallax fixed, we calculate
the model Thiele--Innes coefficients as
\begin{equation}
 \begin{aligned}
 (A,B,F,G)_{\rm mod}(q)
 &=s(q)\,(A,B,F,G)_{\rm DR3},\\
 s(q)&=\frac{a_{\text{phot},\rm mod}(q)}{a_{\text{phot}}}.
 \end{aligned}
 \label{eq:thiele_innes_scaling}
\end{equation}
We determine $q$ by minimizing
\begin{equation}
 \begin{aligned}
 \chi^2_{\rm G}(q)
 ={}&[\boldsymbol{y}_{\rm mod}(q)-\boldsymbol{y}_{\rm DR3}]^{\mathsf T}
 \\ &\quad\times\boldsymbol{C}_{\rm Gaia}^{-1}
 [\boldsymbol{y}_{\rm mod}(q)-\boldsymbol{y}_{\rm DR3}],
 \end{aligned}
 \label{eq:gaia_candidate_selection}
\end{equation}
where $\boldsymbol{y}=(A,B,F,G,e,P,T_{\rm peri},\varpi)^{\mathsf T}$
and $\boldsymbol{C}_{\rm Gaia}$ is the candidate's DR3 covariance matrix.
The selected masses determine the primary's
RV semi-amplitude through
equation~\eqref{eq:component_semiamplitudes};
equation~\eqref{eq:component_velocities} then predicts
its velocity at each observing epoch.

When the secondary is a star with a flux nearly equal to that of the primary, we determine the luminous binary system's $q$ that is consistent with the Gaia astrometry of its blended photocentre. As for the system with a dark companion, this is done by minimizing equation~\eqref{eq:gaia_candidate_selection}. Here, however, both $M_1(q)$ and $\beta_G(q)$ vary
with the trial mass ratio, because both binary components contribute to the $K_s$ magnitude. At each trial mass ratio $q=M_2/M_1$, we solve for $M_1$ and $M_2=qM_1$ such that
\begin{equation}
	10^{-0.4M_{K_s,\mathrm{tot}}}
	=
	10^{-0.4M_{K_s}(M_1)}
	+
	10^{-0.4M_{K_s}(qM_1)},
\end{equation}
where $M_{K_s,\mathrm{tot}}$ is the observed combined
absolute magnitude and $M_{K_s}(M)$ is the absolute
magnitude implied by the relation of \citet{Mann2019}
for a star of mass $M$. For each trial component mass $M_j$, we obtain the effective
temperature and radius $R_j$ from the solar-metallicity,
5-Gyr isochrone of \citet{Baraffe2015},
and calculate the surface gravity $g_j=GM_j/R_j^2$.

\end{multicols}
\clearpage
\begin{figure*}[p]
 \centering
 \includegraphics[width=\textwidth]{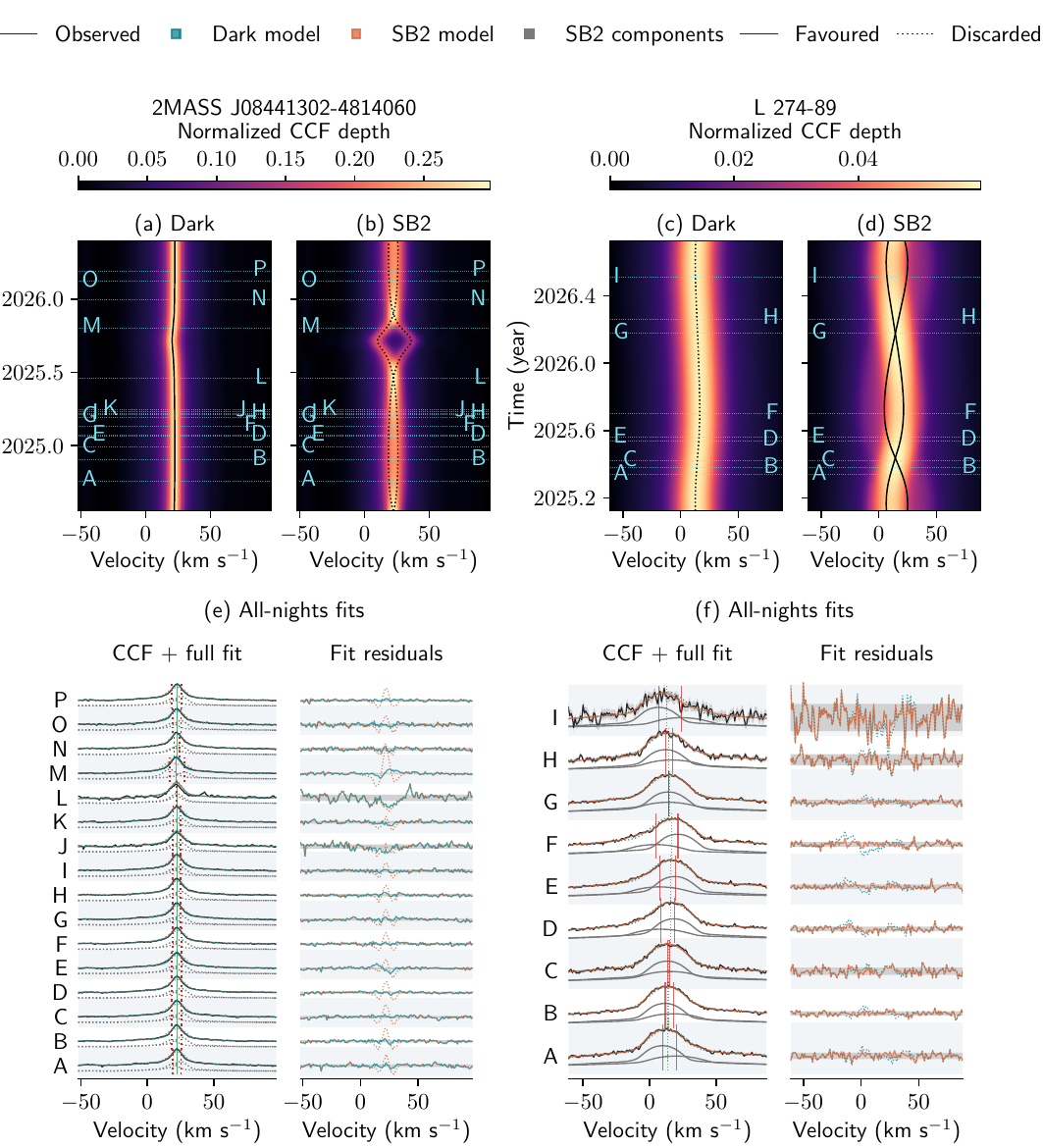}
 \caption{CCF evolution for
 2MASS~J08441302-4814060 (left) and L~274-89 (right). (a--d) Model
 dark-companion and luminous-binary normalized CCF depth heatmaps; cyan letters
 mark observing epochs and match the
 rows below, while black curves trace the predicted velocities.
 (e--f) Observed CCFs (black), model dark (teal) and
 SB2 (orange) fits, and residuals. 
 Residuals are magnified by 2.5; grey curves show the SB2 components,
 offset for clarity. Solid curves denote the favoured scenario
 (dark for 2MASS~J08441302-4814060, SB2 for L~274-89), dotted curves the alternative.
 Green and red row markers identify the velocity center of the star--dark-companion model and the two SB2 velocity
 centers, respectively, using the same line-style convention. Grey bands show the $\pm1\sigma$ CCF uncertainties.}
 \label{fig:ccf_all_epochs}
\end{figure*}
\clearpage
\begin{figure*}[p]
    \centering
    \includegraphics[width=\textwidth]{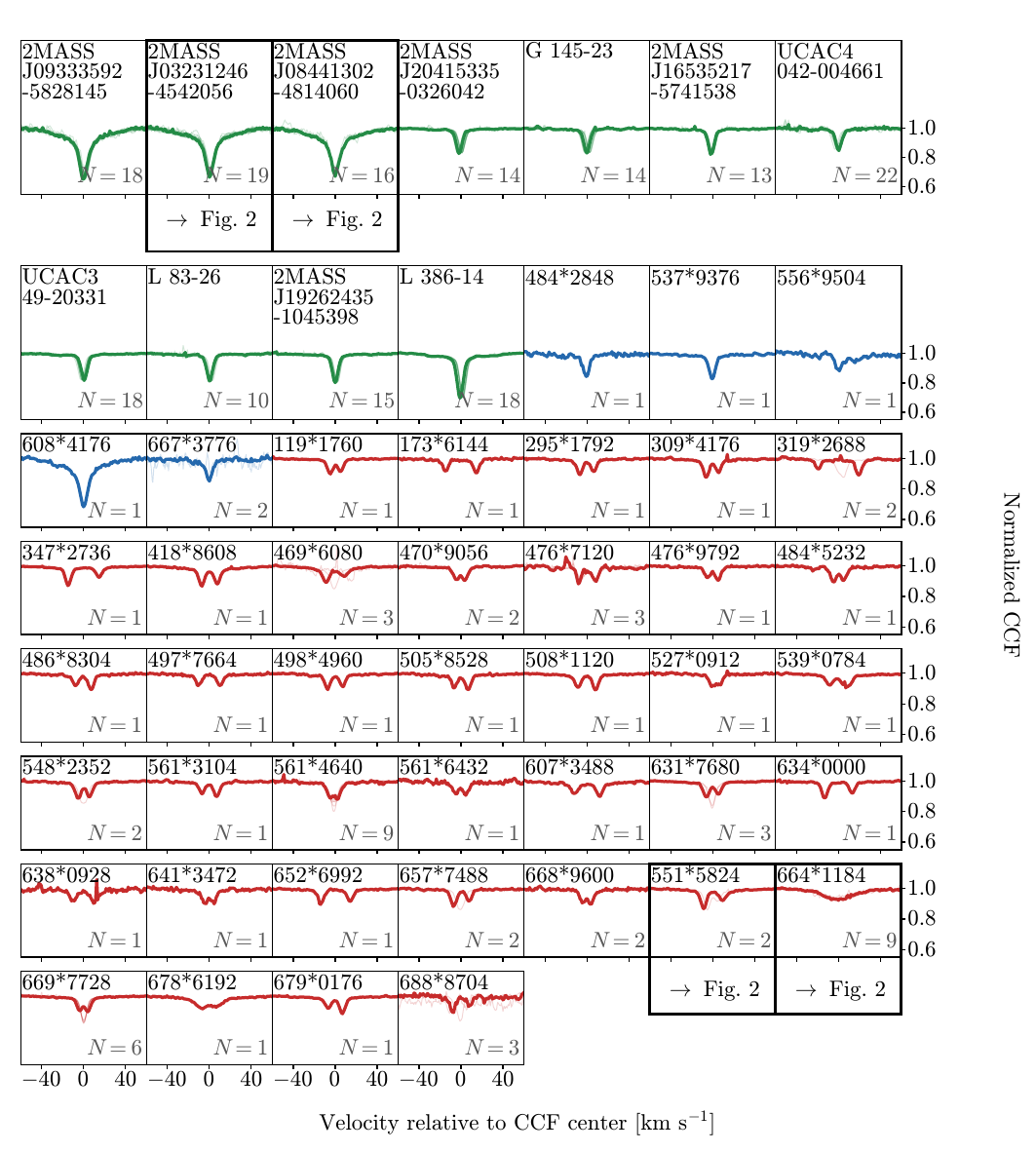}
    \caption{Nightly NIRPS CCFs. Profiles are normalized and shown against
    velocity relative to the CCF center. Colors denote Single (green), SB2
    (red), and Inconclusive (blue). The eleven systems with adopted orbital solutions
    are labelled with their names from Table~\ref{tab:nirps_observations}; the remaining panels use compacted
    Gaia DR3 host-source identifiers. The label $N$ gives the number of
    displayed nightly CCF stacks for each target.}
    \label{fig:ccf_classification_grid}
\end{figure*}
\clearpage
\begin{multicols}{2}
These properties determine the PHOENIX
spectra with flux $F_{\lambda,j}$ \citep{Husser2013}.
The Gaia-band fluxes are then
\begin{equation}
	F_{G,j}\propto R_j^2
	\int F_{\lambda,j}\,
	\lambda T_G(\lambda)\,d\lambda,
\end{equation}
where $j=1,2$ labels the binary components and $T_G(\lambda)$
is the Gaia $G$-band response \citep{Riello2021}.
The secondary light fraction $\beta_G$ follows from
equation~\eqref{lightfrac}.

Its component masses determine both RV
semi-amplitudes through
equation~\eqref{eq:component_semiamplitudes}, and
equation~\eqref{eq:component_velocities} gives the
oppositely directed orbital velocities at each
observing epoch.

We construct synthetic CCFs $\mathcal{T}^{\rm PHX}_j$ from PHOENIX spectra
\citep{Husser2013}, using the correlation masks used by the NIRPS pipeline.
For the stellar binary scenario, we use the atmospheric
parameters determined above for both components. For the star--dark-companion
model, we obtain the primary's temperature
and radius from the same isochrone at its inferred mass,
and calculate $g_1=GM_1/R_1^2$.
We adopt solar metallicity for all atmospheric templates
and hold their parameters fixed during the CCF fits.
The star--dark-companion
model contains one stellar spectrum ($N_\star=1$). For the luminous binary ($N_\star=2$), the component spectra are scaled by their squared radii, and each component's line depths are expressed relative to the total continuum light of both stars. We shift each component CCF according to
its predicted RV $v_j(t_k)$ at each observing night $k$.

Having constructed the star--dark-companion and stellar binary CCF models, we compare them to NIRPS observations. Repeated exposures from one observing night are combined. For each luminous component, we fit the projected rotational
velocity $v_{\mathrm{rot},j}$ and an additional Gaussian velocity
width $\sigma_j$, together with one common velocity offset $\gamma_{\rm CCF}$
and one overall CCF amplitude $\mathcal{A}$ to the NIRPS CCFs. These four global parameters for
the star--dark-companion model, or six for the luminous binary, are shared across
all observing nights; only a linear background, with a separate
intercept $b_{0,k}$ and slope $b_{1,k}$, varies nightly.  For each scenario, we fit substituting both $\eta=+1$ and $\eta=-1$ in eq. \eqref{eq:component_velocities} and retain the fit with the smaller CCF residual sum of squares. We assume a Gaussian instrumental profile with resolving power $R=70{,}000$ and
adopt a fixed linear limb-darkening coefficient $u=0.35$
for rotational broadening.

Thus, the model for nightly epoch $k$ is
\begin{equation}
 \begin{aligned}
 C_{k,\mathrm{mod}}(v)={}&b_{0,k}+b_{1,k}v\\
 &+\mathcal{A}\sum_{j=1}^{N_\star}
 \left[\mathcal{T}^{\rm PHX}_j*\mathcal{R}_j*
 \mathcal{G}_{s_j}\right]\!\left(v-v_j(t_k)\right),\\
 s_j^2={}&\sigma_j^2+\sigma_{\rm inst}^2,\qquad
 \sigma_{\rm inst}=\frac{c}{2\sqrt{2\ln2}\,R},
 \end{aligned}
 \label{eq:synthetic_ccf_profile}
\end{equation}
where $c$ is the speed of light. The symbol $*$ denotes convolution in velocity,
$\mathcal{R}_j$ is a rotational kernel with fitted
$v_{\mathrm{rot},j}$ and fixed limb darkening, and
$\mathcal{G}_{s_j}$ is a Gaussian with standard deviation $s_j$.

\subsection{All-epoch predictions and leave-one-night-out validation}
\label{app:loo}

For 2MASS~J08441302-4814060 and L~274-89, we also test whether each model can reproduce a night's CCF without using that night to constrain its parameters. In fold $k$, the other 15 or
eight nightly profiles, respectively, constrain each star's
projected rotation and additional Gaussian width, the common
velocity offset, and the overall CCF amplitude in
equation~\eqref{eq:synthetic_ccf_profile}. We also refit the
training nights' linear backgrounds and select between the two
astrometric node choices independently under each hypothesis.

We then predict the omitted CCF,
fitting only its linear continuum offset and slope with inverse-variance
weights, separately under each scenario. For each model we compute an
unweighted RMS in normalized CCF depth over the same 150 velocity bins
within $75\,\mathrm{km\,s^{-1}}$ of the median DRS velocity across the observation nights.
The per-night diagnostic is
\begin{equation}
 R_k=\frac{\mathrm{RMS}[C_k-C_{\mathrm{SB2},-k}]}
 {\mathrm{RMS}[C_k-C_{\mathrm{dark},-k}]},
 \label{eq:loeo_ccf_rms}
\end{equation}
where $C_k$ is the observed CCF for night $k$, and
$C_{\mathrm{SB2},-k}$ and $C_{\mathrm{dark},-k}$ are the
predicted CCFs for that night under the luminous-binary
and star--dark-companion models, respectively. The
subscript $-k$ indicates that night $k$ was excluded
when fitting the shared parameters; $R_k<1$ favours
the luminous-binary model.

For validation, each omitted night yields
one RMS ratio $R_k$.

The star--dark-companion model predicts 2MASS~J08441302-4814060 better in all 16 folds,
with median $R_k=1.9$ and range $1.2$--$6.8$.
For L~274-89, the luminous-binary model predicts all nine omitted
nights better, with median $R_k=0.8$ and range
$0.6$--$0.9$.

The two
illustrative nightly profiles per target in Figure~\ref{fig:l27489_binary_ccf}
are selected only for display: the global fits use 19, 16, nine, and
two nightly profiles of 2MASS~J03231246-4542056, 2MASS~J08441302-4814060, L~274-89, and UCAC4~220-018903,
respectively. Figure~\ref{fig:ccf_all_epochs} shows all the nightly
CCF profiles and model-data residuals for 2MASS~J08441302-4814060 and L~274-89.

\section{Methods of the population analysis}
\label{app:population}
\resetapptablenumbers

\subsection{Binning for \texorpdfstring{$f_{\rm u}$}{f\_u}}
\label{app:funresolved}

We group the stars of our sample into $0.5$-pc distance bins.
For each distance bin, we place the 38 Winters reference systems at the bin's central distance and convert each pair's measured projected physical separation into an angular separation.
We use the Gaia EDR3 probability of resolving two stars as separate sources, as a function of angular separation \citep[Fig.~8]{Fabricius+2021}, to predict whether Gaia records each Winters pair as one blended entry or two separate entries.

We calculate the fraction of blended entries in the $0.4$--$0.6\,M_{\odot}$ mass range for each distance bin.
Subsequently, for each star $i$ of our sample,
we use the fraction calculated for its distance bin
as $f_{\mathrm{u},i}$.

\subsection{Scanning-law check of the completeness}
\label{app:scanlaw}

The dependence of completeness on ecliptic latitude in Figure~\ref{fig:rayleigh_completeness}(f) is consistent with the Gaia scanning law, which provides more observations near $|\beta|\simeq45^\circ$ and more uniform scan-angle coverage towards the ecliptic poles \citep{Holl+2023}.

\end{multicols}
\clearpage
\begin{multicols}{2}
\subsection{Fitting the logistic classification model}
\label{app:logistic_classifier}

For each of the 32 follow-up/training systems indexed by $j$, the observed follow-up label was $y_j=1$ for a stellar binary false positive (23 systems) and $y_j=0$ for a dark companion (nine systems). The model predicts the false-positive probability $p_{\mathrm{SB2},j}$ from Equation~\ref{eq:logistic_companion_probability}. We standardized the two predictors using their means and standard deviations in the training sample,
\begin{equation}
    z_{M,j}=\frac{f_{M,j}-\overline{f_M}}{s_{f_M}},
    \qquad
    z_{E,j}=\frac{E_{\mathrm{Way},j}-\overline{E_{\mathrm{Way}}}}{s_{E_{\mathrm{Way}}}}.
\end{equation}
Here, $\overline{f_M}$ and $\overline{E_{\mathrm{Way}}}$ are the training-sample means, and $s_{f_M}$ and $s_{E_{\mathrm{Way}}}$ are the corresponding standard deviations.

Assuming classifications conditional on the model coefficients $(\alpha,\beta_M,\beta_E)$, we used the Bernoulli likelihood
\begin{equation}
    \mathcal{L}(\alpha,\beta_M,\beta_E)=
    \prod_{j=1}^{32}
    p_{\mathrm{SB2},j}^{y_j}
    (1-p_{\mathrm{SB2},j})^{1-y_j}.
    \label{eq:classifier_bernoulli_likelihood}
\end{equation}
Each observed false positive therefore contributes $p_{\mathrm{SB2},j}$ to the likelihood, and each observed dark companion contributes $1-p_{\mathrm{SB2},j}$.
We placed uniform priors $\mathcal{U}(-B,B)$ on the intercept and both slopes. We repeated the fit with $B=10$, $B=50$ and $B=200$ to test the sensitivity to the finite prior limits and adopted $B=200$. The model can accommodate a very steep classification boundary, and the prior limits how steep it can become. Although the coefficient posteriors remained sensitive to the prior limits, the resulting occurrence rate posteriors were similar.

We fitted $\alpha$, $\beta_M$, and $\beta_E$ by sampling their joint posterior with \texttt{emcee} \citep{ForemanMackey+2013} and used the same convergence criteria as for orbital fits (section \ref{Sampling}).

For each of the 13 candidates that Way did not publish $E_{\mathrm{Way}}$ for, we independently drew its $p_{\mathrm{SB2},i}$ from $\mathcal{U}(0,1)$ at every posterior draw in the $N_{\text{BD}}$ calculation.

\section{Three-fit corner plot comparison}
\resetapptablenumbers

\label{app:four_fit_corners}
\label{app:three_fit_corners}

Figures~\ref{fig:appendix_corner_5305690099977167232}--\ref{fig:appendix_corner_5533714311754660864} present corner plot comparisons between RV-only, Astrometry-only and joint fits for the eleven systems. The main panels show the period
$P$, mean anomaly $M_0$ at J2016, eccentricity $e$, argument of periastron
$\omega_1$, RV semi-amplitude $K_1$, photocentre semimajor axis $\alpha$,
inclination $i$, and longitude of the ascending node $\Omega$.
Separate histograms in the upper right show the fitted parallax,
NIRPS and HARPS RV jitters and Gaia covariance factor.

\end{multicols}
\endgroup
\enlargethispage{65pt}
\thispagestyle{empty}
\begingroup
\setcounter{section}{6}
\setcounter{figure}{0}
\makeatletter\def\@captype{figure}\makeatother
\centering
\makebox[\textwidth][c]{\includegraphics[width=203.9mm]{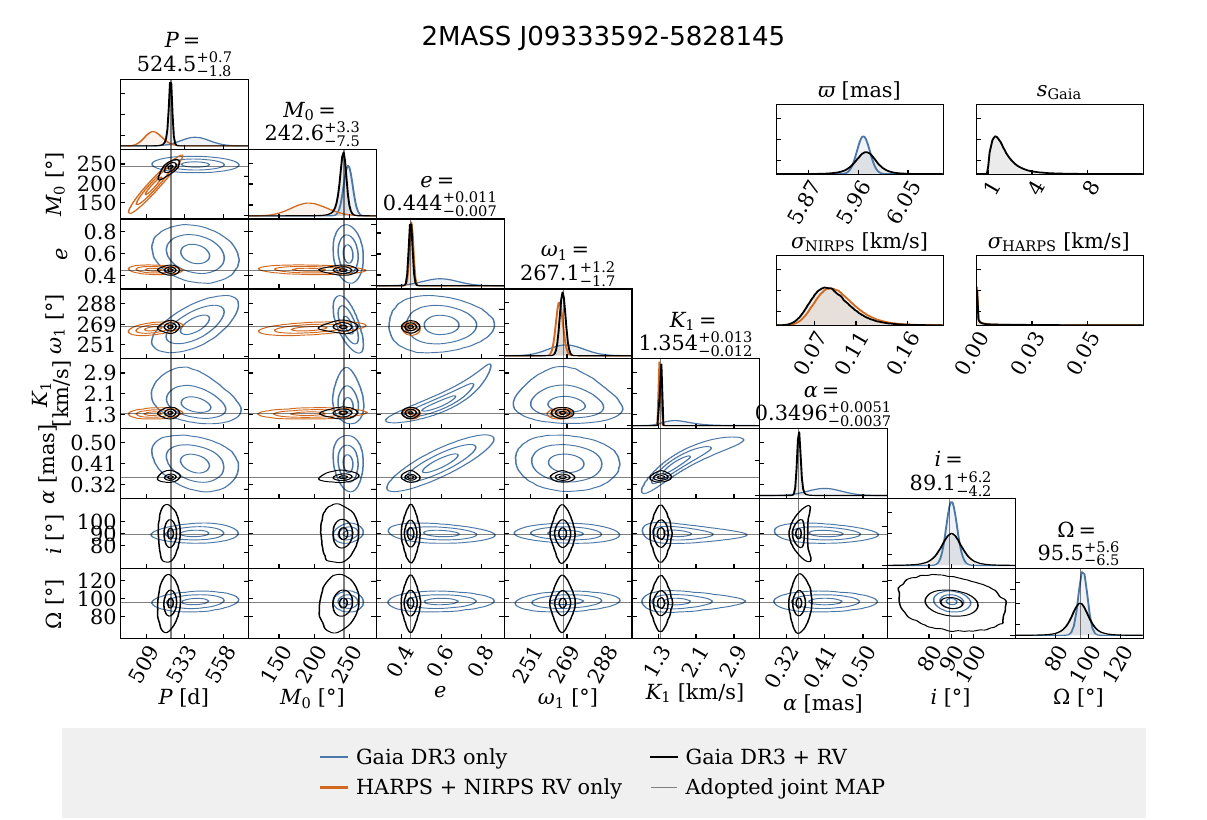}}
\caption{Distributions and contours show the Gaia-only (blue), HARPS and NIRPS RV-only (dark orange), and joint (black) fits for 2MASS~J09333592-5828145. The values above the corner diagonals are the MAP state from the joint fit. From innermost to outermost, the two-dimensional contours correspond to $1\sigma$, $2\sigma$, and $3\sigma$ levels, enclosing 39.4\%, 86.5\%, and 98.9\% probability, respectively. The RV-only fit samples $K_1$ directly; the Gaia-only and adopted joint fits sample $\alpha$ and derive $K_1$ from the astrometric orbit. For the $M_0$ posterior, we propagate each RV-only draw to J2016.0, with the period of that draw.}
\label{fig:appendix_corner_5305690099977167232}
\par
\endgroup
\twocolumn
\flushbottom
\begingroup
\setcounter{section}{6}
\setcounter{figure}{1}
\pagestyle{empty}
\begin{figure*}[p]
\vspace*{-50pt}
\centering
\makebox[\textwidth][c]{\includegraphics[width=203.9mm]{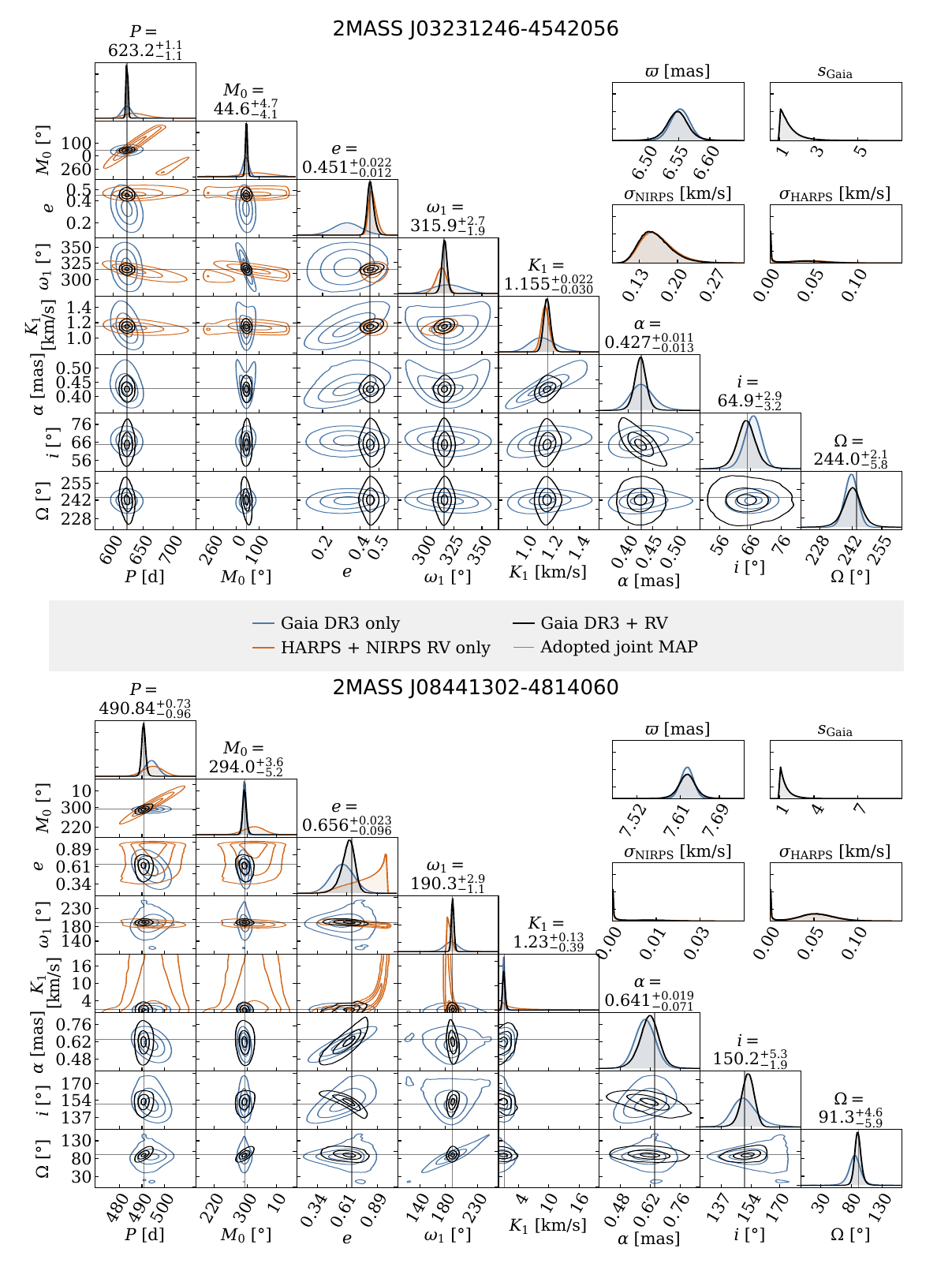}}
\caption{Same as Figure~\ref{fig:appendix_corner_5305690099977167232}, but for 2MASS~J03231246-4542056 (top) and 2MASS~J08441302-4814060 (bottom).}
\label{fig:appendix_corner_4846381143359598592}
\label{fig:appendix_corner_5329058707789307648}
\end{figure*}

\begin{figure*}[p]
\vspace*{-50pt}
\centering
\makebox[\textwidth][c]{\includegraphics[width=203.9mm]{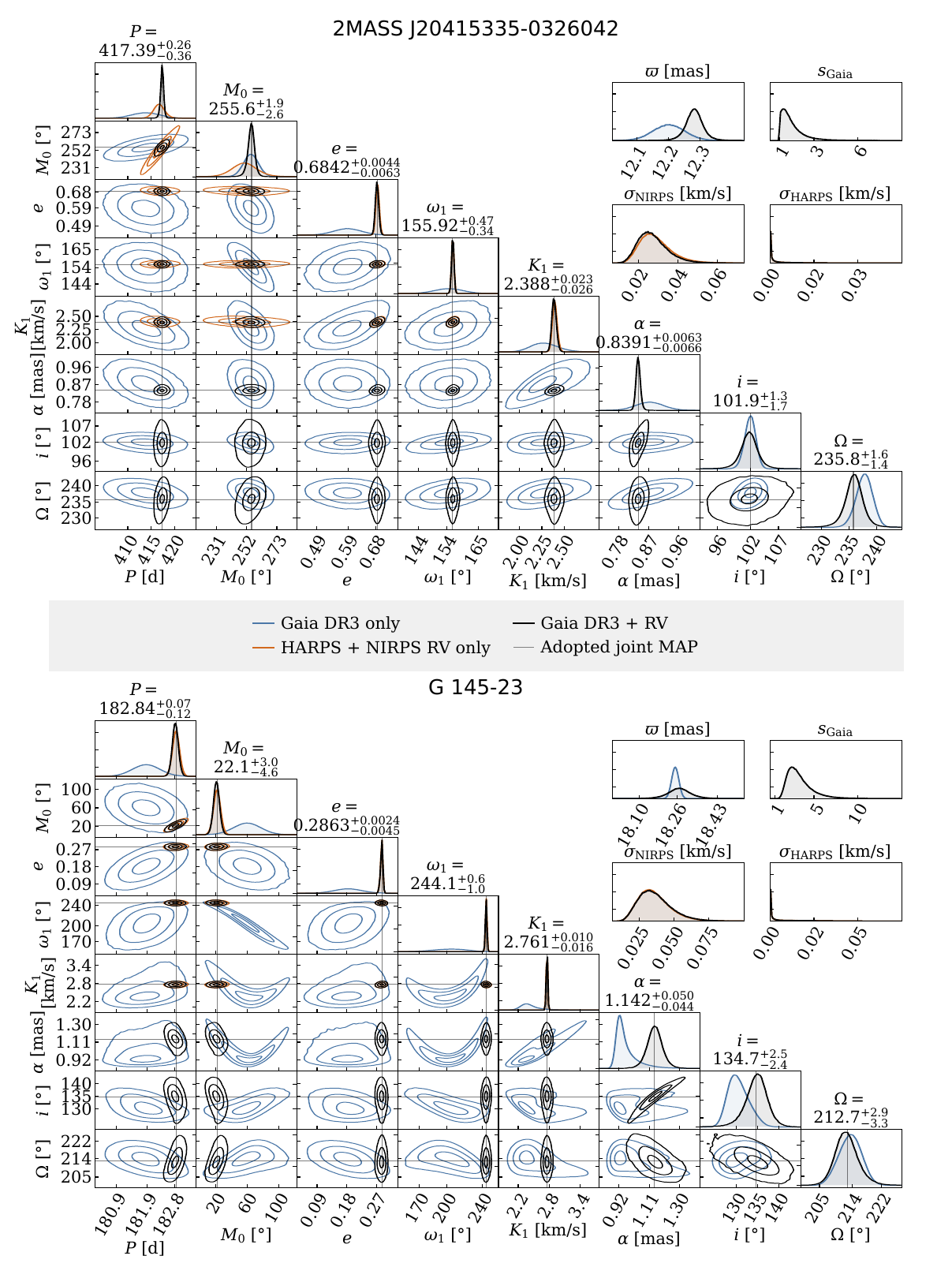}}
\caption{Same as Figure~\ref{fig:appendix_corner_5305690099977167232}, but for 2MASS~J20415335-0326042 (top) and G~145-23 (bottom).}
\label{fig:appendix_corner_4224582166524567040}
\label{fig:appendix_corner_1787984409900989824}
\end{figure*}

\begin{figure*}[p]
\vspace*{-50pt}
\centering
\makebox[\textwidth][c]{\includegraphics[width=203.9mm]{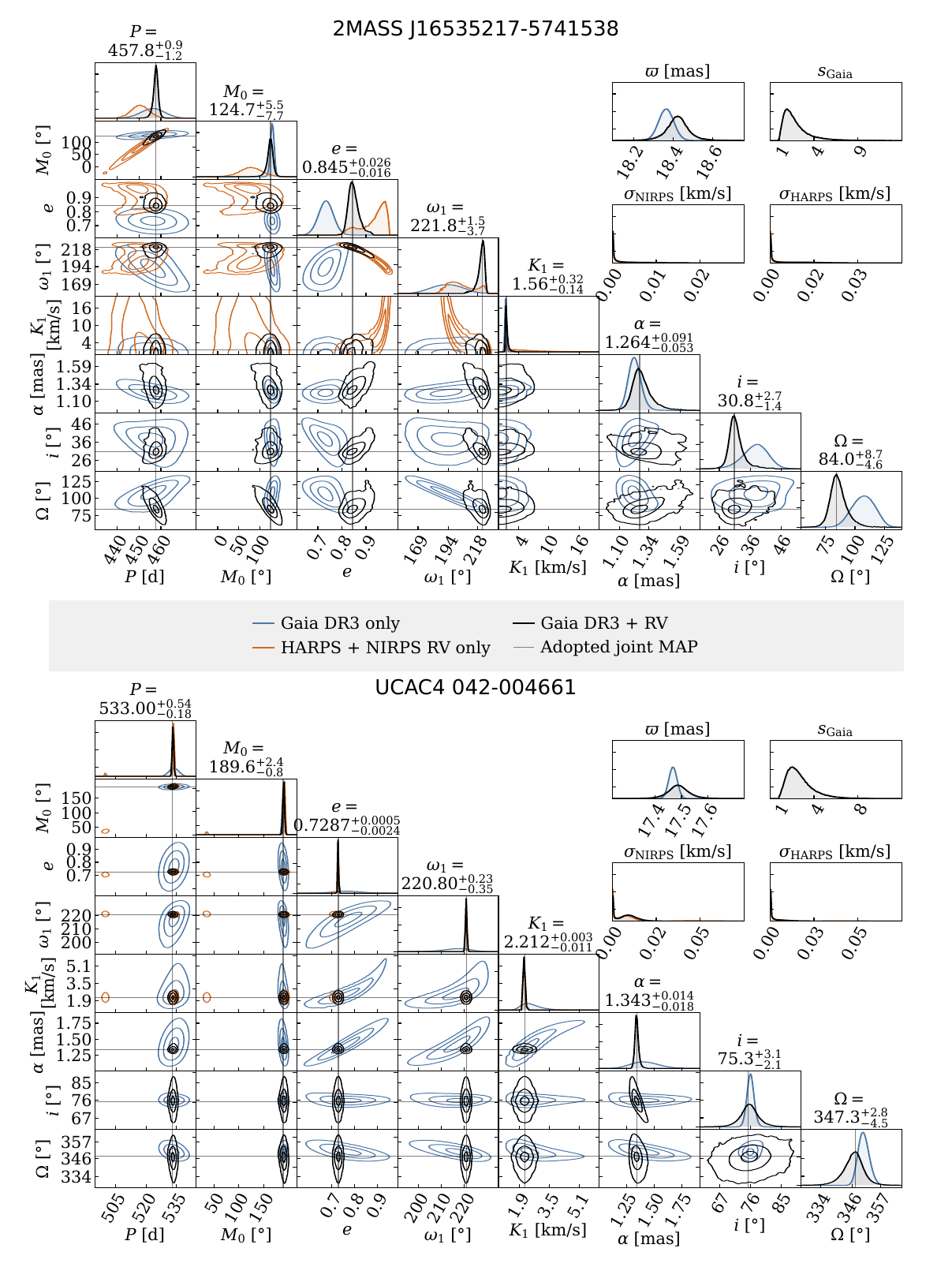}}
\caption{Same as Figure~\ref{fig:appendix_corner_5305690099977167232}, but for 2MASS~J16535217-5741538 (top) and UCAC4~042-004661 (bottom).}
\label{fig:appendix_corner_5927476672953974784}
\label{fig:appendix_corner_5196317344716068224}
\end{figure*}

\begin{figure*}[p]
\vspace*{-50pt}
\centering
\makebox[\textwidth][c]{\includegraphics[width=203.9mm]{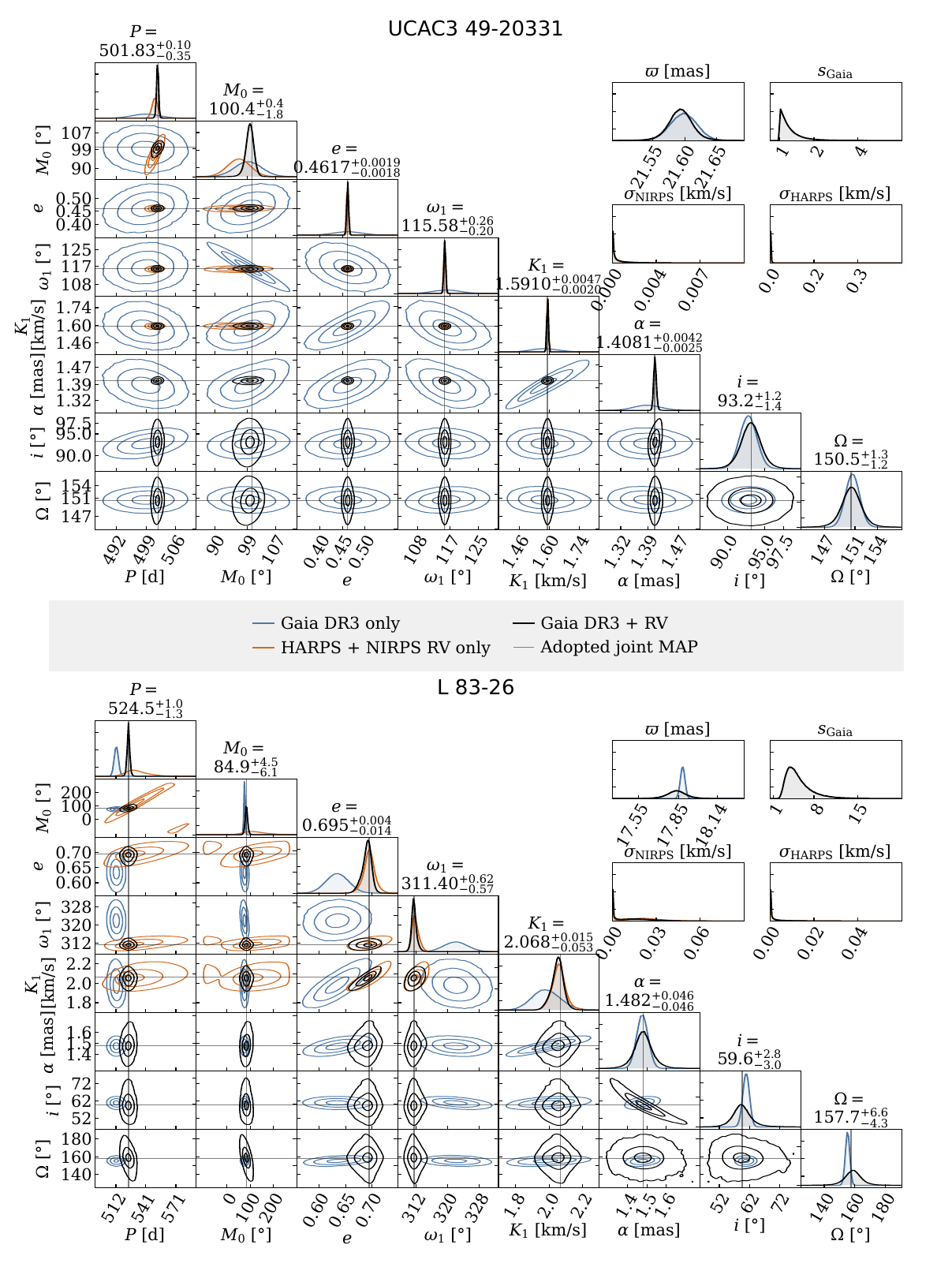}}
\caption{Same as Figure~\ref{fig:appendix_corner_5305690099977167232}, but for UCAC3~49-20331 (top) and L~83-26 (bottom).}
\label{fig:appendix_corner_5284372837240905344}
\label{fig:appendix_corner_6383984898525990272}
\end{figure*}

\begin{figure*}[p]
\vspace*{-50pt}
\centering
\makebox[\textwidth][c]{\includegraphics[width=203.9mm]{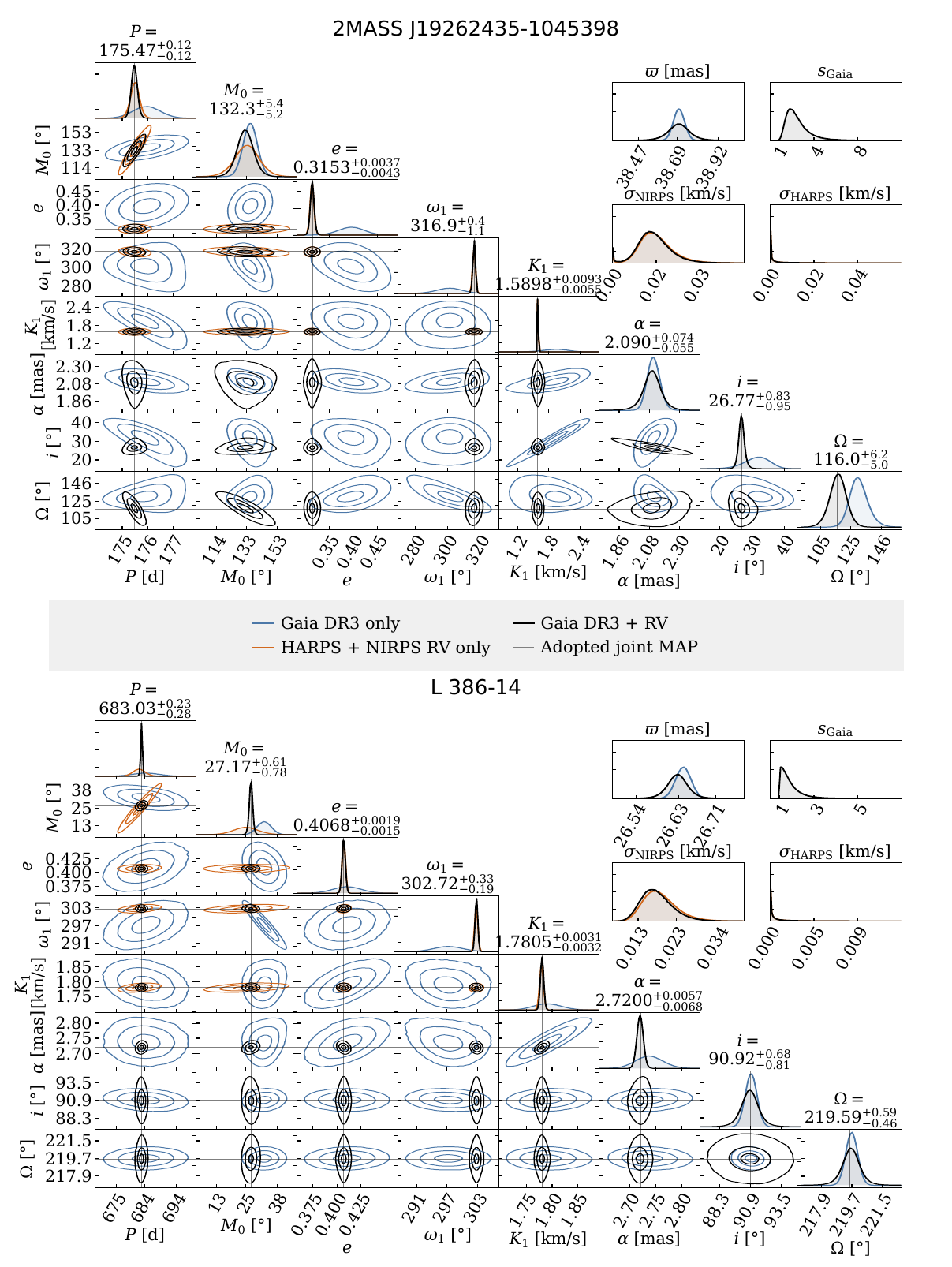}}
\caption{Same as Figure~\ref{fig:appendix_corner_5305690099977167232}, but for 2MASS~J19262435-1045398 (top) and L~386-14 (bottom).}
\label{fig:appendix_corner_4188152322648398592}
\label{fig:appendix_corner_5533714311754660864}
\end{figure*}
\clearpage
\pagestyle{fancy}
\endgroup


\begin{thebibliography}{62}
\expandafter\ifx\csname natexlab\endcsname\relax\def\natexlab#1{#1}\fi

\bibitem[{{Artigau} {et~al.}(2022){Artigau}, {Cadieux}, {Cook}, {Doyon},
  {Vandal}, {Donati}, {et~al.}}]{Artigau2022}
{Artigau}, {\'E}., {Cadieux}, C., {Cook}, N.~J., {et~al.} 2022, \aj, 164, 84

\bibitem[{{Astropy Collaboration} {et~al.}(2022){Astropy Collaboration},
  {Price-Whelan}, {Lim}, {et~al.}}]{Astropy2022}
{Astropy Collaboration}, {Price-Whelan}, A.~M., {Lim}, P.~L., {et~al.} 2022,
  \apj, 935, 167

\bibitem[{{Baraffe} {et~al.}(2015){Baraffe}, {Homeier}, {Allard}, \&
  {Chabrier}}]{Baraffe2015}
{Baraffe}, I., {Homeier}, D., {Allard}, F., \& {Chabrier}, G. 2015, \aap, 577,
  A42

\bibitem[{{Barbato} {et~al.}(2026){Barbato}, {Pinamonti}, {Sozzetti},
  {Desidera}, {D'Orazi}, {Maldonado}, {et~al.}}]{Barbato2026}
{Barbato}, D., {Pinamonti}, M., {Sozzetti}, A., {et~al.} 2026, \aap, 712, A213

\bibitem[{{Bouchy} {et~al.}(2017){Bouchy}, {Doyon}, {Artigau}, {Melo},
  {Hernandez}, {Wildi}, {et~al.}}]{Bouchy+2017}
{Bouchy}, F., {Doyon}, R., {Artigau}, {\'E}., {et~al.} 2017, The Messenger,
  169, 21

\bibitem[{{Bouchy} {et~al.}(2025){Bouchy}, {Doyon}, {Pepe}, {Melo}, {Artigau},
  {Malo}, {et~al.}}]{Bouchy+2025}
{Bouchy}, F., {Doyon}, R., {Pepe}, F., {et~al.} 2025, \aap, 700, A10

\bibitem[{Chevalier {et~al.}(2023)Chevalier, Babusiaux, Merle, \&
  Arenou}]{Chevalier2023}
Chevalier, S., Babusiaux, C., Merle, T., \& Arenou, F. 2023, Astronomy \&
  Astrophysics, 678, A19

\bibitem[{{Christiansen} {et~al.}(2025)}]{Christiansen2025Archive}
{Christiansen}, J.~L. {et~al.} 2025, The Planetary Science Journal, 6, 186

\bibitem[{{Chulkov} \& {Malkov}(2022)}]{ChulkovMalkov2022}
{Chulkov}, D. \& {Malkov}, O. 2022, Monthly Notices of the Royal Astronomical
  Society, 517, 2925

\bibitem[{{El-Badry} {et~al.}(2024){El-Badry}, {Lam}, {Holl}, {Halbwachs},
  {Rix}, {Mazeh}, \& {Shahaf}}]{ElBadry2024GaiaMock}
{El-Badry}, K., {Lam}, C., {Holl}, B., {et~al.} 2024, The Open Journal of
  Astrophysics, 7, 100

\bibitem[{{Fabricius} {et~al.}(2021){Fabricius}, {Luri}, {Arenou},
  {et~al.}}]{Fabricius+2021}
{Fabricius}, C., {Luri}, X., {Arenou}, F., {et~al.} 2021, \aap, 649, A5

\bibitem[{{Foreman-Mackey}(2016)}]{ForemanMackey2016corner}
{Foreman-Mackey}, D. 2016, Journal of Open Source Software, 1, 24

\bibitem[{{Foreman-Mackey} {et~al.}(2013){Foreman-Mackey}, {Hogg}, {Lang}, \&
  {Goodman}}]{ForemanMackey+2013}
{Foreman-Mackey}, D., {Hogg}, D.~W., {Lang}, D., \& {Goodman}, J. 2013, \pasp,
  125, 306

\bibitem[{{Gaia Collaboration} {et~al.}(2023{\natexlab{a}}){Gaia
  Collaboration}, {Arenou}, {Babusiaux}, {Barstow}, {Faigler}, {Jorissen},
  {Kervella}, {Mazeh}, {Mowlavi}, {Panuzzo}, {Sahlmann}, {Shahaf}, {Sozzetti},
  {Bauchet}, {Damerdji}, {Gavras}, {Giacobbe}, {Gosset}, {Halbwachs}, {Holl},
  {Lattanzi}, {Leclerc}, {Morel}, {Pourbaix}, {Re Fiorentin}, {Sadowski},
  {S{\'e}gransan}, {Siopis}, {Teyssier}, {Zwitter}, {Planquart}, {Brown},
  {Vallenari}, {Prusti}, {de Bruijne}, {Biermann}, {Creevey}, {Ducourant},
  {Evans}, {Eyer}, {Guerra}, {Hutton}, {Jordi}, {Klioner}, {Lammers},
  {Lindegren}, {Luri}, {Mignard}, {Panem}, {Randich}, {Sartoretti}, {Soubiran},
  {Tanga}, {Walton}, {Bailer-Jones}, {Bastian}, {Drimmel}, {Jansen}, {Katz},
  {van Leeuwen}, {Bakker}, {Cacciari}, {Casta{\~n}eda}, {De Angeli},
  {Fabricius}, {Fouesneau}, {Fr{\'e}mat}, {Galluccio}, {Guerrier}, {Heiter},
  {Masana}, {Messineo}, {Nicolas}, {Nienartowicz}, {Pailler}, {Riclet}, {Roux},
  {Seabroke}, {Sordo}, {Th{\'e}venin}, {Gracia-Abril}, {Portell}, {Altmann},
  {Andrae}, {Audard}, {Bellas-Velidis}, {Benson}, {Berthier}, {Blomme},
  {Burgess}, {Busonero}, {Busso}, {C{\'a}novas}, {Carry}, {Cellino}, {Cheek},
  {Clementini}, {Davidson}, {de Teodoro}, {Nu{\~n}ez Campos}, {Delchambre},
  {Dell'Oro}, {Esquej}, {Fern{\'a}ndez-Hern{\'a}ndez}, {Fraile}, {Garabato},
  {Garc{\'\i}a-Lario}, {Haigron}, {Hambly}, {Harrison}, {Hern{\'a}ndez},
  {Hestroffer}, {Hodgkin}, {Jan{\ss}en}, {Jevardat de Fombelle}, {Jordan},
  {Krone-Martins}, {Lanzafame}, {L{\"o}ffler}, {Marchal}, {Marrese},
  {Moitinho}, {Muinonen}, {Osborne}, {Pancino}, {Pauwels}, {Recio-Blanco},
  {Reyl{\'e}}, {Riello}, {Rimoldini}, {Roegiers}, {Rybizki}, {Sarro}, {Smith},
  {Utrilla}, {van Leeuwen}, {Abbas}, {{\'A}brah{\'a}m}, {Abreu Aramburu},
  {Aerts}, {Aguado}, {Ajaj}, {Aldea-Montero}, {Altavilla}, {{\'A}lvarez},
  {Alves}, {Anders}, {Anderson}, {Anglada Varela}, {Antoja}, {Baines}, {Baker},
  {Balaguer-N{\'u}{\~n}ez}, {Balbinot}, {Balog}, {Barache}, {Barbato},
  {Barros}, {Bartolom{\'e}}, {Bassilana}, {Becciani}, {Bellazzini},
  {Berihuete}, {Bernet}, {Bertone}, {Bianchi}, {Binnenfeld}, {Blanco-Cuaresma},
  {Blazere}, {Boch}, {Bombrun}, {Bossini}, {Bouquillon}, {Bragaglia},
  {Bramante}, {Breedt}, {Bressan}, {Brouillet}, {Brugaletta}, {Bucciarelli},
  {Burlacu}, {Butkevich}, {Buzzi}, {Caffau}, {Cancelliere}, {Cantat-Gaudin},
  {Carballo}, {Carlucci}, {Carnerero}, {Carrasco}, {Casamiquela}, {Castellani},
  {Castro-Ginard}, {Chaoul}, {Charlot}, {Chemin}, {Chiaramida}, {Chiavassa},
  {Chornay}, \& {Comoretto}}]{GaiaCollaborationArenou+2023}
{Gaia Collaboration}, {Arenou}, F., {Babusiaux}, C., {et~al.}
  2023{\natexlab{a}}, \aap, 674, A34

\bibitem[{{Gaia Collaboration} {et~al.}(2016){Gaia Collaboration}, {Prusti},
  {de Bruijne}, {et~al.}}]{GaiaMission2016}
{Gaia Collaboration}, {Prusti}, T., {de Bruijne}, J. H.~J., {et~al.} 2016,
  \aap, 595, A1

\bibitem[{{Gaia Collaboration} {et~al.}(2021){Gaia Collaboration}, {Smart},
  {Sarro}, {et~al.}}]{GaiaCollaborationSmart+2021}
{Gaia Collaboration}, {Smart}, R.~L., {Sarro}, L.~M., {et~al.} 2021, \aap, 649,
  A6

\bibitem[{{Gaia Collaboration} {et~al.}(2023{\natexlab{b}}){Gaia
  Collaboration}, {Vallenari}, {Brown}, {Prusti}, {de Bruijne}, {Arenou},
  {et~al.}}]{GaiaDR3Summary2023}
{Gaia Collaboration}, {Vallenari}, A., {Brown}, A. G.~A., {et~al.}
  2023{\natexlab{b}}, \aap, 674, A1

\bibitem[{{Giacalone} {et~al.}(2026){Giacalone}, {Howard}, {Gilbert}, {Van
  Zandt}, {Petigura}, \& {Handley}}]{Giacalone2026}
{Giacalone}, S., {Howard}, A.~W., {Gilbert}, G.~J., {et~al.} 2026, \aj, 171, 75

\bibitem[{{Gilbert} {et~al.}(2026){Gilbert}, {Van Zandt}, {Petigura},
  {Giacalone}, {Howard}, \& {Handley}}]{Gilbert2026}
{Gilbert}, G.~J., {Van Zandt}, J., {Petigura}, E.~A., {et~al.} 2026, \aj, 171,
  67

\bibitem[{{Grether} \& {Lineweaver}(2006)}]{GretherLineweaver2006}
{Grether}, D. \& {Lineweaver}, C.~H. 2006, \apj, 640, 1051

\bibitem[{{Halbwachs} {et~al.}(2023){Halbwachs}, {Pourbaix}, {Arenou},
  {Galluccio}, {Guillout}, {Bauchet}, {Marchal}, {Sadowski}, \&
  {Teyssier}}]{Halbwachs+2023}
{Halbwachs}, J.-L., {Pourbaix}, D., {Arenou}, F., {et~al.} 2023, \aap, 674, A9

\bibitem[{{Holl} {et~al.}(2023{\natexlab{a}}){Holl}, {Fabricius}, {Portell},
  {Lindegren}, {Panuzzo}, {Bernet}, {Casta{\~n}eda}, {Jevardat de Fombelle},
  {Audard}, {Ducourant}, {Harrison}, {Evans}, {Busso}, {Sozzetti}, {Gosset},
  {Arenou}, {De Angeli}, {Riello}, {Eyer}, {Rimoldini}, {Gavras}, {Mowlavi},
  {Nienartowicz}, {Lecoeur-Ta{\"\i}bi}, {Garc{\'\i}a-Lario}, \&
  {Pourbaix}}]{Holl+2023}
{Holl}, B., {Fabricius}, C., {Portell}, J., {et~al.} 2023{\natexlab{a}}, \aap,
  674, A25

\bibitem[{{Holl} {et~al.}(2023{\natexlab{b}}){Holl}, {Sozzetti}, {Sahlmann},
  {Giacobbe}, {S{\'e}gransan}, {Unger}, {et~al.}}]{HollOrbit2023}
{Holl}, B., {Sozzetti}, A., {Sahlmann}, J., {et~al.} 2023{\natexlab{b}}, \aap,
  674, A10

\bibitem[{{Husser} {et~al.}(2013){Husser}, {Wende-von Berg}, {Dreizler},
  {Homeier}, {Reiners}, {Barman}, \& {Hauschildt}}]{Husser2013}
{Husser}, T.-O., {Wende-von Berg}, S., {Dreizler}, S., {et~al.} 2013, \aap,
  553, A6

\bibitem[{{Johnson} {et~al.}(2010){Johnson}, {Aller}, {Howard}, \&
  {Crepp}}]{Johnson2010}
{Johnson}, J.~A., {Aller}, K.~M., {Howard}, A.~W., \& {Crepp}, J.~R. 2010,
  \pasp, 122, 905

\bibitem[{{Kipping}(2013)}]{Kipping2013}
{Kipping}, D.~M. 2013, \mnras, 434, L51

\bibitem[{Kounkel {et~al.}(2021)Kounkel, Covey, Stassun,
  {et~al.}}]{Kounkel2021}
Kounkel, M., Covey, K.~R., Stassun, K.~G., {et~al.} 2021, The Astronomical
  Journal, 162, 184

\bibitem[{{Kumar} {et~al.}(2019){Kumar}, {Carroll}, {Hartikainen}, \&
  {Martin}}]{Kumar2019}
{Kumar}, R., {Carroll}, C., {Hartikainen}, A., \& {Martin}, O. 2019, Journal of
  Open Source Software, 4, 1143

\bibitem[{{Lamman} {et~al.}(2020){Lamman}, {Baranec}, {Berta-Thompson}, {Law},
  {Schonhut-Stasik}, {Ziegler}, {et~al.}}]{Lamman2020}
{Lamman}, C., {Baranec}, C., {Berta-Thompson}, Z.~K., {et~al.} 2020, \aj, 159,
  139

\bibitem[{{Lammers} \& {Winn}(2026)}]{LammersWinn2026}
{Lammers}, C. \& {Winn}, J.~N. 2026, \aj, 171, 18

\bibitem[{{Latham} {et~al.}(2002){Latham}, {Stefanik}, {Torres}, {Davis},
  {Mazeh}, {Carney}, {Laird}, \& {Morse}}]{Latham2002}
{Latham}, D.~W., {Stefanik}, R.~P., {Torres}, G., {et~al.} 2002, \aj, 124, 1144

\bibitem[{{Mann} {et~al.}(2019){Mann}, {Dupuy}, {Kraus}, {Gaidos}, {Ansdell},
  {Ireland}, {Rizzuto}, {Hung}, {Dittmann}, {Factor}, {Feiden}, {Martinez},
  {Ru{\'i}z-Rodr{\'i}guez}, \& {Thao}}]{Mann2019}
{Mann}, A.~W., {Dupuy}, T., {Kraus}, A.~L., {et~al.} 2019, \apj, 871, 63

\bibitem[{{Marcussen} \& {Albrecht}(2023)}]{MarcussenAlbrecht2023}
{Marcussen}, M.~L. \& {Albrecht}, S.~H. 2023, \aj, 165, 266

\bibitem[{{Marcussen} {et~al.}(2026){Marcussen}, {Albrecht}, {Kalinowski},
  {Winn}, {Stef{\'a}nsson}, {Larsen}, {Gadeberg}, {Fitzmaurice}, {Schlaufman},
  \& {Mahadevan}}]{MarcussenEtAl2026}
{Marcussen}, M.~L., {Albrecht}, S.~H., {Kalinowski}, K.~K., {et~al.} 2026,
  \aap, 715, A340

\bibitem[{{Mayor} {et~al.}(2003){Mayor}, {Pepe}, {Queloz}, {Bouchy},
  {Rupprecht}, {Lo Curto}, {Avila}, {Benz}, {Bertaux}, {Bonfils}, {Dall},
  {Dekker}, {Delabre}, {Eckert}, {Fleury}, {Gilliotte}, {Gojak}, {Guzman},
  {Kohler}, {Lizon}, {Longinotti}, {Lovis}, {M{\'e}gevand}, {Pasquini},
  {Reyes}, {Sivan}, {Sosnowska}, {Soto}, {Udry}, {van Kesteren}, {Weber}, \&
  {Weilenmann}}]{Mayor+2003}
{Mayor}, M., {Pepe}, F., {Queloz}, D., {et~al.} 2003, The Messenger, 114, 20

\bibitem[{Mignon {et~al.}(2025)Mignon, Delfosse, Meunier,
  {et~al.}}]{Mignon2025}
Mignon, L., Delfosse, X., Meunier, N., {et~al.} 2025, \aap, 700, A146

\bibitem[{{Pennell} {et~al.}(2026){Pennell}, {M{\"u}ller-Horn}, {Shahaf},
  {Rix}, {El-Badry}, \& {Li}}]{Pennell+2026}
{Pennell}, B., {M{\"u}ller-Horn}, J., {Shahaf}, S., {et~al.} 2026, arXiv
  e-prints [\eprint[arXiv]{2611.21800}]

\bibitem[{{Penoyre} {et~al.}(2020){Penoyre}, {Belokurov}, {Wyn Evans},
  {Everall}, \& {Koposov}}]{2020MNRAS.495..321P}
{Penoyre}, Z., {Belokurov}, V., {Wyn Evans}, N., {Everall}, A., \& {Koposov},
  S.~E. 2020, \mnras, 495, 321

\bibitem[{{Perryman} {et~al.}(2014){Perryman}, {Hartman}, {Bakos}, \&
  {Lindegren}}]{Perryman2014}
{Perryman}, M., {Hartman}, J., {Bakos}, G.~{\'A}., \& {Lindegren}, L. 2014,
  \apj, 797, 14

\bibitem[{{Pinamonti} {et~al.}(2026){Pinamonti}, {Sozzetti}, {Barbato},
  {Desidera}, {Biazzo}, {Bonomo}, {et~al.}}]{Pinamonti2026}
{Pinamonti}, M., {Sozzetti}, A., {Barbato}, D., {et~al.} 2026, \aap, 707, A67

\bibitem[{Ribas {et~al.}(2023)Ribas, Reiners, Zechmeister,
  {et~al.}}]{Ribas2023}
Ribas, I., Reiners, A., Zechmeister, M., {et~al.} 2023, \aap, 670, A139

\bibitem[{Riello {et~al.}(2021)Riello, De~Angeli, Evans, {et~al.}}]{Riello2021}
Riello, M., De~Angeli, F., Evans, D.~W., {et~al.} 2021, Astronomy \&
  Astrophysics, 649, A3

\bibitem[{{Sahlmann} \& {G{\'o}mez}(2025)}]{SahlmannGomez2025}
{Sahlmann}, J. \& {G{\'o}mez}, P. 2025, \mnras, 537, 1130

\bibitem[{{Santerne} {et~al.}(2016){Santerne}, {Moutou}, {Tsantaki}, {Bouchy},
  {H{\'e}brard}, {Adibekyan}, {Almenara}, {Amard}, {Barros}, {Boisse},
  {Bonomo}, {Bruno}, {Courcol}, {Deleuil}, {Demangeon}, {D{\'i}az}, {Guillot},
  {Havel}, {Montagnier}, {Rajpurohit}, {Rey}, \& {Santos}}]{Santerne2016}
{Santerne}, A., {Moutou}, C., {Tsantaki}, M., {et~al.} 2016, \aap, 587, A64

\bibitem[{{Skrutskie} {et~al.}(2006){Skrutskie}, {Cutri}, {Stiening},
  {Weinberg}, {Schneider}, {Carpenter}, {Beichman}, {Capps}, {Chester},
  {Elias}, {et~al.}}]{Skrutskie2006}
{Skrutskie}, M.~F., {Cutri}, R.~M., {Stiening}, R., {et~al.} 2006, \aj, 131,
  1163

\bibitem[{{Sozzetti} {et~al.}(2023){Sozzetti}, {Pinamonti}, {Damasso},
  {Desidera}, {Biazzo}, {Bonomo}, {et~al.}}]{Sozzetti2023}
{Sozzetti}, A., {Pinamonti}, M., {Damasso}, M., {et~al.} 2023, \aap, 677, L15

\bibitem[{{Squicciarini} \& {Bonavita}(2022)}]{SquicciariniBonavita2022}
{Squicciarini}, V. \& {Bonavita}, M. 2022, \aap, 666, A15

\bibitem[{{Stef{\'a}nsson} {et~al.}(2025){Stef{\'a}nsson}, {Mahadevan}, {Winn},
  {Marcussen}, {Kanodia}, {Albrecht}, {et~al.}}]{Stefansson2025}
{Stef{\'a}nsson}, G., {Mahadevan}, S., {Winn}, J.~N., {et~al.} 2025, The
  Astronomical Journal, 169, 107

\bibitem[{{Stevenson} {et~al.}(2023){Stevenson}, {Haswell}, {Barnes}, \&
  {Barstow}}]{Stevenson+2023}
{Stevenson}, A.~T., {Haswell}, C.~A., {Barnes}, J.~R., \& {Barstow}, J.~K.
  2023, \mnras, 526, 5155

\bibitem[{{Unger} {et~al.}(2023){Unger}, {S{\'e}gransan}, {Barbato}, {Delisle},
  {Sahlmann}, {Holl}, \& {Udry}}]{Unger2023}
{Unger}, N., {S{\'e}gransan}, D., {Barbato}, D., {et~al.} 2023, \aap, 680, A16

\bibitem[{{van de Kamp}(1975)}]{VanDeKamp1975}
{van de Kamp}, P. 1975, \araa, 13, 295

\bibitem[{{Van Zandt} {et~al.}(2026){Van Zandt}, {Gilbert}, {Petigura},
  {Giacalone}, {Howard}, \& {Handley}}]{VanZandt2026}
{Van Zandt}, J., {Gilbert}, G.~J., {Petigura}, E.~A., {et~al.} 2026, \aj, 171,
  267

\bibitem[{{Vehtari} {et~al.}(2021){Vehtari}, {Gelman}, {Simpson}, {Carpenter},
  \& {B{\"u}rkner}}]{Vehtari2021}
{Vehtari}, A., {Gelman}, A., {Simpson}, D., {Carpenter}, B., \& {B{\"u}rkner},
  P.-C. 2021, Bayesian Analysis, 16, 667

\bibitem[{{Venner} {et~al.}(2026){Venner}, {Huang}, {Latham}, {Quinn},
  {Bieryla}, {Vanderburg}, \& {Wittenmyer}}]{Venner2026}
{Venner}, A., {Huang}, C.~X., {Latham}, D.~W., {et~al.} 2026, \aj, 171, 279

\bibitem[{{Way} {et~al.}(2026){Way}, {L{\'e}pine}, {Gagn{\'e}}, \&
  {Medan}}]{Way+2026}
{Way}, Z., {L{\'e}pine}, S., {Gagn{\'e}}, J., \& {Medan}, I. 2026, The
  Astronomical Journal, 171, 252

\bibitem[{{Wenger} {et~al.}(2000){Wenger}, {Ochsenbein}, {Egret}, {Dubois},
  {Bonnarel}, {Borde}, {Genova}, {Jasniewicz}, {Lalo{\"e}}, {Lesteven}, \&
  {Monier}}]{Wenger2000}
{Wenger}, M., {Ochsenbein}, F., {Egret}, D., {et~al.} 2000, \aaps, 143, 9

\bibitem[{{Wildi} {et~al.}(2017){Wildi}, {Bouchy}, {Doyon}, {Artigau}, {Blind},
  {Reshetov}, {et~al.}}]{Wildi+2017}
{Wildi}, F., {Bouchy}, F., {Doyon}, R., {et~al.} 2017, in Society of
  Photo-Optical Instrumentation Engineers (SPIE) Conference Series, Vol. 10400,
  Techniques and Instrumentation for Detection of Exoplanets VIII, 1040018

\bibitem[{{Winn}(2022)}]{Winn2022}
{Winn}, J.~N. 2022, The Astronomical Journal, 164, 196

\bibitem[{{Winterhalder} {et~al.}(2024){Winterhalder}, {Lacour}, {M{\'e}rand},
  {et~al.}}]{Winterhalder2024}
{Winterhalder}, T.~O., {Lacour}, S., {M{\'e}rand}, A., {et~al.} 2024, \aap,
  688, A44

\bibitem[{{Winters} {et~al.}(2019){Winters}, {Henry}, {Jao}, {Subasavage},
  {Chatelain}, {Slatten}, {Riedel}, {Silverstein}, \& {Payne}}]{Winters+2019}
{Winters}, J.~G., {Henry}, T.~J., {Jao}, W.-C., {et~al.} 2019, \aj, 157, 216

\bibitem[{{Wu} {et~al.}(2025){Wu}, {Hadden}, {Dewberry}, {El-Badry}, \&
  {Matzner}}]{Wu2025Eccentricities}
{Wu}, Y., {Hadden}, S., {Dewberry}, J., {El-Badry}, K., \& {Matzner}, C.~D.
  2025, \apjl, 982, L34

\bibitem[{{Zechmeister} {et~al.}(2018){Zechmeister}, {Reiners}, {Amado},
  {Azzaro}, {Bauer}, {B{\'e}jar}, {Caballero}, {Guenther}, {Hagen}, {Jeffers},
  {Kaminski}, {K{\"u}rster}, {Launhardt}, {Montes}, {Morales}, {Quirrenbach},
  {Reffert}, {Ribas}, {Seifert}, {Tal-Or}, \&
  {Wolthoff}}]{Zechmeister2018SERVAL}
{Zechmeister}, M., {Reiners}, A., {Amado}, P.~J., {et~al.} 2018, \aap, 609, A12

\end{thebibliography}
\end{document}